\documentclass[aps,prd,twocolumn,nofootinbib,floatfix,superscriptaddress,longbibliography=false]{revtex4-2}

\usepackage[utf8]{inputenc}
\usepackage{amsmath,amssymb}
\usepackage{graphicx}
\usepackage{hyperref}
\usepackage{physics}
\usepackage{booktabs}
\usepackage{xcolor}
\usepackage{svg}
\DeclareMathAlphabet\mathbfcal{OMS}{cmsy}{b}{n}

\begin{document}

\title{Quantifying Weighted Morphological Content of Large-Scale Structures via Simulation-Based Inference}

\author{M. H. Jalali Kanafi}
\email{m_jalalikanafi@sbu.ac.ir}
\affiliation{Department of Physics, Shahid Beheshti University, Tehran 1983969411, Iran}
\affiliation{School of Physics, Institute for Research in Fundamental Sciences (IPM), P. O. Box, Tehran 19395-5531, Iran}

\author{S. M. S. Movahed}
\affiliation{Department of Physics, Shahid Beheshti University, Tehran 1983969411, Iran}
\affiliation{School of Astronomy, Institute for Research in Fundamental Sciences (IPM), P. O. Box, Tehran 19395-5531, Iran}
\affiliation{Department of Mathematics and Statistics, The University of Lahore, 1-KM Defence Road, Lahore 54000, Pakistan}
\email{m.s.movahed@ipm.ir}

\date{\today}

\begin{abstract}
In this work, we perform a simulation-based forecasting analysis to compare the cosmological constraining power of higher-order summary statistics of the large-scale structure (LSS)—the Minkowski Functionals (MFs) and a class weighted morphological measure known as the Conditional Moments of Derivatives (CMD)—with that of the redshift-space halo power spectrum multipoles (PS), with a particular focus on their sensitivity to nonlinear and anisotropic features in redshift space. Our analysis relies on halo catalogs from the Big Sobol Sequence (BSQ) simulations at redshift $z=0.5$, employing a likelihood-free inference framework implemented via neural posterior estimation.
At the fiducial Quijote cosmology $(\Omega_{\mathrm{m}}=0.3175,\,\sigma_8=0.834)$ and for a Gaussian smoothing scale of $R=15\,h^{-1}\mathrm{Mpc}$, CMD provide systematically tighter constraints than MFs. Combining MFs and CMD into a joint estimator improves the precision by $27\%^{+9\%}_{-5\%}$ for $\sigma_8$ and $26\%^{+7\%}_{-5\%}$ for $\Omega_{\mathrm{m}}$ relative to MFs alone, highlighting the complementary anisotropy‑sensitive information captured by the CMD in contrast to the scalar morphological content encapsulated by the MFs.
We compare the combined statistic MFs+CMD with the PS at matched effective scales ($k_{\max}\simeq0.16\,h\,\mathrm{Mpc^{-1}}$) under three halo-selection conditions: (i) all halos, (ii) fixed number density, and (iii) mass-selected ($M>3\times10^{13}\,h^{-1}M_\odot$). In the mass-selected configuration, the (weighted) morphological estimator outperforms the power spectrum by $45\%^{+20\%}_{-9\%}$ for $\sigma_8$ and $43\%^{+10\%}_{-7\%}$ for $\Omega_{\mathrm{m}}$.
We also extend the simulation-based forecast analysis across a continuous range of cosmological parameters and multiple smoothing scales for (weighted) morphological measures.
While the absolute uncertainties depend on parameter values and smoothing scale, the relative constraining power of the summary statistics remains nearly constant.
\end{abstract}

\maketitle

\section{Introduction}

The large-scale structure (LSS) of the Universe encodes a wealth of information about fundamental cosmological parameters and the physics governing structure formation. To extract this information, traditional inference pipelines have typically relied on compressing observational data into low-order summary statistics—most commonly the power spectrum in harmonic space or two-point correlation function in configuration space—and comparing them to analytical models within the paradigm of conventional  inference methods ~\cite{Tegmark1997, Peebles1980, Cole2005, PhysRevD.103.083533, PhysRevD.105.023520, Trotta2017}.

While this approach is computationally efficient and theoretically well-motivated in the linear regime, it suffers from two key limitations:  First, it discards a substantial portion of the available information by focusing solely on second-order correlations. However, as cosmic structures evolve under gravity and enter the nonlinear regime, their statistical properties become increasingly non-Gaussian, and key features of the matter distribution—especially on intermediate and small scales—escape capture by two-point statistics~\cite{Bernardeau2002, Scoccimarro2000}. Second, conventional (traditional) inference methods typically require explicit modeling of the likelihood function, which may be inaccurate or entirely intractable for complex summary statistics or strongly non-Gaussian observables ~\cite{Cranmer2020, Alsing2019}.
 
Mentioned challenges are further exacerbated by the advent of ongoing and next-generation galaxy surveys—such as 
	DESI\footnote{\url{https://www.desi.lbl.gov/}}~\cite{DESI2016}, 
	PFS\footnote{\url{http://pfs.ipmu.jp/}}~\cite{tamura2016prime},
	the Roman Space Telescope\footnote{\url{http://wfirst.gsfc.nasa.gov/}}~\cite{spergel2015wide, wang2022high}
	and Euclid\footnote{\url{https://www.euclid-ec.org/}}~\cite{Euclid2016}—
	which provide highly precise measurements of the LSS and matter over-density tracers.
	The unprecedented statistical power and access to mildly and fully nonlinear regimes offered by these surveys intensify the need for advanced statistical tools capable of capturing non-Gaussian signals, as well as modern inference frameworks that can robustly extract cosmological information beyond the assumptions of explicit likelihood-based methods.

In recent years, significant efforts have been focused to addressing both of these challenges. On one side, a diverse range of advanced summary statistics have been developed to encapsulate the non-Gaussian and nonlinear features of the LSS. Among these, the
higher order correlation functions such as bispectrum has emerged as a natural extension of the power spectrum and provides access to mode coupling information and non-Gaussian features~\cite{Sefusatti2007, GilMarin2016}.
The higher-order moments, which serve as the tools for extraction of  non-linear properties have also been utilized (e.g. 
\cite{fry1985cosmological,bouchet1993moments, bernardeau1993skewness,croton20042df, cappi2015vimos, sabiu2019graph, philcox2022probing}).

	Marked power spectrum statistics enhance conventional clustering observables by assigning weights to galaxies according to their local environmental characteristics, which in turn increase sensitivity to nonlinear structures~\cite{White2016, Massara_2023, Cowell2024}.
	Wavelet-based techniques have also gained traction as a multiscale strategy for identifying localized, anisotropic, or filamentary features in cosmological fields~\cite{valogiannis2022towards, eickenberg2022wavelet}. By decomposing the density field into spatial-frequency components, wavelets offer scale-dependent sensitivity to both Gaussian and non-Gaussian signals from intermediate and small-scale structures.
	An alternative real-space approach is density-split clustering, which partitions the density field into quantiles and analyzes correlations within or between these partitions~\cite{paillas2024cosmological, paillas2023constraining, cuesta2024sunbird}. 
	
	Morphological descriptors generally examine aspects such as shape, size, boundaries, and connectedness of the underlying field, which has a long-standing history~\cite{hadwiger2013vorlesungen}. Considering the excursion set associated with LSS tracers and imposing additional constraints, we can define the critical sets in the form of local maxima (peaks and clusters), local minima (troughs and voids), saddles, genus, and skeleton. The aforementioned measures encapsulate geometrical and topological components of LSS. ~\cite{matsubara2003,gay2012non,codis2013non,sousbie20083d,sousbie2009fully,sousbie2011persistent}.		
	Additionally, Minkowski Valuations and persistent homology offer a distinct class of measures for characterizing the non-Gaussian features of the LSS. Minkowski Functionals (MFs), as prototypical examples of Minkowski Valuations, quantify various aspects of isodensity surfaces, including volume, surface area, mean curvature, Euler characteristic, and have been widely used to study the nonlinear evolution and morphology of cosmic structures~\cite{mecke1993robust, schmalzing1995minkowski,matsubara1996statistics,matsubara2003, codis2013non,matsubara2022minkowski,jiang2022effects, Liu2022ProbingMN, Liu2023ProbingMN}.	
	Minkowski Tensors generalize this framework to clarify anisotropic information, providing additional sensitivity to shape and directional features in the matter distribution~\cite{appleby2018minkowski, appleby2019ensemble, Appleby2023, liu2025probing}.	
	Weighted morphology, which is an extension class of unweighted morphology, has been introduced and applied for probing the  anisotropic signals in redshift-space distortion~\cite{kanafi2024probing}, and CMB anisotropy~\cite{afzal2025cosmic}.
	Persistent homology also tracks the birth and death of topological features—such as connected components, filaments, and voids—across multiple density thresholds, and has gained considerable attention in multiscale analysis of the LSS~\cite{pranav2017topology, Xu2019, 2021MNRAS.507.2968W, biagetti2022fisher, yip2024cosmology, jalali2024imprint, abedi2024impact, prat2025dark}.		

In addition to the extension of high-order summary statistics, the development of simulation-based inference (SBI) frameworks has been achieved, which eliminates the necessity for explicit analytical likelihood modeling by relying on forward simulations. Depending on the approach, SBI methods either learn the likelihood function implicitly (e.g. via likelihood ratio estimation) or directly approximate the posterior distribution using neural density estimators~\cite{Cranmer2020, Alsing2019}. A key advantage of SBI is its flexibility to perform inference using arbitrary summary statistics— the lacking straightforward and unambiguous theoretical relation between summary statistics and cosmological parameters —making it particularly powerful when applied to complex, non-Gaussian observables with intricate or nontrivial parameter-to-data mappings. In recent years, these methods have been successfully deployed across a wide range of cosmological and astrophysical applications.

SBI has enabled precise estimation of cosmological parameters from weak gravitational lensing observations. Dedicated SBI pipelines have been developed for the KiDS-1000 cosmic shear data from the Kilo-Degree Survey~\cite{10.1093/mnras/stad2262, von2025kids}. SBI has also been applied to Year-1 data from the Hyper Suprime-Cam (HSC) survey, combined with non-Gaussian summary statistics—including MFs and counts of peaks and minima—to achieve tighter constraints on $S_8$ compared to traditional two-point correlation functions~\cite{novaes2025cosmology}. Applications to the Dark Energy Survey (DES) have extended SBI techniques by leveraging neural compression of weak-lensing map statistics to perform likelihood-free inference~\cite{jeffrey2021likelihood, Gatti2024, jeffrey2025dark}. Recent developments include optimal neural summarization and standardized benchmark pipelines for LSST-scale datasets, further underscoring SBI’s potential for high-precision inference in forthcoming lensing surveys~\cite{lanzieri2025optimal, zeghal2025simulation}.

In the context of galaxy clustering, SBI has been applied both to summary statistics and directly at the field level. The SimBIG framework exemplifies this versatility: field-level inference is performed via convolutional neural networks on galaxy density fields~\cite{PhysRevD.109.083536}, while analyses based on summary statistics employ marked power spectra~\cite{massara2024sc}, bispectra~\cite{hahn2024cosmological}, and skew spectra~\cite{Hou2024} to capture non-Gaussian information beyond two-point correlations. Field-level inference has also been explored for model selection, revealing the potential of SBI to distinguish between cosmological models including those with dynamical dark energy~\cite{mancini2024field}.
Integration of SBI with the Effective Field Theory of Large-Scale Structure (EFTofLSS) has enabled constraints on $\sigma_8$ from biased tracers using a joint analysis of power spectra and bispectra~\cite{tucci2024eftoflss}.
Further developments include systematic assessments of the sensitivity of SBI outcomes to halo-finding algorithms and halo occupation distribution (HOD) prescriptions, identifying critical factors in the forward-modeling pipeline~\cite{modi2025sensitivity}. 
Galaxy cluster abundance has also emerged as a powerful observable within the SBI framework. Applications to both synthetic and real galaxy cluster catalogs have demonstrated that accurate constraints on $\Omega_m$ and $\sigma_8$ can be attained without relying on explicit likelihood functions~\cite{reza2024constraining, zubeldia2025extracting}.

SBI has also been employed in observables where analytical modeling is challenging or unavailable. Applications to void lensing have demonstrated the ability of SBI to extract cosmological information in the absence of a tractable likelihood~\cite{su2025cosmological}. In the context of $21\,cm$ cosmology, SBI has also shown more accurate and computationally efficient inference performance compared to standard likelihood-based methods~\cite{prelogovic2023exploring}.

Recent developments have further focused on methodological aspects of SBI in cosmology. Within the \textit{Big Sobol Sequence} (BSQ) simulation suite, neural scaling laws were developed to forecast the simulation budget required for high-fidelity inference~\cite{bairagi2025many}.
The \textit{cosmoSWAG} framework, based on Bayesian neural networks, improved posterior calibration and robustness to simulation mismatch~\cite{lemos2023robust}. The impact of observational systematics—including redshift uncertainties and survey masks—on field-level SBI has also been quantified, demonstrating its viability for use with realistic galaxy catalogs~\cite{de2025field}.

\begin{figure*}[t]
	\centering
	\includegraphics[scale=0.6]{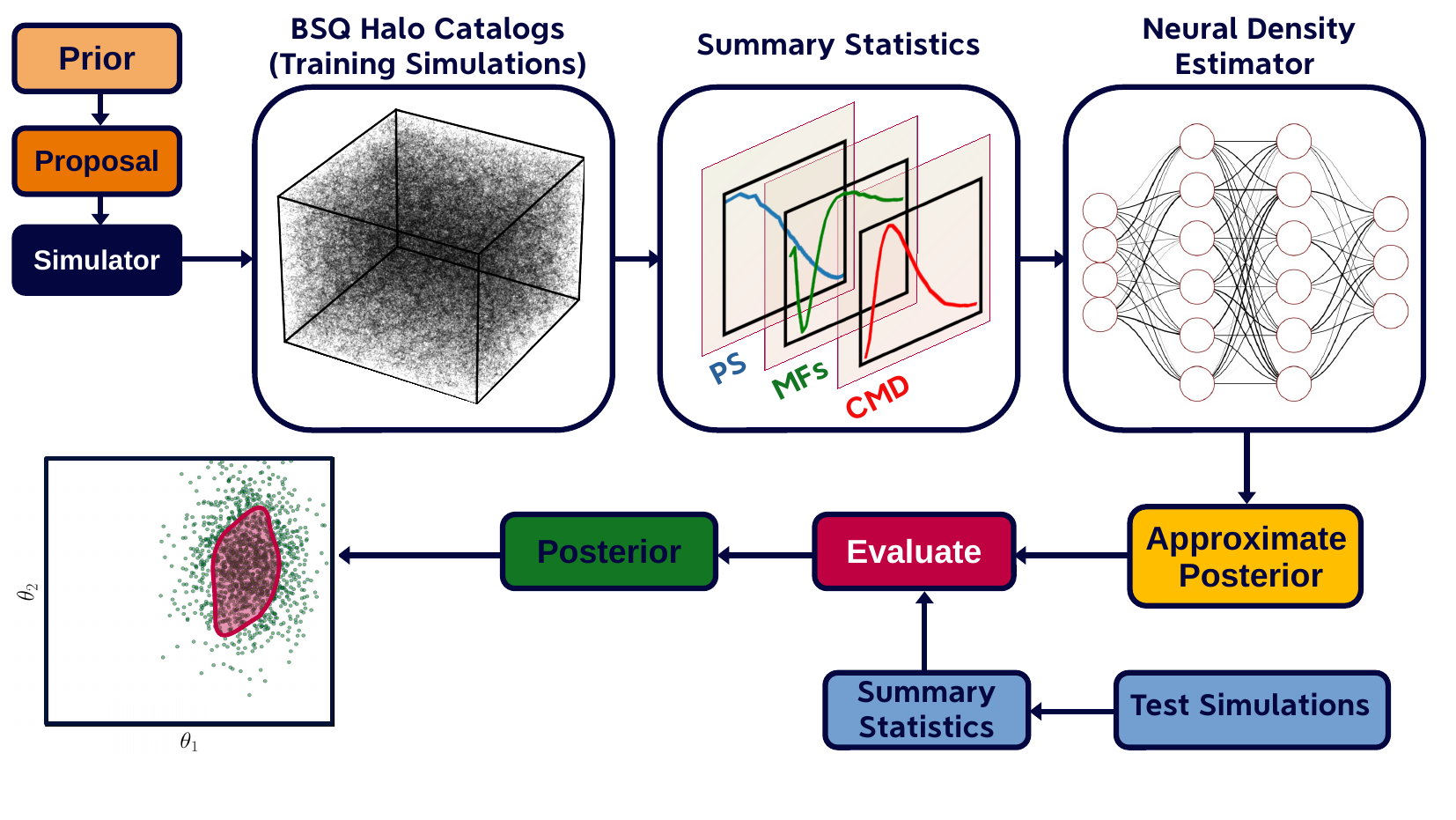}
	\caption{Schematic overview of the simulation-based forecasts pipeline used in this work. Starting from broad priors on cosmological parameters, we explore the parameter space defined by the \textit{Big Sobol Sequence} (BSQ) simulation suite at $z=0.5$, making use of its redshift-space halo catalogs. 
	From these catalogs, we compute three types of summary statistics:  Power Spectrum (PS), Minkowski functionals (MFs) and the Conditional Moments of Derivative (CMD). These statistics are feed to train neural density estimators that approximate the posterior distribution of parameters. The trained models are evaluated on independent test simulations from the same suite to assess the relative and joint constraining capability of the considered descriptors.}
	\label{fig:fig_1}
\end{figure*}

In the astrophysical domain, SBI has been applied to various subjects: inferring the total mass of the Milky Way–Andromeda system from dynamical measurements~\cite{lemos2021sum}, constraining the properties of warm dark matter using stellar streams~\cite{hermans2021towards}, and probing the population of dark matter sub-halos through strong gravitational lensing~\cite{brehmer2019mining, coogan2024effect}. Moreover, SBI frameworks have facilitated efficient inference of halo occupation models~\cite{hahn2017approximate}. A growing number of studies have also demonstrated the power of SBI in modeling the parameters of strongly lensed systems~\cite{legin2021simulation, wagner2021hierarchical}. The real-time inference of compact binary coalescence parameters using neural posterior estimation has been done in ~\cite{Dax2021}.
The  \texttt{peregrine}, a sequential SBI framework that achieves high-fidelity inference from gravitational waveforms with only a fraction of the simulation cost required by nested sampling has been developed in ~\cite{Bhardwaj2024}. Similar strategies have been used to study population-level inference~\cite{PhysRevD.106.083014} and to constrain stochastic gravitational wave backgrounds using likelihood-free methods~\cite{Alvey2024}. These applications underscore the versatility of SBI for analyzing complex and high-volume data expected from current and upcoming observatories such as LIGO\footnote{\url{https://www.ligo.caltech.edu/}}, 
	Virgo\footnote{\url{https://www.virgo-gw.eu/}}, 
	LISA\footnote{\url{https://www.lisamission.org/}}, 
	Einstein Telescope\footnote{\url{https://www.et-gw.eu/}}, and Cosmic Explorer\footnote{\url{https://cosmicexplorer.org/}}.
	
In this work, we present a simulation-based forecasting analysis aimed at evaluating and comparing the constraining capability of different (weighted) morphological descriptors of the LSS, with a particular focus on their sensitivity to nonlinear and anisotropic features in redshift-space. Our analysis is based on halo catalogs extracted from the BSQ simulations\footnote{\url{https://quijote-simulations.readthedocs.io/en/latest/bsq.html}} \cite{Quijote_sims} at $z = 0.5$, and leverages simulation-based inference (SBI) as a likelihood-free framework for cosmological parameter estimation.
Specifically, we compare three classes of summary statistics:
\\
1) The redshift-space power spectrum multipoles (PS), which capture the Gaussian and two-point information content of the halo density field and serve as a standard benchmark in large-scale structure analyses;
\\
2) Minkowski Functionals (MFs), which fall under the broader category of \textit{morphology} and encode both geometrical and topological information of excursion sets of the density field;
\\
3) Conditional Moments of Derivatives (CMD), a directional and anisotropy-sensitive statistic introduced in \cite{kanafi2024probing,afzal2025cosmic}, belonging to a class of \textit{weighted morphological} measures designed to probe local anisotropies and higher-order information beyond what is accessible through traditional two-point or isotropic morphological descriptors.

Unlike Fisher matrix-based forecasts, which assess parameter sensitivity around a fixed fiducial cosmology, our approach employs SBI to learn the full posterior over the parameter space defined by broad priors. The analysis is purely simulation-driven: a subset of BSQ realizations is used for training the SBI model, and another subset (test simulations) is reserved for evaluating forecasts.

The key research questions addressed in this study are:\\
(I) How does incorporating directional weighted morphological information through CMD—alongside scalar morphology captured by MFs—affect the precision of cosmological parameter estimation in redshift space, particularly for $\Omega_m$ and $\sigma_8$?
\\
(II) How does the smoothing scale affect the constraining power of the proposed (weighted) morphological statistics, and how does this behavior relate to the transition from linear to mildly non-linear regimes?\\
(III) How sensitive are the forecasted constraints to the choice of fiducial cosmological parameters, and to what extent does the inferred constraining power depend on the fiducial point?\\
(IV) How does the constraining power of (weighted) morphological statistics compare to that of PS at matched effective scales, and what role does anisotropic information play in this comparison?\\
(V) How sensitive are the inferred cosmological constraints to variations in halo sampling, including fixed number density selections and minimum halo mass thresholds, and what do these tests reveal about the physical versus abundance-driven origin of the information content?\\

To the best of our knowledge, this is the first attempt to incorporate both standard morphological descriptors and novel weighted morphological statistics into a simulation-based inference pipeline for cosmological forecasting from halo fields in redshift-space (see the pipeline presented in Figure~\ref{fig:fig_1}). The framework that we propose, enables a principled comparison of the information content of different summary statistics, shedding light on the relative and joint efficacy of geometrical, topological, and directional descriptors in extracting cosmological information beyond Gaussian assumptions.

The remainder of this paper is organized as follows:
Section~\ref{sec:data}  describes the BSQ simulation suite and the construction of halo catalogs.
In Section~\ref{sec:summary_statistics}, we introduce the theoretical background of the summary statistics employed in this work, including the PS, MFs and  CMD, and outline their role as complementary measures of the large‑scale structure in redshift-space. 
Section~\ref{sec:npe} presents the  SBI framework implemented via Neural Posterior Estimation (NPE) and details the posterior calibration tests. 
Section~\ref{sec:results} discusses the cosmological forecasting results, and Section~\ref{sec:summary} summarizes the main results and outlines prospects for future works.

\section{From Prior Information to  Synthetic Data}
\label{sec:data}

In this study, we utilize the Big Sobol Sequence (BSQ) simulations, a subset of the Quijote simulation suite~\cite{Quijote_sims}, specifically designed for machine learning applications in cosmology. The BSQ comprises 32,768 N-body simulations, each evolving $512^3$ dark matter particles within a periodic comoving volume of $(1000~h^{-1}\mathrm{Mpc})^3$. These simulations are instrumental in exploring the non-linear regime of structure formation and provide a comprehensive dataset for training and validating simulation-based inference models.

Each BSQ simulation varies five key cosmological parameters—namely the matter density parameter $\Omega_{\mathrm{m}}$, the baryon density parameter $\Omega_{\mathrm{b}}$, the dimensionless Hubble parameter $h$, the scalar spectral index $n_s$, and the amplitude of matter fluctuations $\sigma_8$—sampled using a sobol sequence to ensure a uniform and comprehensive coverage of the parameter space. The varied parameters and their respective prior ranges are summarized in Table~\ref{tab:cosmo_params}. All remaining parameters are maintained constant throughout the simulations: the total neutrino mass, $M_\nu = 0.0$ eV, the dark energy equation of state parameter, $w = -1$, corresponding to a cosmological constant, and the curvature parameter, $\Omega_{\mathrm{K}} = 0$, assuming a spatially flat universe. We denote all desired cosmological parameters varying throughout analysis by $\{\boldsymbol{\theta}\}:\{\Omega_{\rm m}, \Omega_{\rm b},h,n_{\rm s},\sigma_8\}$ known as parameter set. 

\begin{table}[h]
\caption{The range of cosmological parameters in the BSQ simulations. The shape of prior for all parameters over the considered range is Top-Hat. }
\label{tab:cosmo_params}
\begin{ruledtabular}
\begin{tabular}{lc}
Parameter & Range \\
\hline
$\Omega_{\mathrm{m}}$ & $[0.10,\ 0.50]$ \\
$\Omega_{\mathrm{b}}$ & $[0.02,\ 0.08]$ \\
$h$                   & $[0.50,\ 0.90]$ \\
$n_{\rm s}$                 & $[0.80,\ 1.20]$ \\
$\sigma_8$            & $[0.60,\ 1.00]$ \\
\end{tabular}
\end{ruledtabular}
\end{table}
Initial conditions for the simulations are generated at redshift $z = 127$ using second-order Lagrangian perturbation theory (2LPT). The simulations are then evolved to $z = 0$ using the TreePM code GADGET-III, which combines a short-range tree algorithm with a long-range particle-mesh method to efficiently compute gravitational interactions~\cite{Springel2005TheCS}. This setup ensures accurate modeling of the non-linear evolution of cosmic structures.

Furthermore, the BSQ simulations provide multiple halo catalogs for each realization, generated using two commonly adopted algorithms: the Friends-of-Friends (FoF) method and the Rockstar phase-space halo finder. In this study, we exclusively utilize the FoF halo catalogs, which offer a simple and robust definition of halos based solely on spatial proximity. The FoF algorithm identifies groups of dark matter particles by linking together all particles separated by less than a specified fraction of the mean inter-particle spacing. 

In this work, we specifically select the halo catalogs corresponding to the redshift $z = 0.5$. This redshift lies within the intermediate-redshift regime probed by several current and upcoming spectroscopic/photometric surveys. For example, the Dark Energy Spectroscopic Instrument (DESI) targets luminous red galaxies (LRGs) in the range $0.4 < z < 1.0$, making our choice of snapshot broadly relevant for analyses of large-scale structure at comparable epochs.
\begin{figure}[t]
    \centering
    \includegraphics[width=\linewidth]{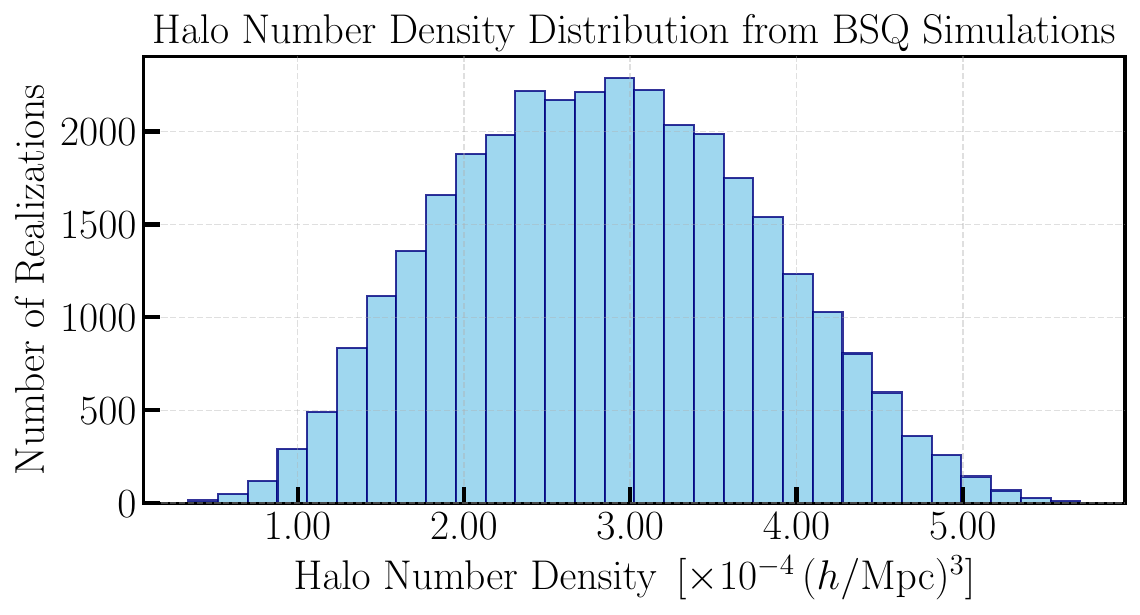}
    \caption{Distribution of halo number densities across BSQ simulation realizations at redshift $z = 0.5$. The histogram reflects variations in halo abundance due to different cosmological parameter combinations. No minimum halo mass threshold is applied.}
    \label{fig:fig_2}
\end{figure}

We also include all halos recognized by the FoF algorithm without imposing any minimum mass threshold. Since the cosmological parameters change significantly across the BSQ simulation suite (see Table~\ref{tab:cosmo_params}), the resulting halo number density consequently exhibits variation. Across different realizations, the halo number density spans the range $\sim[0.34, 5.70] \times 10^{-4}\, h^3\,\mathrm{Mpc}^{-3}$, as shown in Figure.~\ref{fig:fig_2}. This variation reflects the sensitivity of the halo abundance to the underlying cosmological model.

Even though this variation introduces additional complexity, it also reflects the behavior of real data, where the abundance and bias of the tracers depend on the underlying cosmology. Therefore, our approach avoids artificially fixing the number density and instead leverages this cosmological dependence to enrich the inference process. Although the resulting halo number densities span a finite range, they remain broadly consistent with the galaxy number densities anticipated in surveys like DESI and Euclid. As such, our setup serves not as a direct forecast for a particular survey, but rather as a diagnostic framework to study how large-scale morphological information responds to changes in cosmology.

In this work, we construct redshift-space halo catalogs by adjusting the positions of halos based on their peculiar velocities along the line of sight (LOS), which is aligned with one of the axes of the simulation boxes. This approach allows us to incorporate redshift-space distortions, enabling a more realistic comparison with spectroscopic observations.  These catalogs form the input for the summary statistic computations described in Sec.~\ref{sec:summary_statistics}. It is worth mentioning that this study may extend this analysis by including observational systematics to evaluate the reliability and robustness of morphological inference under more realistic survey conditions.

\section{From Data to Summary Statistics}

\label{sec:summary_statistics}
We analyze the LSS traced by dark matter halos in cosmological simulations to extract information about the underlying cosmological parameters. Working with halos rather than galaxies allows us to avoid complications arising from galaxy bias~\cite{desjacques2018large}, redshift uncertainties, and survey selection effects, thereby isolating the physical impact of cosmology. Nevertheless, to approximate observational conditions—particularly those relevant to spectroscopic galaxy surveys—we construct the halo density field and compute summary statistics in redshift-space, where peculiar velocities induce anisotropies in the observed distribution ~\cite{kaiser1987clustering,hamilton1998linear,scoccimarro2004redshift}.

In what follows, we first describe the construction of the redshift-space halo density field and then introduce the summary statistics proposed in our analysis, which include both standard clustering measures (PS) and higher-order (weighted) morphological descriptors.

\subsection{Density Field in Redshift-Space}

To construct the large-scale distribution of simulated halos, we begin by mapping the real-space halo positions into redshift-space by accounting for their peculiar velocities. Redshift-space distortions (RSDs) arise from peculiar velocities along the LOS, shifting the apparent radial positions of tracers and introducing anisotropic features in the observed clustering~\cite{hamilton1998linear}. On large scales, coherent infall enhances the clustering signal along the LOS (the Kaiser effect)~\cite{kaiser1987clustering}, while on small scales, virial motions within collapsed structures lead to elongation and suppression of the clustering (the Fingers-of-God effect)~\cite{scoccimarro2004redshift}.

We include RSDs by displacing real-space halo positions, $\mathbf{r}$, along  the LOS (chosen as the $z$-axis) using:
\begin{equation}
\mathbf{s} = \mathbf{r} + \frac{v_z}{aH} \hat{z},
\end{equation}
where $v_z$ is the peculiar velocity component along the LOS, $a$ is the scale factor, and $H$ is the Hubble parameter.

The redshift-space position, $\mathbf{s}$, is then interpolated onto a regular $360^3$ Cartesian grid covering a volume of $(1000\,h^{-1}\,\mathrm{Mpc})^3$ using the Cloud-in-Cell (CIC) mass assignment scheme~\cite{hockney2021computer}, as implemented in the \texttt{Pylians} library~\cite{Pylians}.
This yields a gridded halo number density field, $n(\mathbf{x})$, from which the dimensionless density contrast is computed as:
\begin{equation}\label{eq:density}
\delta(\mathbf{x}) \equiv \frac{n(\mathbf{x}) - \bar{n}}{\bar{n}},
\end{equation}
with $\bar{n}$ denoting the mean halo number density. The resulting redshift-space density field acts as the basis for all subsequent statistical analysis. Specifically, the RSD-induced anisotropies embedded in $\delta(\mathbf{x})$ are essential for evaluating the response of (weighted) morphological descriptors to distortions along the LOS.

\subsection{Power Spectrum}
The redshift-space power spectrum is a cornerstone observable in LSS analysis~\cite{Peebles1980,hamilton1998linear}, encapsulating the amplitude of two-point correlations as a function of scale and orientation with respect to the LOS. Due to the presence of peculiar velocities, the observed clustering becomes anisotropic in redshift-space, which is naturally captured by decomposing the power spectrum into angular multipoles.

We begin by computing the two-dimensional power spectrum, $P(k, \mu)$, from the Fourier transform of the redshift-space density contrast field, $\delta(\mathbf{x})$ (Eq.~(\ref{eq:density})), where $k = |\mathbf{k}|$ is the wavenumber and $\mu \equiv \mathbf{k}\!\cdot\!\hat{z}/k$. 
The angular dependence is projected onto Legendre polynomials to obtain the multipole moments~\cite{feldman1993power}:
\begin{equation}
	P_\ell(k) = \frac{2\ell + 1}{2} \int_{-1}^{1} P(k, \mu) \, \mathcal{L}_\ell(\mu) \, d\mu,
\end{equation}
where $\mathcal{L}_\ell$ is the Legendre polynomial of order $\ell$. Here, we focus on the first three non-zero moments: the monopole ($\ell=0$), quadrupole ($\ell=2$), and hexadecapole ($\ell=4$), which together encode the isotropic clustering and the anisotropic effects induced by redshift-space distortions.

The numerical estimation of $P(k, \mu)$ and its multipoles is performed using the \texttt{Pylians} library, which bins the Fourier modes over spherical shells in $(k, \mu)$ space. 
In this work, we restrict the analysis to modes with $k \leq k_{\max}\simeq 0.16\,h\,\mathrm{Mpc}^{-1}$. 
This choice is primarily motivated by consistency with the real-space smoothing scale adopted in the morphological analysis, ensuring that the power spectrum and (weighted) morphological probe comparable physical scales. 
At these wavenumbers, the clustering remains in the linear to mildly non-linear regimes, where the power spectrum multipoles are robustly measured and less affected by small-scale non-linearities, shot noise, and resolution-driven artifacts.

These power spectrum multipoles are regarded as the first class of summary statistics in our inference pipeline (Figure.~\ref{fig:fig_1}). As they efficiently capture the Gaussian and two-point information contents of the halo distribution, they provide a natural benchmark for assessing the performance of more complex (weighted) morphological statistics in terms of cosmological parameter constraints.

\subsection{Morphology: Minkowski Functionals}
Minkowski Functionals (MFs) serve as essential morphological tools in understanding the geometrical and topological properties of  random fields, particularly in cosmology~\cite{mecke1993robust,schmalzing1995minkowski,schmalzing1997beyond}. In this section, we explore the mathematical definitions of these functionals for three-dimensional density fields. To define the MFs for $\delta(\mathbf{x})$ (Eq.~(\ref{eq:density})), we construct   the smoothed density field, $\delta_s(\mathbf{x})$, as follows:
\begin{equation}
\delta_s(\mathbf{x}) = \int d\mathbf{x}' \, W_R(|\mathbf{x} - \mathbf{x}'|)\; \delta(\mathbf{x}'),
\end{equation}
where \( W_R(|\mathbf{x}|) \) is a smoothing kernel with radius \( R \). For our analysis, we utilize a Gaussian kernel defined by:
\begin{equation}
W_R(|\mathbf{x}|) = \frac{e^{-|\mathbf{x}|^2/(2R^2)}}{(2\pi)^{3/2} R^3}.
\end{equation}
A practical approach for defining MFs incorporates the notion of excursion set, which identifies the regions where the smoothed density contrast surpasses the specified threshold $\vartheta$, and it is defined as:
\begin{equation}
Q_\vartheta = \{ \mathbf{x} \in \mathbb{R}^3 \,|\, \delta_s(\mathbf{x}) \geq \vartheta \}.
\end{equation}
The boundary of the excursion set, denoted by $\partial Q_\vartheta$, represents the isodensity surface at the threshold $\vartheta$. The geometrical and topological properties of these surfaces are quantified using the four MFs, which provide a comprehensive description of the structure and morphology of the density field.
In three-dimensional space, we denote the four MFs as \( V_n(\vartheta) \) for \( n = 0, 1, 2, 3 \).
The zeroth functional, \( V_0(\vartheta) \), measures the fractional volume of the excursion set, \( Q_\vartheta \), and is given by:
\begin{equation}
V_0(\vartheta) = \frac{1}{V} \int_{Q_\vartheta} d^3x,
\end{equation}
where \( V \) is the total volume of the simulation box (survey domain), and \( d^3x \) denotes the infinitesimal volume element. The first functional, \( V_1(\vartheta) \), quantifies the total surface area of the isodensity boundary, \( \partial Q_\vartheta \), and is expressed as:
\begin{equation}
V_1(\vartheta) = \frac{1}{6V} \int_{\partial Q_\vartheta} dA,
\end{equation}
with \( dA \) being the infinitesimal surface area element. 
 The second and third functionals encode curvature information of the isodensity surface \( \partial Q_\vartheta \). Specifically, the second functional, \( V_2(\vartheta) \), measures the integrated mean curvature:
\begin{equation}
V_2(\vartheta) = \frac{1}{6\pi V} \int_{\partial Q_\vartheta} K\, dA,
\end{equation}
and the third functional, \( V_3(\vartheta) \), measures the integrated Gaussian curvature:
\begin{equation}
V_3(\vartheta) = \frac{1}{4\pi V} \int_{\partial Q_\vartheta} G\, dA,
\end{equation}
where \( K = k_1 + k_2 \) is the mean curvature and \( G = k_1 k_2 \) is the Gaussian curvature, expressed in terms of the principal curvatures \( k_1 \) and \( k_2 \) at each point on the surface which characterize the local bending of the surface in orthogonal directions. They are quantifying  geometrical and topological features of associated excursion set. One can compute mentioned curvatures directly from the smoothed density field, \( \delta_s(\mathbf{x}) \), and its derivatives. Letting \( \nabla_i \delta_s \) and \( \nabla_i \nabla_j \delta_s \) denote the first and second derivatives of the field, accordingly, the mean and Gaussian curvatures take the differential forms:
\begin{equation}
K(\mathbf{x}) = \frac{1}{|\nabla \delta_s|} \left( \nabla_i \nabla_j \delta_s \left[ \delta^{ij} - \frac{\nabla^i \delta_s \, \nabla^j \delta_s}{|\nabla \delta_s|^2} \right] \right),
\end{equation}
\begin{equation}
G(\mathbf{x}) = \frac{\det(\nabla_i \nabla_j \delta_s)}{|\nabla \delta_s|^2}.
\end{equation}
These expressions, in conjunction with the mathematical identity that relates surface integrals over isodensity contours to volume integrals (see e.g.~\cite{schmalzing1997beyond}),
\begin{equation}
\int_{\partial Q_\vartheta} f(\mathbf{x}) \, dA = \int d^3x \; \delta_D(\delta_s(\mathbf{x}) - \vartheta) \, |\nabla \delta_s| \, f(\mathbf{x}),
\end{equation}
enable a reformulation of the MFs as volume integrals over the entire domain.
Specifically, they take the form:
\begin{align}
V_0(\vartheta) &= \frac{1}{V} \int_V d^3x \; \Theta(\delta_s(\mathbf{x}) - \vartheta), \\
V_1(\vartheta) &= \frac{1}{6V} \int_V d^3x \; \delta_D(\delta_s(\mathbf{x}) - \vartheta) \, |\nabla \delta_s|, \label{eq:v1}\\
V_2(\vartheta) &= \frac{1}{6\pi V} \int_V d^3x \; \delta_D(\delta_s(\mathbf{x}) - \vartheta) \, |\nabla \delta_s| \, K(\mathbf{x}), \\
V_3(\vartheta) &= \frac{1}{4\pi V} \int_V d^3x \; \delta_D(\delta_s(\mathbf{x}) - \vartheta) \, |\nabla \delta_s| \, G(\mathbf{x}),
\end{align}
where \( \Theta \) is the Heaviside step function and \( \delta_D \) is the Dirac delta function. Above equations establish a direct connection between the MFs and the differential structure of the smoothed field, and are particularly suited for numerical implementation on gridded data.

\subsection{Weighted Morphology: Conditional Moment of Derivative}
The MFs provide a complete set of scalar morphological characterizers for generic fields under the assumptions of additivity, motion invariance, and conditional continuity. According to the {\it Hadwiger’s theorem}~\cite{hadwiger2013vorlesungen,mecke1993robust}, any functional \( \mathcal{F}(Q) \) defined on a body \( Q \subset \mathbb{R}^d \) that satisfies these properties can be uniquely expressed as a linear combination of the \( d + 1 \) MFs:
\begin{equation}
\mathcal{F}(Q) = \sum_{n=0}^{d} c_n V_n(Q),
\end{equation}
where \( c_n \) is a real coefficient independent of the body \( Q \) (see~\cite{mecke1993robust} for details). 
While this property ensures the completeness of MFs in characterizing the scalar morphological content of a typical field, they remain intrinsically insensitive to directional information or the orientation of anisotropies present in the field. Nevertheless, several studies have shown that their amplitudes may exhibit mild sensitivity to the level of anisotropy~\cite{matsubara1996statistics,codis2013non,jiang2022effects}.

This limitation becomes particularly relevant in redshift-space analyses, where the large-scale structure is distorted along the LOS due to peculiar velocities. These RSDs introduce anisotropic features in the observed density field, which cannot be fully captured by scalar functionals. To address this issue, the extensions of the standard MFs have been proposed—most notably the Minkowski Tensors—which are designed to probe the anisotropic morphology of isodensity contours~\cite{beisbart2002vector,appleby2019ensemble}.

For example, translation-invariant Minkowski Tensors of rank two provide directional extensions of scalar MFs by encoding anisotropic shape information. A representative case is the surface tensor defined over the isodensity contour as:
\begin{equation}
W^{0,2}_1(\vartheta) = \frac{1}{V} \int_{\partial Q_\vartheta} \hat{\mathbf{n}} \otimes \hat{\mathbf{n}} \, dA,
\end{equation}
where \( \hat{\mathbf{n}} \) is the unit normal vector to the isodensity surface at threshold \( \vartheta \), and \( \otimes \) denotes the tensor product. Using the standard surface-to-volume transformation, each component of this tensor can be equivalently written in terms of the smoothed density field as:
\begin{equation}
W^{0,2}_{1,ij}(\vartheta) = \frac{1}{V} \int_V d^3x \; \delta_D(\delta_s(\mathbf{x}) - \vartheta) \, \frac{\nabla_i \delta_s \, \nabla_j \delta_s}{|\nabla \delta_s|}.
\end{equation}
Here, the surface normal is given by the normalized gradient of the field, \( \hat{n}_i = \nabla_i \delta_s / |\nabla \delta_s| \), where \( i \in \{x, y, z\} \) denotes the spatial component of the unit normal vector. 

A key characteristic of both MFs and their tensorial generalizations is that their computation depends solely on the geometry and topology of excursion sets without requiring direct access to the underlying field values. These quantities are constructed entirely from the shape and topology of isodensity contours, making them agnostic to internal variations of the field within those contours. While this geometric abstraction confers a degree of generality and robustness, it also restricts the capacity of these descriptors to probe features that are sensitive to local field amplitudes or gradients.

To address this limitation, we consider a complementary approach wherein morphological measures are explicitly coupled to the field itself. In this formulation, scalar or tensorial weights derived from the field value or its derivatives are incorporated into the integrand, effectively embedding field-level information into the morphological analysis. These weighted descriptors retain the geometric basis of excursion set-based statistics but extend their sensitivity to variations in field intensity and structure, thereby providing access to a richer set of physical information—particularly in contexts involving non-Gaussianity, anisotropy, or RSDs.

As a concrete example of weighted morphological statistics, we have introduced the Conditional Moments of Derivative (CMD) in~\cite{kanafi2024probing,afzal2025cosmic}. In this essay, we confine ourselves to carry out the first derivative of field to reduce the uncertainty in numerical analysis as much as possible. Hereafter,  we  consider the  conditional moments of first derivative denoted by CMD. This measure quantifies the local behavior of the field gradient restricted to isodensity surfaces and takes the form of directional moments, allowing sensitivity to anisotropic features while retaining scalar structure. Therefore, the CMD is defined as~\cite{kanafi2024probing}:
\begin{equation}
\label{eq:cmd}
N_{\mathrm{CMD}}^{(m)}(\vartheta, i) = \frac{1}{V} \int_V d^3x \; \delta_D\left( \delta_s(\mathbf{x}) - \vartheta \right) | \nabla_i \delta_s(\mathbf{x}) |^m,
\end{equation}
where \( |\nabla_i \delta_s(\mathbf{x})| \) denotes the absolute value of the first derivative of the smoothed field along the \( i \)-th spatial direction, and \( m \) controls the order of the moment. The CMD can be expressed in terms of surface integral over the boundary of excursion set as:
\begin{equation}
N_{\mathrm{CMD}}^{(m)}(\vartheta, i) = \frac{1}{V} \int_{\partial Q_\vartheta} |\nabla \delta_s(\mathbf{x})|^{m-1} \, |\hat{n}_i|^m \, dA.
\end{equation}
This representation makes explicit the weighted nature of the CMD statistics, such that the integrand involves a scalar weight
$|\nabla \delta_s(\mathbf{x})|^{m-1}$, which depends directly on the local gradient magnitude of the underlying field. In contrast to the traditional MFs and their tensorial extensions—which can be evaluated purely from the geometry of excursion sets—CMD incorporates explicit field-level information through this weighting as mentioned before.  The remaining directional factor, \( |\hat{n}_i|^m \), governed by the local orientation of the isodensity surface relative to the chosen spatial axis \( i \), is fully determined by the geometry of the excursion set and does not require direct access to field values. Nevertheless, its inherent directionality, modulated by the exponent \( m \), enhances the sensitivity of CMD to anisotropic features in the field. 

Finally, the summary statistics employed in this work are given by $\{\mathbfcal{D}\} : \{ 	\mathbf{PS^{(\ell=0,2,4)}}, \bold {V}_{\diamond}, \bold{CMD}, \bold {MFs}, \bold {MFs} + \bold {CMD} \}$, whose detailed combinations are presented in Table~\ref{tab:summary_stats}.

\begin{table}[h]
	\centering
	\caption{
		Various summary statistics configurations used in this work.
		The superscript $R$ in the notation indicates the smoothing scale, with $R \left[~h^{-1}\mathrm{Mpc}\right] \in \{15, 20, 25\}$.}
	\label{tab:summary_stats}
	\begin{tabular}{l}
		\hline
		\textbf {Summary statistics} \\
		\hline
		$\mathbf{PS^{(\ell=0,2,4)}}: \{P_0(k), P_2(k), P_4(k)\}$\\
		$\bold {V}^{(R)}_{\diamond}:\{V^{(R)}_{\diamond} \mid \diamond \in \{0,1,2,3\} \}$ \\
		$\bold {CMD}^{(R)}: \{ N_{\mathrm{CMD}}^{(R)}(\hat{x}),\, N_{\mathrm{CMD}}^{(R)}(\hat{z}) \}$ \\
		$\bold {MFs}^{(R)}:$ $\{V_0^{(R)},\, V_1^{(R)},\, V_1^{(R)}, \, V_3^{(R)}\}$ \\
		$\bold {MFs}^{(R)} + \bold {CMD}^{(R)}: \{V_0^{(R)},\, N_{\mathrm{CMD}}^{(R)}(\hat{x}),\, N_{\mathrm{CMD}}^{(R)}(\hat{z}), \, V_2^{(R)},\, V_3^{(R)} \}$ \\
		\hline
	\end{tabular}
\end{table}


\section{Simulation-Based Inference Framework}
\label{sec:npe}
Traditional approaches to cosmological inference rely on explicitly specifying the likelihood function, $p(\mathbfcal{D}|\boldsymbol{\theta})$~\cite{trotta2008bayes}, a procedure that becomes increasingly challenging—and in many cases intractable—when confronted with complex, non-Gaussian summary statistics (observables) such as (weighted) morphological descriptors (Table \ref{tab:summary_stats}). Simulation-based inference (SBI) circumvents this requirement by leveraging forward models to connect model parameters directly to observables, thereby enabling parameter estimation without an explicit likelihood~\cite{Cranmer2020}. Generally, two widely used approaches in SBI are known as \textit{neural posterior estimation} (NPE)  and \textit{neural likelihood estimation} (NLE). In this section, we focus on the former framework.    

\subsection{Neural Posterior Estimation}

\begin{figure}[t]
	\centering
	\includegraphics[scale=0.45]{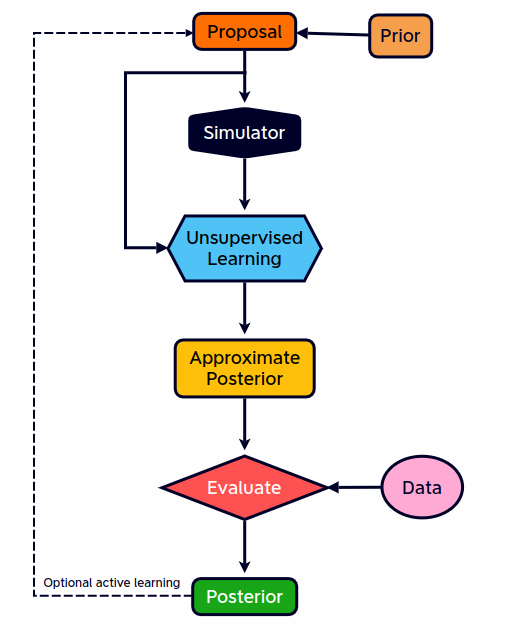}
	\caption{Schematic representation of the neural posterior estimation (NPE) workflow, inspired by~\cite{Cranmer2020}. Starting from a prior over model parameters $\boldsymbol{\theta}$, proposal samples are passed through a simulator to generate mock observations $\mathbfcal{D}$. An unsupervised learning stage maps these data to an approximate posterior, which is evaluated against target data.}
	\label{fig:fig_3}
\end{figure}

We adopt the NPE framework, in which a conditional neural density estimator is trained on simulated data to approximate the posterior distribution $p(\boldsymbol{\theta}|\mathbfcal{D})$~\cite{papamakarios2016fast} (see Figure.~\ref{fig:fig_3}). Our implementation employs halo catalogs from the BSQ simulations at redshift $z = 0.5$, which form a Sobol-sampled grid in the five-parameter $\Lambda$CDM space and provide uniform prior coverage across cosmological parameters. The overall pipeline, schematically illustrated in Figure.~\ref{fig:fig_1}, proceeds sequentially as follows:
\\
(I) In the first step, the cosmological parameter set $\{\boldsymbol{\theta}\}:\{\Omega_{\rm m}, \Omega_{\rm b},h,n_{\rm s},\sigma_8\}$ is drawn from independent uniform priors within the ranges listed in Table~\ref{tab:cosmo_params}, forming the proposal distribution for simulation. The BSQ suite then produces halo catalogs in redshift-space at $z = 0.5$.
\\
(II) The halo positions are subsequently converted into redshift-space and interpolated onto a $360^3$ Cartesian grid, ensuring that nonlinear clustering and the anisotropies induced by RSDs are retained.
\\
(III) From each gridded density field, the components of the summary statistics, $\mathbfcal{D}$, defined in Table~\ref{tab:summary_stats} are then computed, capturing the (weighted) morphological information of halo catalogs.
\\
(IV) This sequence yields matched samples $(\mathbfcal{D}, \boldsymbol{\theta})$. In this work, we employ a total of $30{,}000$ simulated realizations, of which $25{,}000$ are used for training, $3{,}000$ for validation, and $2{,}000$ are reserved as test data for forecasting analyses.
\\
(V) In the next step, the conditional neural density estimator, $q_\phi(\boldsymbol{\theta}|\mathbfcal{D})$, is optimized to learn an accurate approximation of the target posterior.
\\
(VI) Finally, the inferred model is applied to independent test datasets excluded from the training stage, enabling quantitative evaluation of inference performance and forecasting capability.

To estimate posterior, we employ conditional \textit{normalizing flows},
which approximate the true posterior distribution
 $p(\boldsymbol{\theta}|\mathbfcal{D})$ by a flexible parametric model,
$q_\phi(\boldsymbol{\theta} | \mathbfcal{D})$, where $\phi$ denotes the set of learnable parameters
of the flow~\cite{papamakarios2021normalizing}.
In this framework, a latent variable, $\mathbf{z}$, is first drawn from a standard
multivariate Gaussian $\mathcal{N}_{\mathbf{z}}(\mathbf{0}, \mathbf{I})$,
and transformed into a cosmological parameter set $\boldsymbol{\theta}$,
via an invertible mapping $f_\phi$, conditioned on the summary statistics $\mathbfcal{D}$:
\begin{equation}
\boldsymbol{\theta} = f_\phi(\mathbf{z}; \mathbfcal{D}).
\end{equation}
When $f_\phi$ is expressed as the composition of $L$ simpler bijections,
$f_\phi = f_L \circ \cdots \circ f_1$, the change-of-variables formula gives:
\begin{equation}
	\log q_\phi(\boldsymbol{\theta}|\mathbfcal{D}) = \log p_0(\mathbf{z}) 
	- \sum_{l=1}^L \log \left| \det \left( \frac{\partial f_l}{\partial \mathbf{z}_{l-1}} \right) \right|,
\end{equation}
where the Jacobian determinants account for the local change in volume
induced by each transformation.

In this work, we implement $f_\phi$ as a \textit{Neural Spline Flow} (NSF)~\cite{durkan2019neural} 
using the \texttt{sbi} library~\cite{tejero2020sbi},
which replaces the typical affine transformations with monotonic rational--quadratic splines.
Conditioned on $\mathbfcal{D}$, the spline parameters are predicted by a neural network, 
providing enhanced flexibility to capture sharp and non-Gaussian features in the posterior.

We partition the simulation-derived catalogs into training and validation sets with a
$\sim$ 90/10 split and train our neural density estimator by maximizing the score function:
\begin{equation}
	\mathcal{L}(\phi) = \sum_{t \in \mathcal{T}} \log q_\phi(\boldsymbol{\theta}^{(t)} |\mathbfcal{D}^{(t)}),
\end{equation}
where the index $t$ runs over the training set, $\mathcal{T}$.
Maximizing $\mathcal{L}$ is equivalent to minimizing the Kullback–Leibler divergence
between the modeled distribution, $q_\phi(\boldsymbol{\theta} | \mathbfcal{D})$, and $p(\boldsymbol{\theta} | \mathbfcal{D})$.
After each training epoch, the score function is also computed on the validation set
to monitor generalization performance. Training is terminated early
if this validation score does not improve for 20  epochs, thereby
mitigating overfitting.

To further enhance the robustness of our estimated posterior, 
we construct an ensemble of density estimators by generating a pool of 300 NSF models. 
These models are built by sampling architectural and training hyperparameters 
using the \texttt{Optuna} \cite{akiba2019optuna} package, which efficiently explores 
the search space (see Table~\ref{tab:nsf_hyperparams}).
Each candidate is trained on the same dataset, and we retain 
the five highest-performing models based on their final validation score function. 
The resulting ensemble posterior, $q^{\mathrm{ens}}_\phi(\boldsymbol{\theta} | \mathbfcal{D})$, is obtained by averaging the predictive densities of the 
selected flows, yielding a more stable and reliable approximation of 
$p(\boldsymbol{\theta}| \mathbfcal{D})$.

In combination with physically motivated summary statistics, this approach retains the nonlinear information content of the LSS field without relying on approximate analytical likelihoods, making it well-suited for high-precision cosmological inference.

\begin{table}[ht]
	\centering
	\caption{Search space for NSF hyperparameters explored with \texttt{Optuna}.
		The range for hidden features is defined by $h_{\mathrm{min}}$ and $h_{\mathrm{max}}$, 
		where $h_{\mathrm{min}} = \max(32, [\texttt{input\_dim} / 2])$ and 
		$h_{\mathrm{max}} = \max(64, \texttt{input\_dim} \times 4)$.
		Here, \texttt{input\_dim} denotes the number of features in the summary statistics.}
	\label{tab:nsf_hyperparams}
	\begin{tabular}{p{4.2cm} p{4.2cm}}
		\hline
		\textbf{Hyperparameter} & \textbf{Search range} \\
		\hline
		Hidden features & $[h_{\mathrm{min}},\ h_{\mathrm{max}}]$  \\
		Number of transforms & $[4, 12]$  \\
		Number of bins & $[4, 16]$ \\
		Batch size & $[1, 128]$\\
		Dropout probability & $[0.0, 0.5]$ \\
		Learning rate & $[5\times 10^{-5}, 10^{-2}]$ \\
		\hline
	\end{tabular}
\end{table}

\begin{figure*}[t]
	\centering
	\includegraphics[width=0.7\linewidth]{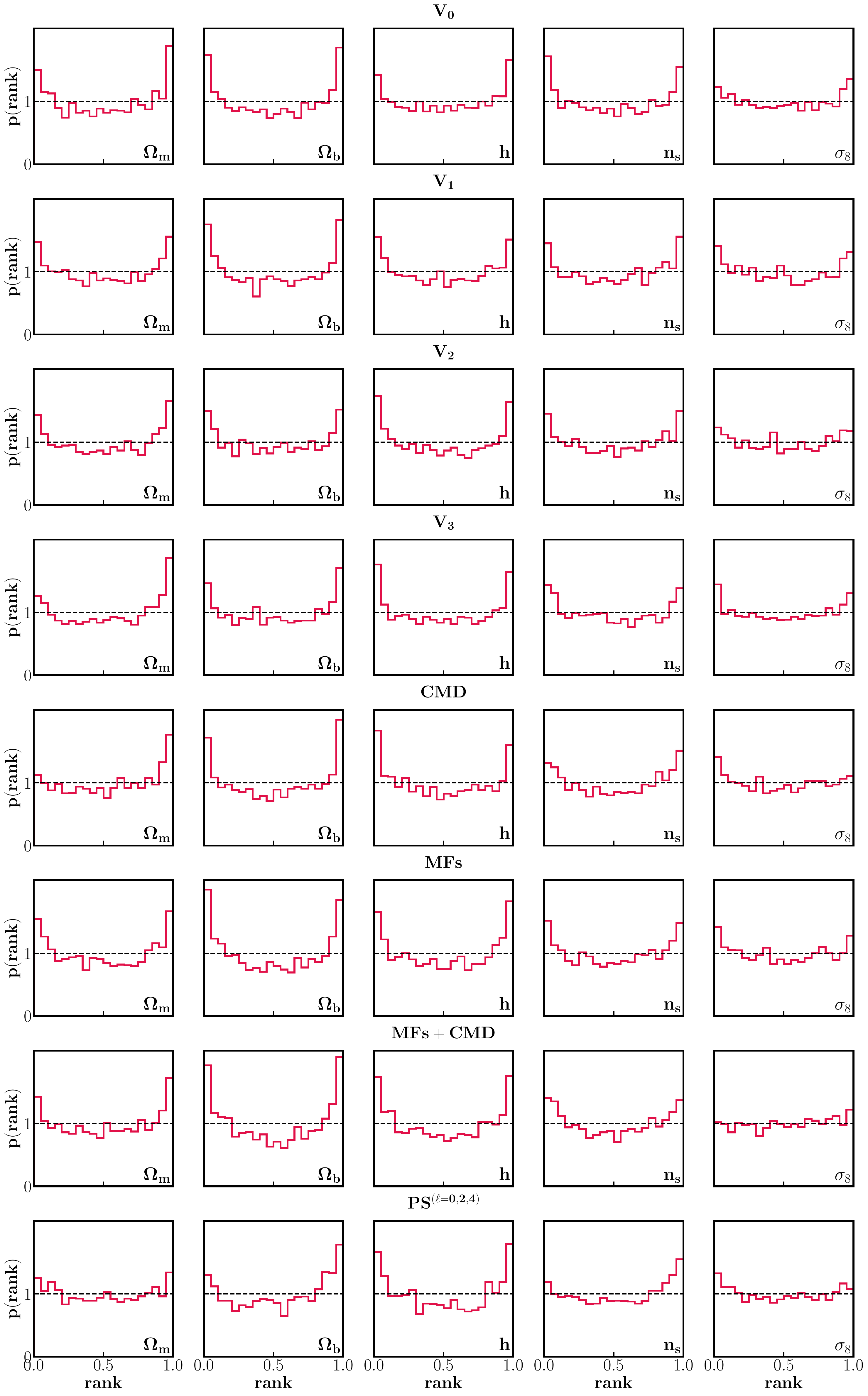}
	\caption{
			Normalized rank distributions for the five $\Lambda$CDM parameters $\{\boldsymbol{\theta}\}:\{\Omega_{\mathrm{m}}$, $\Omega_{\mathrm{b}}$, $h$, $n_{\mathrm{s}}$, $\sigma_{8}\}$ across the ten summary statistics configurations listed in Table~\ref{tab:summary_stats}. 
			Each row corresponds to a distinct summary statistics configuration denoted by its title, and each column to one of the cosmological parameters. The horizontal dashed line represents the expectation for perfectly calibrated posteriors (uniform distribution).}
	\label{fig:fig_4}
\end{figure*}

\subsection{Posterior Validation}

An essential step in our simulation-based forecast pipeline is to check the calibration of the inferred posteriors, $q^{\mathrm{ens}}_\phi(\boldsymbol{\theta} | \mathbfcal{D})$, ensuring that their credible intervals provide the intended statistical coverage.
For this purpose, we implement the simulation-based calibration method, focusing on the use of rank statistics \cite{talts2018validating}. This test is performed on the validation set that was also used for early stopping during the training process.

For each realization in validation set, we estimate the corresponding posterior and draw $5 \times 10^3$ independent samples from it.
We then compute the rank of true parameter value, defined as the fraction of posterior samples that lie below it. Repeating this over all validation realizations yields a rank distribution for each parameter. 
Figure~\ref{fig:fig_4} shows the normalized rank distributions for different cosmological parameters and various summary statistics configurations listed in Table~\ref{tab:summary_stats}. In the case of perfectly calibrated posteriors, the curves would coincide with the horizontal dashed line, representing an ideal uniform rank distribution. Systematic $\bigcup$-shaped deviations reveal overconfidence, where the credible intervals are narrower than warranted by the data. Conversely, $\bigcap$-like
patterns indicate under-confidence, with the inferred posteriors broader than the true distributions—an outcome that may be mitigated through more effective training or improved summary statistics.
We note that the interpretation of rank deviations depends critically on whether a parameter is effectively constrained by the data or remains prior-dominated.

Building on the results presented in Figure~\ref{fig:fig_4}, the rank distributions for $\sigma_{8}$ are generally close to the uniform baseline, indicating that the posterior coverage is largely accurate across most configurations.
The mentioned distribution for $\Omega_{\mathrm{m}}$ also exhibits predominantly flat profiles, with only mild tilts in specific summary statistics combinations.
For $n_{\rm s}$, $h$, and $\Omega_{\mathrm{b}}$, the rank distribution for several configurations exhibit $\bigcup$-shaped profiles characteristic of overconfident posteriors. Such behavior is not surprising in SBI, particularly when the effective information content for a given parameter is limited.
Accordingly, these parameters are only weakly constrained by the considered summary statistics, and their inferred posteriors largely reproduce the corresponding priors.
In this regime, U-shaped rank distributions naturally emerge when the true parameter values of the validation simulations lie near the boundaries of the prior range, resulting in rank values clustering close to 0 or 1.
Therefore, the observed U-shaped rank profiles for $n_{\rm s}$, $h$, and $\Omega_{\mathrm{b}}$ should not be interpreted as evidence for biased inference, but rather as an expected consequence of prior-dominated posteriors.
Furthermore, as the posterior bounds for these weakly constrained parameters are comparable in scale to the prior ranges, the achieved overconfidence is not a cause for concern in our analysis.
For this reason, we focus our quantitative conclusions on the well-constrained parameters, namely $\sigma_8$ and $\Omega_{\mathrm{m}}$ in the remainder of this work. 
More generally, we note that such behavior is consistent with recent findings showing that sharply bounded priors can induce generic edge effects in simulation-based inference, which may degrade the performance of neural posterior estimators when inferred posteriors intersect or extend beyond prior limits \cite{tirapongprasert2026learning}.

\section{Results}
\label{sec:results}

Based on our simulation-based forecast pipeline (Figure~\ref{fig:fig_1}), we intend to evaluate the constraining capability of two sets of summary statistics — morphology and weighted morphology — as well as their various combinations (Table~\ref{tab:summary_stats}), on the desired cosmological parameters, $\boldsymbol{\theta}$. This assessment is conducted over a broad region of parameter space as delineated by the prior, rather than being restricted to a single fiducial cosmology. Thus, the results indicate the performance across a wide range of physically plausible cosmologies, and the notion of forecast precision introduced below should be interpreted as a function of location within the parameter space rather than as a single global average.

\begin{figure}[t]
	\centering
	\includegraphics[width=\linewidth]{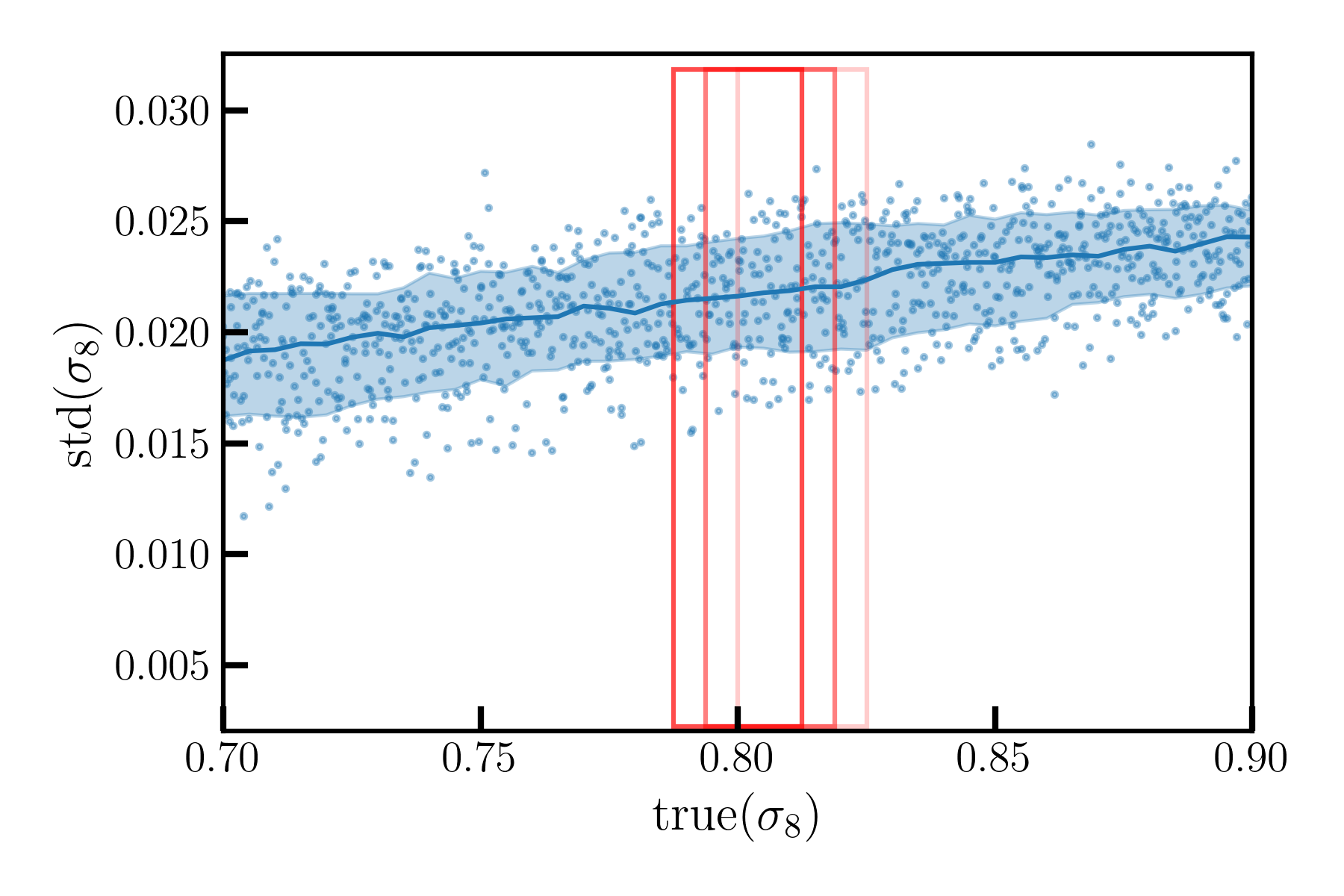}
	\caption{The standard deviation of estimated $\sigma_8$ posteriors across 2000 test simulations (dots) for $\sigma_8\in [0.7-0.9]$. The blue solid line represents the median trend within the red sliding windows. The shaded region encompasses the 16th-84th percentile interval. Here we used the $\bold{MFs}^{(15)}$ as the descriptor. }
	\label{fig:fig_5}
\end{figure}

To this end, we use 2000 independent mock realizations drawn from the BSQ simulation set, which have not been included in the training or validation stages of the posterior estimation. These realizations cover the full prior domain of the five-dimensional cosmological parameter vector. Subsequently, using the trained ensemble model,
$q^{\mathrm{ens}}_\phi(\boldsymbol{\theta} \mid \mathbfcal{D})$,
we estimate the posterior distribution for every realization and each collection of summary statistics. From these inferred distributions, we draw 5000 samples to compute forecast–related metrics. In particular, the standard deviation of the parameters of interest,
\(\mathrm{std}(\boldsymbol{\theta}_i)\),
derived from the sampled ensemble outputs, serves as our primary metric for assessing forecast precision, and is computed separately for each individual test realization.

To build intuition for the interpretation of the forecast precision across the parameter space, Figure~\ref{fig:fig_5} illustrates the case of \(\sigma_{8}\) for a representative choice of summary statistics. Each blue point corresponds to the precision metric, \(\mathrm{std}(\sigma_{8})\), computed from the ensemble samples for an individual realization, and is shown as a function of the true parameter value. The scatter therefore reflects the variation in forecast precision across different cosmologies. 
The blue solid curve traces the median of the precision metric obtained within a sliding window across the prior range, while the shaded region marks the 16th–84th percentile interval, quantifying the variation among realizations with similar true values.  
Importantly, this construction provides a local estimate of forecast uncertainty conditioned on the true value of the parameter, rather than an average over all test simulations.
This diagnostic example is considered as a visual reference for understanding the subsequent figures presenting other parameters and summary statistics configurations.

\begin{figure}[t]
	\centering
	\includegraphics[width=\linewidth]{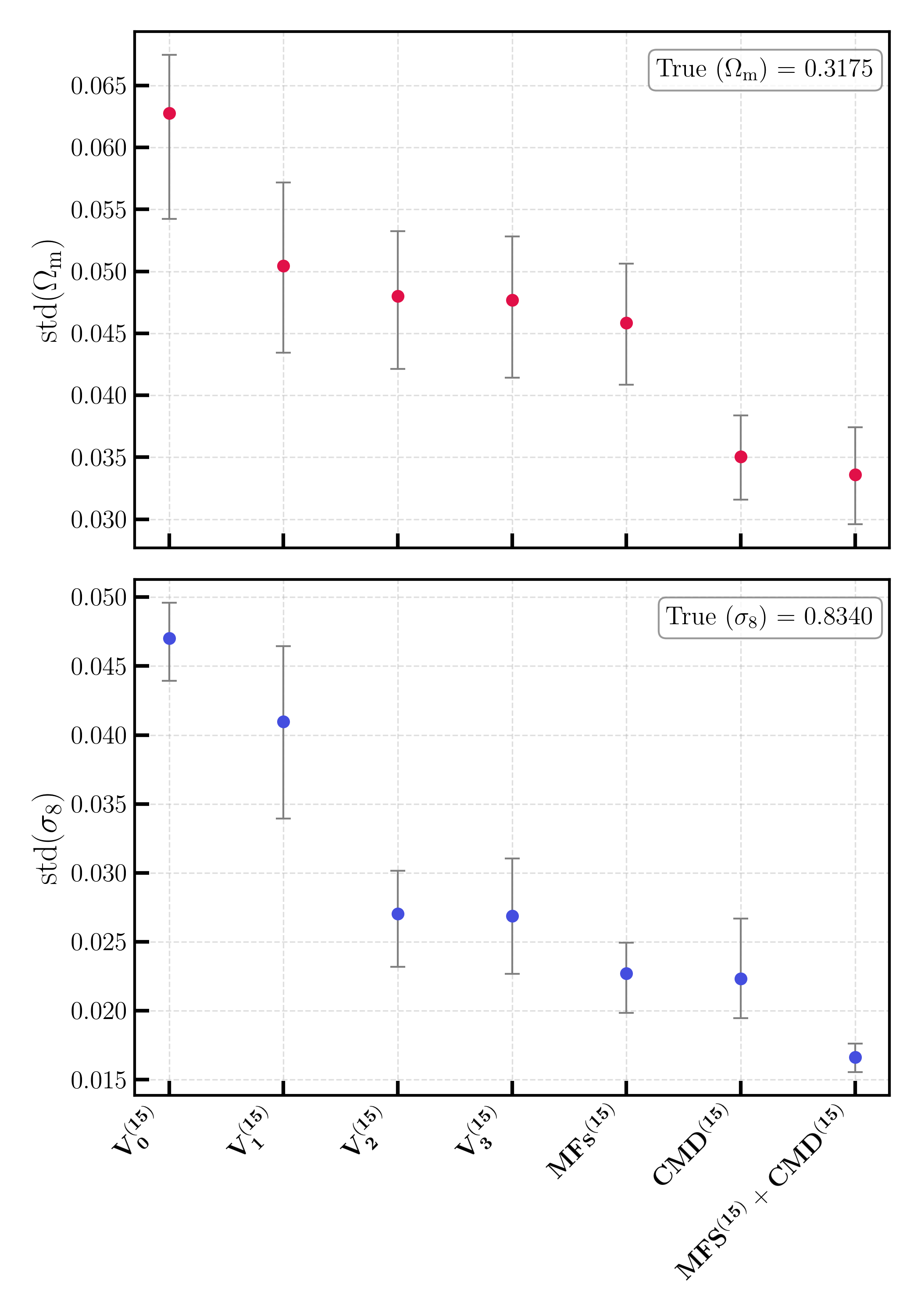}
	\caption{
		Forecast uncertainties, $\mathrm{std}(\theta_i)$, of $\Omega_{\mathrm{m}}$ (top) and $\sigma_8$ (bottom) obtained from different morphological summary statistics at $R=15\,h^{-1}\mathrm{Mpc}$, evaluated at the fiducial cosmology of the Quijote simulations, $(\Omega_{\mathrm{m}}=0.3175,\sigma_8 = 0.834)$.  
		Each point represents the median forecast precision estimated from BSQ realizations falling within the corresponding parameter bins, while vertical bars indicate the 16th–84th percentile range among realizations. 
		These realizations are selected using the same sliding-window procedure illustrated in Figure~\ref{fig:fig_5}, centered on the quoted fiducial parameter values.
	}
	\label{fig:fig_6}
\end{figure}

\subsection{Morphology vs. Weighted Morphology}
Building on the visual reference provided by Figure~\ref{fig:fig_5}, we now turn to a systematic comparison of forecast precision across multiple cosmological parameters and summary statistics. 
First, we focus on the (weighted) morphological measures in redshift-space, adopting a fixed smoothing scale of $R = 15\,h^{-1}\mathrm{Mpc}$. 
Figure~\ref{fig:fig_6} presents the forecast uncertainties, $\mathrm{std}(\theta_i)$, for $\Omega_{\mathrm{m}}$ (top) and $\sigma_8$ (bottom) at the fiducial cosmology of the Quijote simulation ($\Omega_{\mathrm{m}} = 0.3175$ and $\sigma_8 = 0.834$)~\cite{Quijote_sims}, which is very close to the \textit{Planck} results. 
We emphasize that these fiducial uncertainties do not correspond to a single test realization, but instead summarize the typical forecast precision of an ensemble of nearby cosmologies sharing the same value of the parameter under consideration.

In the following, we first justify the pronounced gap in the constraining power between $\bold {V}_{0}^{(15)}$ and the other MFs components illustrated in the upper panel of Figure~\ref{fig:fig_6}. In real space, the mathematical form of the MFs is primarily governed by the ratio of spectral-moments\footnote{In the context of an isotropic random field $\delta(\mathbf{x})$, the spectral moments $\sigma_j$ are defined by
	$	\sigma_j^2 = \frac{1}{2\pi^2} \int_0^{\infty} k^{2j+2} P(k)\, W^2(kR)\, \mathrm{d}k$,
	where $P(k)$ denotes the power spectrum and $W(kR)$ is the smoothing kernel of scale $R$.}, $\sigma_1/\sigma_0$, which determines the amplitude of the Gaussian component, and by non‑Gaussian correction terms introduced through nonlinear gravitational evolution~\cite{matsubara2022minkowski}. When RSDs are incorporated, both components are modified: the Gaussian amplitude is rescaled by a factor that depends on the growth rate of structure, $f(\Omega_{\mathrm{m}})$, while the non‑Gaussian corrections gain an additional dependence from the coupling between velocity distortions and nonlinear gravitational effects~\cite{matsubara1996statistics,codis2013non}. Consequently, in redshift-space, the MFs exhibit three distinct sources of $\Omega_{\mathrm{m}}$ dependence: (i) the intrinsic spectral‑moment ratio defined in real space, (ii) the amplitude modulation induced by RSDs, and finally (iii) the RSDs‑modified non‑Gaussian corrections. Among the four MFs, these effects manifest in different ways.
Specifically, $\bold{V}_1$, $\bold{V}_2$, and $\bold{V}_3$ collectively capture information related to the field’s gradient and curvature,  where all three contributions are present. 
In contrast, $\bold{V}_0$ lacks both the spectral‑moment and amplitude‑modulation terms, retaining only the weak non‑Gaussian dependence. As a result, $\bold{V}_0$ exhibits substantially weaker constraining power on $\Omega_{\mathrm{m}}$ compared to the other MFs components — a trend that is clearly visible in the upper panel of Figure~\ref{fig:fig_6}. 

The scalar nature of MFs limits their sensitivity to the directional anisotropy of RSD along and perpendicular to the LOS.  In contrast, the $\boldsymbol{\mathrm{CMD}}$ statistics, by construction, preserves directional information, allowing its posterior to respond more directly to the RSD‑induced anisotropy and thereby yielding stronger constraint on $\Omega_{\mathrm{m}}$. In addition to the directional dependency, the amplitude of $\boldsymbol{\mathrm{CMD}}$  is proportional to the spectral-moments in real space by $\sigma_1^2/\sigma_0$ (see Eq.~22 of~\cite{kanafi2024probing}), which leads to a further enhancement of its sensitivity to $\Omega_m$ compared to scalar MFs.

Our results also demonstrate that the constraining power of $\bold{V}_{2}$ and $\bold{V}_{3}$ on $\sigma_8$ surpasses that of $\bold{V}_{0}$ and $\bold{V}_{1}$  (lower panel of Figure~\ref{fig:fig_6}). 
The dependence of the MFs on $\sigma_8$ arises through two main channels: 
(i) the spectral-moments, and (ii) the non‑Gaussian corrections that trace the nonlinear growth of structure. 
Since $\sigma_8$ mainly rescales the overall amplitude of the matter power spectrum, its impact on $\sigma_1$ and $\sigma_0$ almostly cancels in the ratio $(\sigma_1/\sigma_0)^{\,n}$ for $\bold{V}_n$~\cite{matsubara2022minkowski}. Consequently, the effective sensitivity of entire MFs components to $\sigma_8$ originates primarily from the non‑Gaussian corrections. 
Among these components, $\bold{V}_{2}$ and $\bold{V}_{3}$ probe the geometry and curvature of the isodensity surfaces and are therefore more responsive to small‑scale fluctuations and nonlinearities than $\bold{V}_{0}$ and $\bold{V}_{1}$, which only quantify the volume or surface area enclosed by the isodensity contours. 
This explains their comparatively stronger constraints on $\sigma_8$. 
On the other hand, the amplitude of the $\boldsymbol{\mathrm{CMD}}$ scales with the ratio $\sigma_1^2/\sigma_0$ (see~\cite{kanafi2024probing,afzal2025cosmic}), introducing an explicit dependence on the spectral amplitude itself. 
This asymmetry in the exponents of the spectral-moments enhances the explicit sensitivity of $\boldsymbol{\mathrm{CMD}}$ to $\sigma_8$, leading to its superior constraining power compared to the scalar MFs components.

Quantitatively, at the adopted fiducial values in Figure~\ref{fig:fig_6}, utilizing $\bold{CMD}^{(15)}$ improves constraint on the $(\Omega_{\rm m}, \sigma_8)$ with respect to $\bold{V}^{(15)}_{0}$, $\bold{V}^{(15)}_{1}$, $\bold{V}^{(15)}_{2}$, $\bold{V}^{(15)}_{3}$ as
 $\sim$ $(44\%^{+6\%}_{-6\%}, 52\%^{+9\%}_{-6\%})$, $(30\%^{+8\%}_{-5\%}, 45\%^{+10\%}_{-6\%})$, $(27\%^{+6\%}_{-5\%}, 17\%^{+10\%}_{-8\%})$ and $(26\%^{+7\%}_{-6\%}, 17\%^{+20\%}_{-10\%})$, respectively. In addition, we assess a joint configuration in which $\bold{CMD}$ replaces $\bold{V}_{1}$ within the full $\bold{MFs}$ set, motivated by their structural similarity as mentioned before (compare Eqs.~\ref{eq:v1} and~\ref{eq:cmd}). The $\bold{MFs}^{(15)} + \bold{CMD}^{(15)}$ setup improves constraints by $(26\%^{+7\%}_{-5\%}, 27\%^{+9\%}_{-5\%})$ for $(\Omega_{\rm m}, \sigma_8)$ compared to the standard $\bold{MFs}^{(15)}$ combination, underscoring the complementary cosmological information provided by anisotropy-sensitive weighted morphology relative to purely scalar measures.

\begin{figure}[t]
	\centering
	\includegraphics[width=\linewidth]{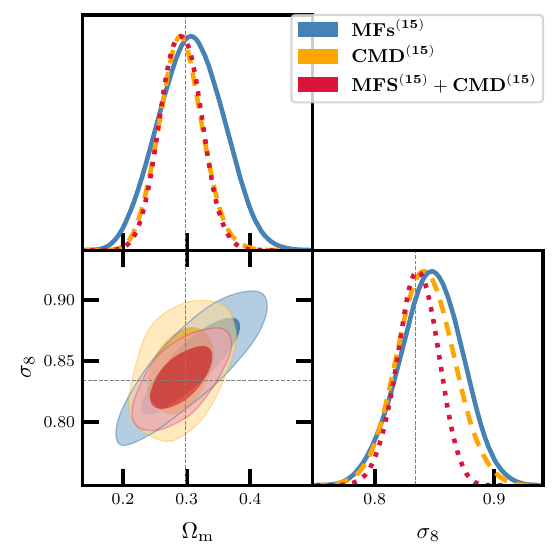}
\caption{
	Comparison of the  cosmological constraints obtained from the (weighted) morphological statistics at $R=15\,h^{-1}\mathrm{Mpc}$. 
	Shown are the two-dimensional posterior distributions for $(\Omega_{\mathrm{m}},\sigma_8)$ from $\mathbf{MFs}^{(15)}$ (blue), $\mathbf{CMD}^{(15)}$ (orange), and their joint combination $\mathbf{MFs}^{(15)}+\mathbf{CMD}^{(15)}$ (red). 
	All contours correspond to the $68\%$ and $95\%$ credible regions, estimated at the fiducial cosmology (gray dashed lines). 
}
	\label{fig:fig_7}
\end{figure} 
 
Following the quantitative ranking presented in Figure~\ref{fig:fig_6}, 
it is therefore instructive to examine the two–dimensional posterior in $\Omega_{\mathrm{m}}-\sigma_8$ plane, which reveals not only the relative forecast precision but also the orientation and extent of parameter degeneracies. The corner plot shown in Figure~\ref{fig:fig_7} makes this comparison explicit by illustrating how different morphological summary statistics either mitigate or amplify the degeneracy directions. For the construction of these contours, we selected a single BSQ test realization whose true cosmological parameters lie close to the Quijote fiducial cosmology. From the posterior of each statistics, we randomly drew $5 \times 10^3$ samples to visualize the corresponding $68\%$ and $95\%$ credible regions. As shown in Figure~\ref{fig:fig_7}, the relative sizes of the contours are consistent with the quantitative results reported in Figure~\ref{fig:fig_6}. 
All posteriors follow an almost similar degeneracy direction, reflecting the correlation between $\Omega_{\mathrm{m}}$ and $\sigma_8$. 
However, the area of the credible regions systematically decreases when moving from $\mathbf{MFs}^{(15)}$ (blue) to $\mathbf{CMD}^{(15)}$ (orange), and further to their joint combination $\mathbf{MFs}^{(15)}+\mathbf{CMD}^{(15)}$ (red). 
This visible contraction of the contours confirms that incorporating directional weighted morphological information significantly improves forecast precision for both parameters, in full agreement with the trend observed in Figure~\ref{fig:fig_6}.

\begin{figure*}[t]
	\centering
	\includegraphics[scale=0.6]{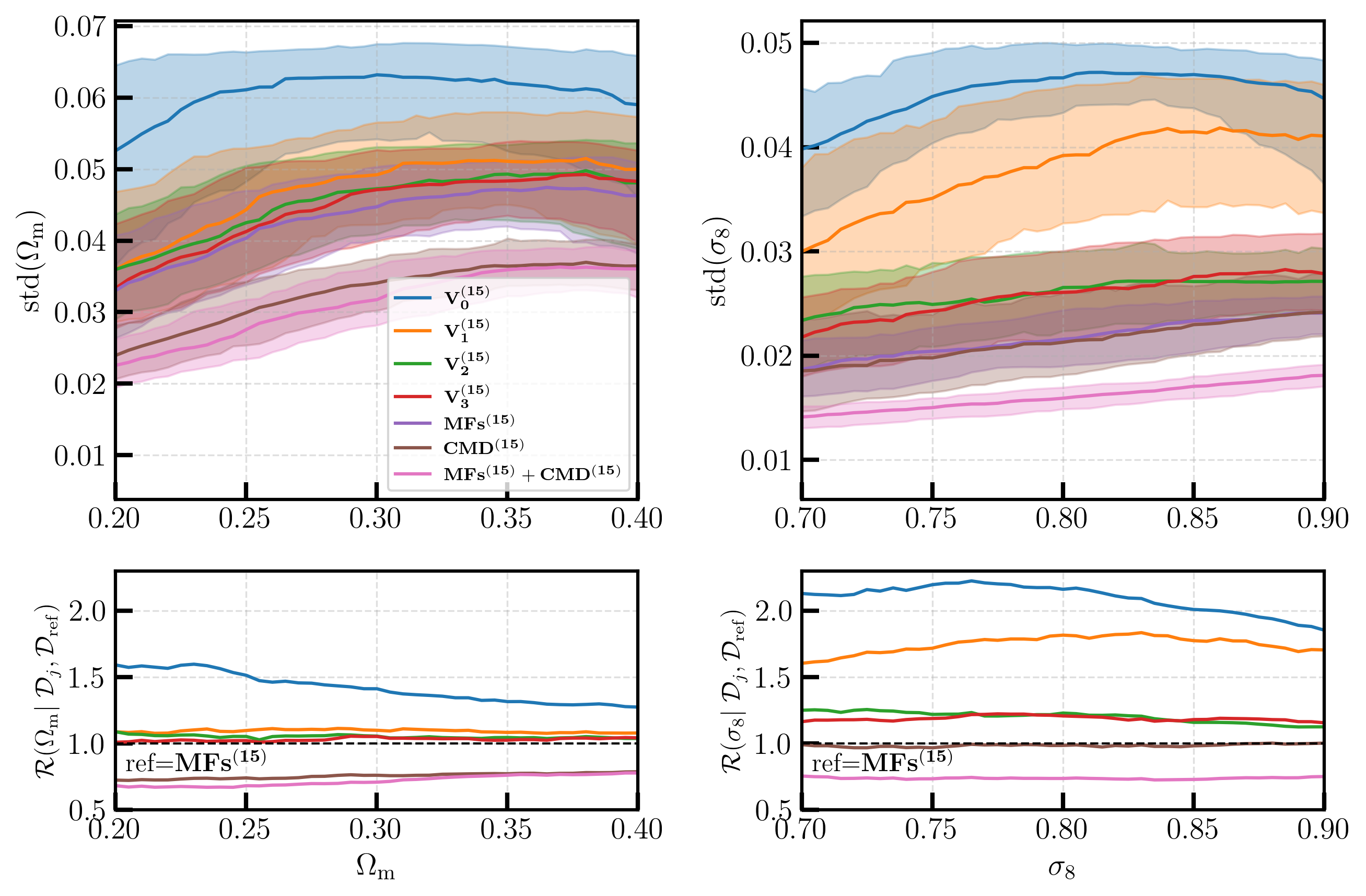}
\caption{
	Forecast uncertainties of $\Omega_{\mathrm{m}}$ (left) and $\sigma_8$ (right) derived from different morphological summary statistics at $R=15\,h^{-1}\mathrm{Mpc}$. 
	Solid curves and shaded regions are constructed following the procedure illustrated in Figure~\ref{fig:fig_5}, where median trends of $\mathrm{std}(\theta_i)$ are obtained within sliding parameter boxes and shaded bands represent the 16th–84th percentile scatter among BSQ realizations. 
	Upper panels show the variation of forecast precision across the parameter space, and lower panels present ratios relative to the reference statistics $\mathbf{MFs}^{(15)}$.
}
	\label{fig:fig_8}
\end{figure*}

\begin{figure*}[t]
	\centering
	\begin{minipage}{0.7\linewidth}
		\includegraphics[width=\linewidth]{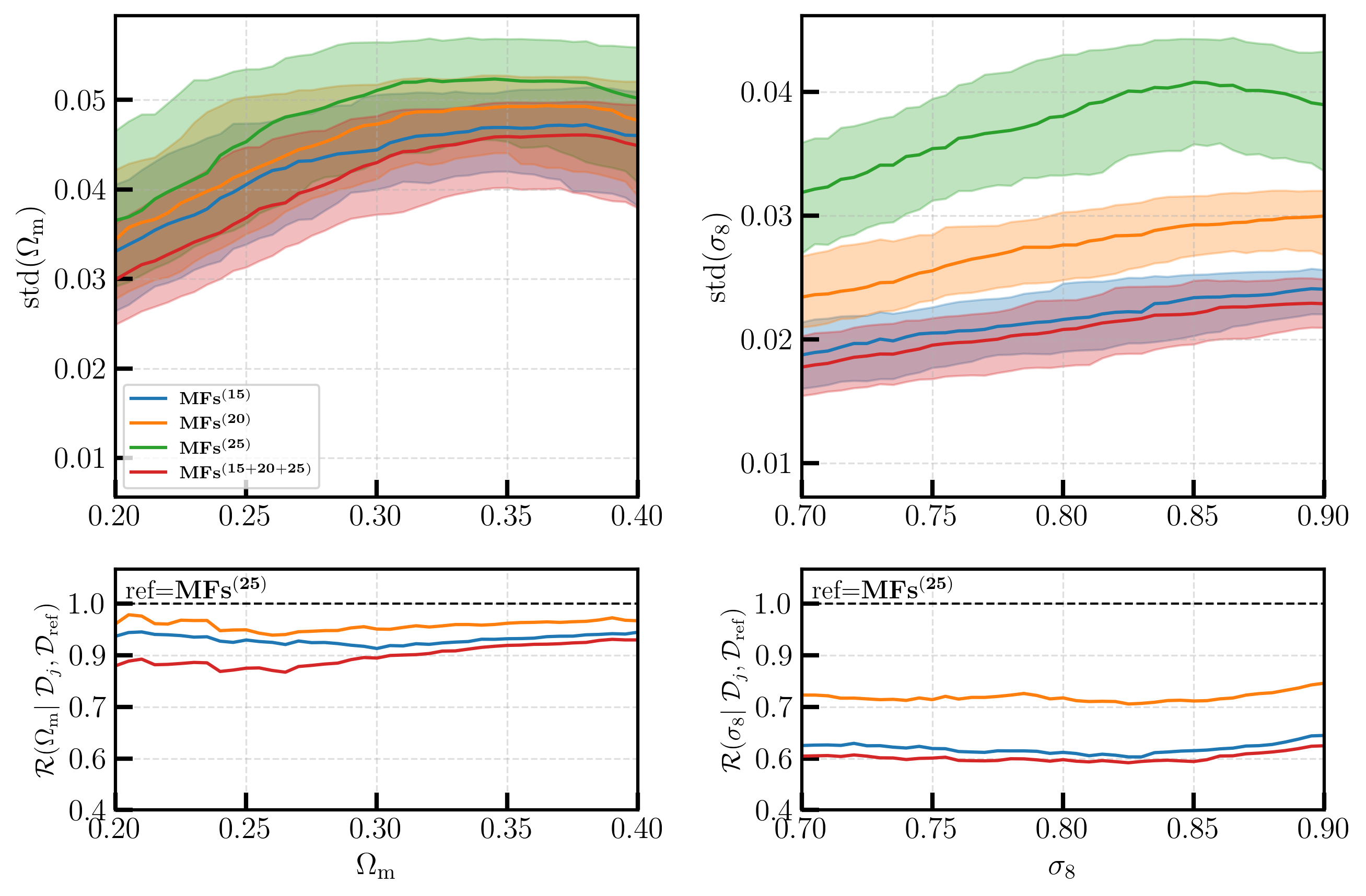}
		\put(-180,240){\textbf{(a)}}
	\end{minipage}
	\par\vspace{0.3em}
	\begin{minipage}{0.7\linewidth}
		\includegraphics[width=\linewidth]{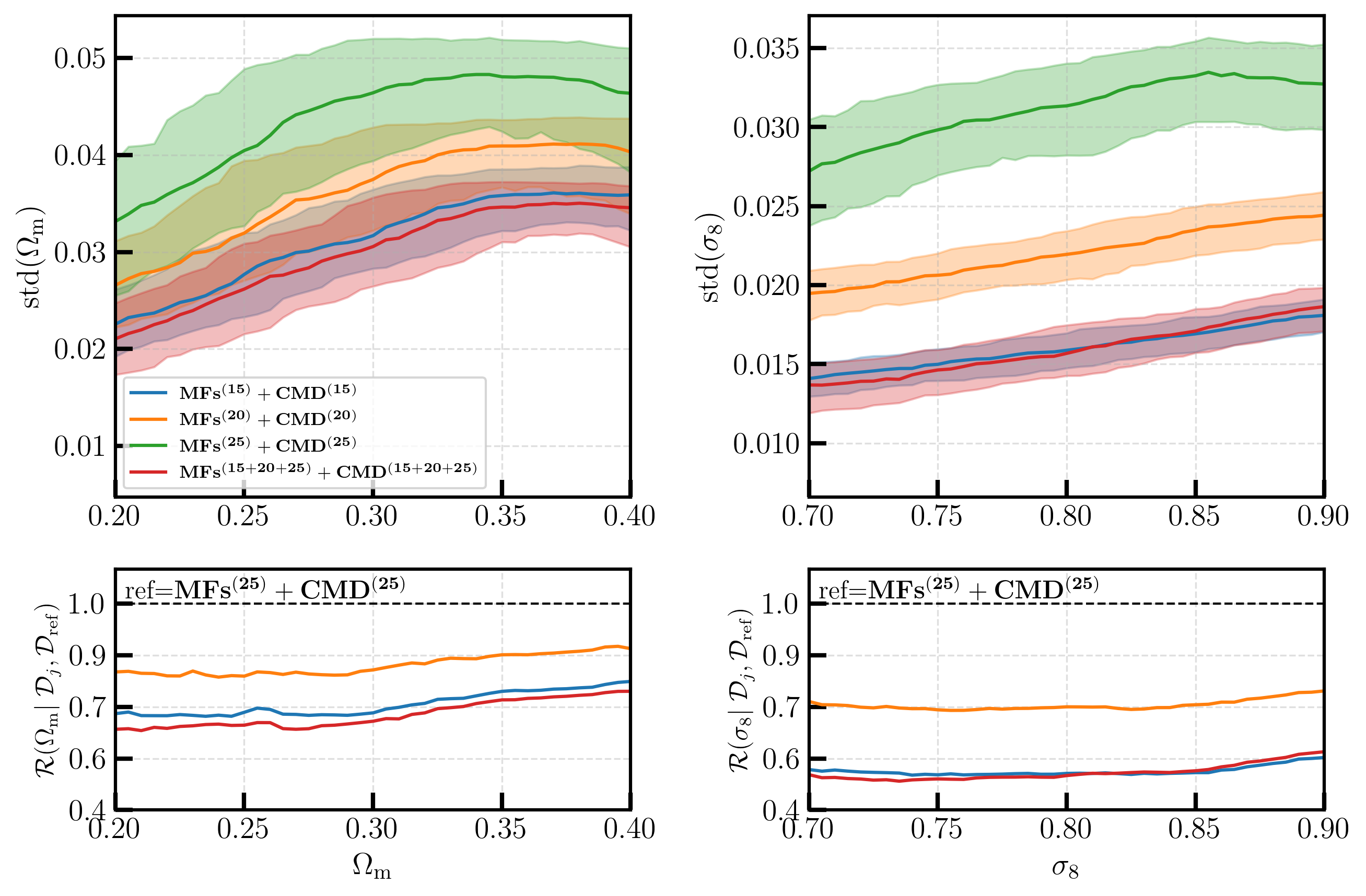}
		\put(-180,240){\textbf{(b)}}
	\end{minipage}
	\caption{
		Forecast precision dependence on the smoothing scale for two morphological descriptors: $\bold{MFs}$ (panel~$\bold{a}$) and $\bold{MFs} + \bold{CMD}$ (panel~$\bold{b}$). 
		The plotting conventions and definitions follow those described in Figure~\ref{fig:fig_8}. 
		Here, we adopt three smoothing lengths, $R = 15$, $20$, and $25\,h^{-1}\mathrm{Mpc}$, with the largest scale ($R=25\,h^{-1}\mathrm{Mpc}$) used as the reference configuration for comparing the relative constraining power. 
		Black dashed lines in the lower sub-panels mark this reference level.
	}
	\label{fig:fig_9}
\end{figure*}

In the next step, we go beyond the single fiducial cosmology and perform forecasts across a continuous range of parameter values, aiming to assess how the constraining capability of different summary statistics varies throughout the cosmological parameter space. 
Figure~\ref{fig:fig_8} indicates the std($\Omega_{\mathrm{m}}$) (left panel) and std($\sigma_8$) (right panel) as a function of various fiducial values. Each curve corresponds to the summary statistics employed in the fiducial forecast of Figure~\ref{fig:fig_6} for smoothing scale of $R = 15\,h^{-1}\mathrm{Mpc}$.  In the upper panels, the trends (solid curves) and shaded intervals are constructed following the same procedure described in Figure~\ref{fig:fig_5}. As one can see, the qualitative hierarchy in constraining power among the summary statistics—previously identified in the fixed fiducial analysis (Figure~\ref{fig:fig_6})—persists throughout the explored domain. These panels also illustrate how the constraining power evolves across the entire parameter space.  It is worth noting that, for both $\Omega_{\mathrm{m}}$ and $\sigma_8$, and for all our summary statistics, the level of uncertainty ($\mathrm{std}(\theta_i)$) depends noticeably on the underlying cosmology. Based on the solid curves, the highest relative variation for $\Omega_{\mathrm{m}}$ corresponds to the combined statistics  $\bold {MFs}^{(15)} + \bold {CMD}^{(15)}$ ($\sim61\%$) for given domain, whereas the smallest variation is found for $\bold{V}^{(15)}_{0}$ ($\sim20\%$). For $\sigma_8$, the strongest variation occurs for $\bold{V}^{(15)}_{1}$ ($\sim39\%$) and the most stable behavior for $\bold{V}^{(15)}_{2}$ ($\sim16\%$). Such variations in the sensitivity and forecasted uncertainties of the summary statistics are  expected, given their inherently non-linear dependence on the cosmological parameters. The magnitude of the uncertainty reflects how the derivatives of the statistics with respect to the parameters, $\partial\mathbfcal{D}/\partial\boldsymbol{\theta}_i$, vary across the parameter space. We explicitly verified this behavior using Gaussian predictions for the MFs components, calculated with the power spectra obtained by \texttt{CAMB} software~\cite{Lewis:1999bs}. Specifically, the theoretical form of first MFs in the Gaussian regime implies that the std$(\Omega_{\mathrm{m}})$ grows monotonically with increasing $\Omega_{\mathrm{m}}$, in agreement with our numerical results. 
The exact amplitude of the growth, however, depends on several secondary factors, including contributions from non-Gaussian corrections,  halo bias, and the intrinsic noise in the summary statistics themselves.

A further qualitative insight can be drawn from the relative size of the shaded regions surrounding each curve in Figure~\ref{fig:fig_8}. For a given summary statistics, the width of its shaded band remains almost constant across the scanned parameter range, indicating that the intrinsic scatter of the forecast uncertainties is not strongly affected by the specific value of the cosmological parameter. When comparing the amplitudes of these bands across the different statistics, the same hierarchy observed in their constraining power is reproduced, with the narrowest bands obtained for the combined measure $\bold{MFs}^{(15)}+\bold{CMD}^{(15)}$. The relative narrowness of the shaded region for $\bold{MFs}^{(15)}+\bold{CMD}^{(15)}$ highlights its superior stability and lower realization‑to‑realization variance, implying that the mentioned measure provides more robust constraining capability across cosmological models.

We also define the ratio of median forecast uncertainties for each summary statistics, 
$\mathrm{Med}\!\left[\mathrm{std}(\boldsymbol{\theta}_i)\mid\mathcal{D}_{j}\right]$, 
relative to a given reference descriptor, as
\begin{equation}
	\mathcal{R}(\boldsymbol{\theta}_i \mid \boldsymbol{\mathcal{D}}_{j},\boldsymbol{\mathcal{D}}_{\mathrm{ref}})\equiv
	\frac{\mathrm{Med}\!\left[\mathrm{std}(\boldsymbol{\theta}_i)\mid\mathbfcal{D}_{j}\right]}
	{\mathrm{Med}\!\left[\mathrm{std}(\boldsymbol{\theta}_i)\mid\mathbfcal{D}_{\mathrm{ref}}\right]},
\end{equation}
This quantity highlight the comparative constraining power of summary statistics across the parameter domain. In the lower panels of Figure~\ref{fig:fig_8}, we display $\mathcal{R}$  when the reference summary statistics is adopted by ${\mathbf{MFs}^{(15)}}$  for $\Omega_{\rm m}$ (left side) and  for $\sigma_{8}$ (right side). While the absolute strength of the constraints depends on the parameter values, as seen in the upper panels, the relative constraining power among the summary statistics remains nearly constant over the considered range of model parameters, in analogy with the results reported for the fixed fiducial cosmology in Figure~\ref{fig:fig_6}.

As a quantitative example, the combined measure $\bold {MFs}^{(15)} + \bold {CMD}^{(15)}$ provides an improvement of roughly 22--32\% for $\Omega_m$ and 24--28\% for $\sigma_8$ compared to $\bold {MFs}^{(15)}$, confirming their complementary roles in enhancing precision. Such resilience against variation of fiducial values indicates that the comparative performance of the morphological descriptors is almost not affected by changes in cosmological parameters. 

To elucidate the dependence of the constraining power on the smoothing scale, we evaluate the forecast uncertainties of $\Omega_{\mathrm{m}}$ and $\sigma_8$ for different values of $R$. Figure~\ref{fig:fig_9} presents the results for two cases: $\bold{MFs}$ (panel $\bold{a}$) and $\bold{MFs}+\bold{CMD}$ (panel $\bold{b}$). In each panel, the statistics computed at $R=25\,h^{-1}\mathrm{Mpc}$ serves as the reference case for normalization of the relative precision, $\mathcal{R}(\boldsymbol{\theta}_i \mid \boldsymbol{\mathcal{D}}_{j},\boldsymbol{\mathcal{D}}_{\mathrm{ref}})$. 
For the $\bold{MFs}$ (top panel), the uncertainty on $\Omega_{\mathrm{m}}$ varies only mildly—by less than $10\%$ between $R=15$ and $R=25\,h^{-1}\mathrm{Mpc}$—whereas the precision on $\sigma_8$ changes by nearly $40\%$. This dissimilarity between $\Omega_{\mathrm{m}}$ and $\sigma_8$ originates from the distinct physical sources of information encoded in the $\bold {MFs}$. The constraining power on $\sigma_8$ primarily stems from non–Gaussian corrections to the morphological measures, which are strongly suppressed at large scales. Whereas, as we mentioned earlier, the sensitivity to $\Omega_{\mathrm{m}}$ arises through three channels: (i) the real–space spectral ratio $\sigma_1/\sigma_0$, (ii) the amplitude correction induced by RSDs (including Kaiser and Finger-of-God effects), and (iii) non–Gaussian corrections. Among these, the Kaiser effect persists across smoothing scales and thus maintains a nearly constant contribution, while the spectral ratio term decreases only modestly with $R$ in the range $[15,25]\,h^{-1}\mathrm{Mpc}$ (see e.g.~\cite{matsubara2022minkowski}), explaining the limited $\sim 10\%$ weakening of the $\Omega_{\mathrm{m}}$ constraint.
Although increasing $R$ also reduces the effective shot-noise through spatial averaging, within this range of smoothing scales this effect is subdominant compared to the loss of non--Gaussian and small--scale structural information, which ultimately drives the degradation of the constraints.

The lower panel of Figure~\ref{fig:fig_9} shows the corresponding analysis for the $\bold{MFs}+\bold{CMD}$ statistics. The overall behavior of the constraining power with changing the smoothing scale is almost similar to that of the $\bold{MFs}$, but the degradation for $\Omega_{\mathrm{m}}$ is slightly stronger. This is because the expected amplitude of $\bold{CMD}$ depends on the ratio $\sigma_1^2/\sigma_0$, whose suppression with increasing smoothing scale $R$ is considerably stronger than that of the $\sigma_1/\sigma_0$ governing the $\bold{MFs}$. As a result, the Kaiser correction and spectral-moment contributions that trace $\Omega_{\mathrm{m}}$ are more strongly enhanced in $\bold{CMD}$ with decreasing $R$, causing its effective information content to increase more rapidly than for the purely scalar $\bold{MFs}$.

\subsection{Comparison with Power Spectrum}
\label{sec:comparison_ps}

	To further evaluate the statistical efficiency of our (weighted) morphological estimators, we compare the constraining power of the combined statistic $\mathbf{MFs}^{(15)} + \mathbf{CMD}^{(15)}$ with that of the halo power spectrum measured in redshift space, which we adopt as a benchmark. The power spectrum analysis is performed both for the monopole-only case ($\ell=0$) and for the joint combination of monopole, quadrupole, and hexadecapole ($\ell=0,2,4$), allowing us to explicitly assess the impact of anisotropic information encoded in redshift-space distortions.

	To ensure a fair comparison, the power spectrum is restricted to the same effective scale range as the morphological analysis, corresponding to a minimum Gaussian smoothing radius of $R=15\,h^{-1}\mathrm{Mpc}$. This choice is primarily driven by practical considerations related to halo sampling density and the stability of simulation-based inference. In particular, suppressing shot-noise and retaining a sufficiently large number of realizations with adequate tracer density becomes increasingly challenging at smaller smoothing scales\footnote{
		Previous studies have shown that an optimal smoothing scale for genus and MFs analyses can be expressed as $R=\bar{d}/\sqrt{2}$, where $\bar{d}=\bar{n}_h^{-1/3}$ denotes the mean tracer separation~\cite{vogeley1994topological, gott1989topology, Liu2023ProbingMN}. Given the halo number density distribution of the BSQ simulations, adopting $R=15\,h^{-1}\mathrm{Mpc}$ retains $\sim3.2\times10^{4}$ realizations with $\bar{n}_h\gtrsim10^{-4}\,h^3\mathrm{Mpc^{-3}}$, whereas reducing the smoothing scale to $R=10\,h^{-1}\mathrm{Mpc}$ leaves only $\sim8\times10^{3}$ realizations, significantly degrading the robustness of SBI.}.

\begin{figure}[t]
	\centering
	\includegraphics[width=\linewidth]{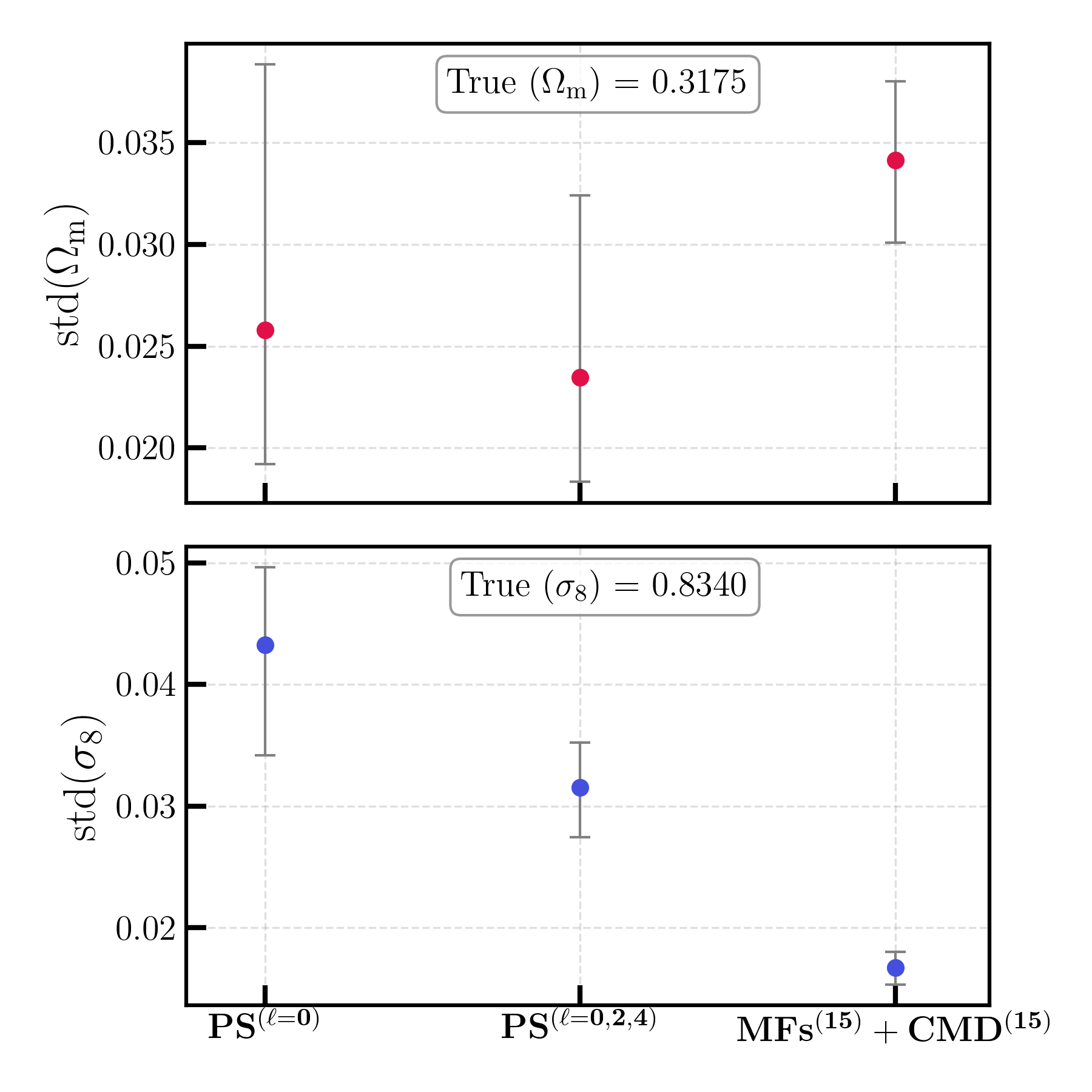}
	\caption{
			Similar to Figure~\ref{fig:fig_6},  but here comparing $\mathbf{MFs}^{(15)}+\mathbf{CMD}^{(15)}$, evaluated at $R=15\,h^{-1}\mathrm{Mpc}$, with the halo power spectrum in redshift space, measured up to $k_{\max}\simeq0.16\,h\,\mathrm{Mpc^{-1}}$. Power spectrum results are presented for both the monopole-only case ($\mathbf{PS^{(\ell=0)}}$) and the joint multipole combination ($\mathbf{PS^{(\ell=0,2,4)}}$). All constraints are evaluated at the fiducial cosmology $\Omega_{\mathrm{m}}=0.3175$ (top panel) and $\sigma_8 = 0.834$ (bottom panel).
	}
	\label{fig:fig_10}
\end{figure}

	Following the prescription $R\,k_{\max}=2.5$ (utilized in~\cite{Liu2023ProbingMN}), the adopted smoothing scale corresponds to an effective Fourier-space cutoff of $k_{\max}\simeq0.16\,h\,\mathrm{Mpc^{-1}}$, placing the analysis in a mildly non-linear regime. Consequently, the present comparison is not intended to probe the deeply non-linear scales where morphological statistics are expected to yield their strongest gains, but rather to assess the additional constraining power provided by anisotropy-sensitive and weighted morphological information relative to the power spectrum in redshift space at matched effective scales.

	Figure~\ref{fig:fig_10} summarizes the resulting parameter constraints. We first consider the power spectrum alone. Including higher-order multipoles beyond the monopole leads to a modest improvement in constraining power for $\Omega_{\mathrm{m}}$, with uncertainties reduced by $\sim11\%^{+12\%}_{-13\%}$, consistent with only a marginal gain. In contrast, the improvement for $\sigma_8$ is substantially stronger, reaching $\sim25\%^{+14\%}_{-2\%}$. This trend reflects the additional constraining information carried by redshift-space anisotropies, particularly for parameters that are closely tied to the overall clustering amplitude.

	We next compare the combined morphological statistic $\mathbf{MFs}^{(15)}+\mathbf{CMD}^{(15)}$ to the power spectrum. For $\sigma_8$, the morphological estimator yields a substantial improvement of $\sim60\%^{+7\%}_{-5\%}$ relative to the  $\mathbf{PS^{(\ell=0)}}$, and remains significantly more constraining even when compared to the full multipole combination $\mathbf{PS^{(\ell=0,2,4)}}$ ($\sim47\%^{+6\%}_{-4\%}$). In contrast, constraints on $\Omega_{\mathrm{m}}$ obtained from the power spectrum are slightly tighter on these mildly non-linear scales; however, the difference with respect to the morphological constraints is not statistically significant given the uncertainties shown in the figure.

	Overally, this comparison demonstrates that weighted morphological statistics are particularly sensitive to the clustering amplitude, achieving constraints on $\sigma_8$ that surpass those of the redshift-space power spectrum at matched effective scales, while providing constraints on $\Omega_{\mathrm{m}}$ that are broadly comparable.

\subsection{Impact of Halo Number Density and Mass Resolution}

Throughout the analyses presented so far, we have relied on the full FoF halo catalogs from the BSQ simulations, without applying any minimum halo mass cut or enforcing a fixed tracer number density across realizations. Variations in halo abundance can modify the effective shot-noise contribution to the summary statistics, but they also constitute part of the physical information encoded in the halo number-density contrast field, particularly through its amplitude and spatial fluctuations. In addition, the inclusion of low-mass halos may generate numerical artifacts due to limited mass resolution. Both effects could, in principle, influence the resulting cosmological constraints.

To disentangle these effects, we perform two complementary tests. First, we enforce a fixed tracer abundance by selecting $10^5$ of the most massive halos in each realization. While this procedure equalizes the number density and controls abundance-driven sampling effects, it can also alter the variance and contrast of the underlying density field. Second, we apply a constant mass threshold across all simulations, retaining only halos with masses above $M_{\rm min}=3\times10^{13}\,h^{-1}M_\odot$, which directly probes the sensitivity of our results to marginally resolved low-mass objects.

\begin{figure}[t]
	\centering
	\includegraphics[width=\linewidth]{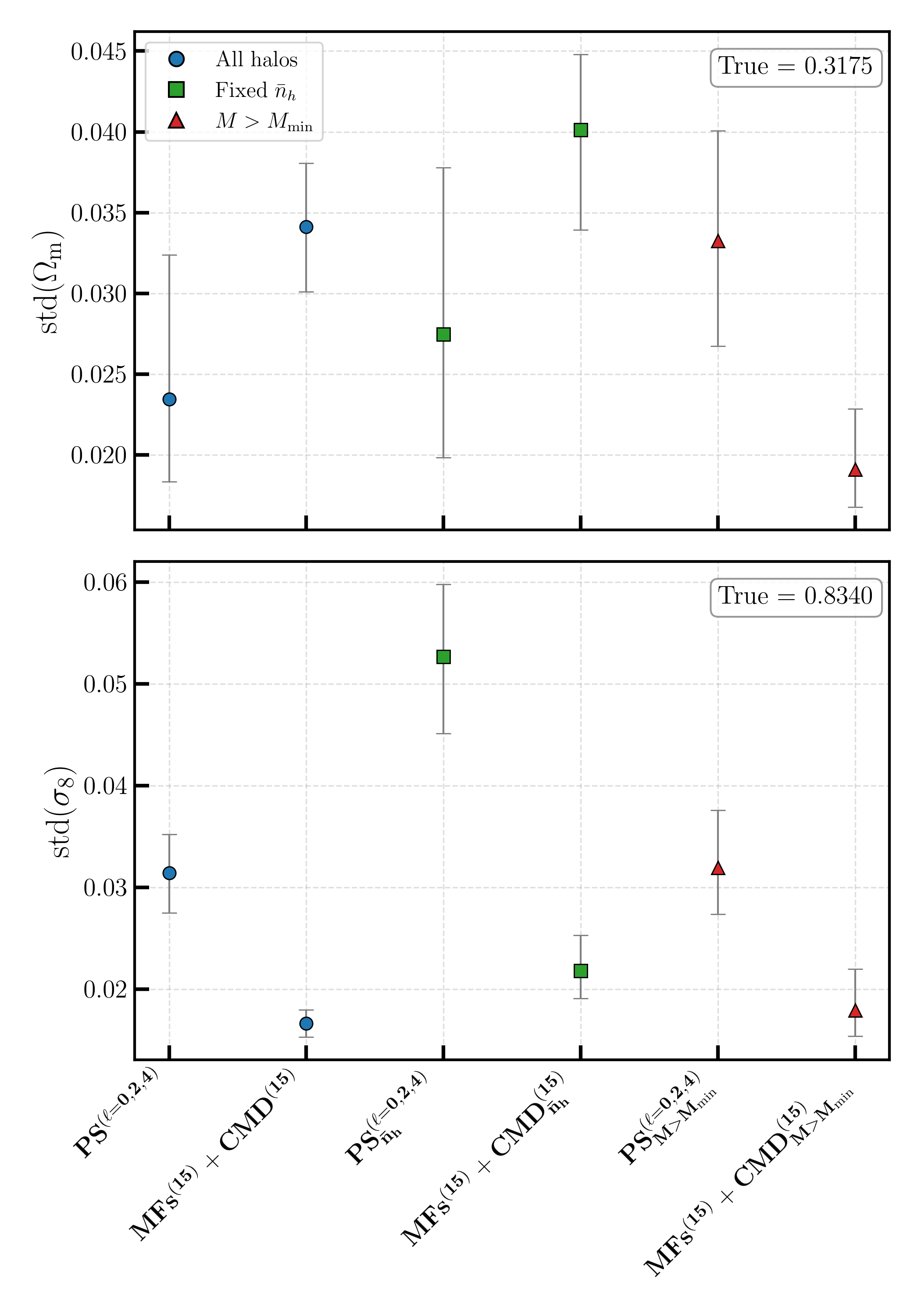}
	\caption{Same as Figure~\ref{fig:fig_10}, but for different halo selections. 
Blue circles show the baseline analysis including all halos.
Green squares correspond to the fixed number density sample with $\bar n_h= 10^{-4}\,h^3\mathrm{Mpc^{-3}} $,
while red triangles denote the fixed mass threshold sample with
$M > M_{\min}=3\times10^{13}\,h^{-1}M_\odot$. For each halo selection, constraints are shown for the power spectrum multipoles,
$\mathbf{PS^{(\ell=0,2,4)}}$, and for the combined weighted morphological statistics, $\mathbf{MFs}^{(15)}+\mathbf{CMD}^{(15)}$.} 
	\label{fig:fig_11}
\end{figure}

Following the analysis pipeline described in previous subsection, we apply the same prescription to both configurations introduced above, treating them on equal footing with the baseline analysis. The resulting parameter constraints are summarized in Figure~\ref{fig:fig_11}, which compares $\mathbf{MFs}^{(15)}+\mathbf{CMD}^{(15)}$ and $\mathbf{PS^{(\ell=0,2,4)}}$ for three cases: the baseline analysis including all halos (blue circles), the fixed number density sample with $\bar n_h= 10^{-4}\,h^3\mathrm{Mpc^{-3}} $ (green squares), and the fixed mass threshold sample with $M>M_{\min}$ (red triangles).

We first consider the fixed number density configuration. Imposing a constant tracer abundance leads to a clear reduction in constraining power for both summary statistics. This reflects not only the suppression of abundance-driven variations across realizations, but also a modification of the halo number-density contrast field on which the statistics are evaluated. However, the magnitude of this degradation differs substantially between the two approaches. For the $\mathbf{PS^{(\ell=0,2,4)}}$, fixing the number density results in a weakening of the constraints, with uncertainties on $\sigma_8$ increasing by $54\%^{+22\%}_{-24\%}$ and those on $\Omega_{\mathrm{m}}$ by $17\%^{+25\%}_{-30\%}$ relative to the baseline all-halos case. This loss of sensitivity, particularly for $\sigma_8$, indicates that a significant fraction of the power spectrum’s constraining power on these scales originates from abundance-driven information that is effectively suppressed when the tracer density is controlled.

In other hand hand, the combined  $\mathbf{MFs}^{(15)}+\mathbf{CMD}^{(15)}$ statistic exhibits a weaker dependence on fixing the number density. The corresponding constraints degrade by $30\%^{+15\%}_{-17\%}$ for $\sigma_8$ and by $14\%^{+14\%}_{-13\%}$ for $\Omega_{\mathrm{m}}$, demonstrating that (weighted) morphological estimators retain most of their cosmological sensitivity even when abundance fluctuations are suppressed. As a result, the relative performance of (weighted) morphology improves under controlled sampling conditions: while the all-halos analysis yields a $\sim47\%^{+6\%}_{-4\%}$ improvement over the power spectrum for $\sigma_8$, this gain increases to $\sim57\%^{+3\%}_{-7\%}$ in the fixed number density case. This behavior confirms that the constraining power of weighted morphological statistics is primarily driven by their sensitivity to the non-Gaussian geometry and coherent anisotropic structure of the density contrast field, rather than by shot-noise variations associated with tracer abundance.

We next turn to the fixed mass threshold configuration, in which only halos above a constant minimum mass are retained. Unlike the fixed number density test, this selection simultaneously reduces the tracer abundance and enhances the effective bias of the remaining halo population, thereby modifying both the shot-noise properties and the physical information encoded in the halo number density contrast field.
For the $\mathbf{PS^{(\ell=0,2,4)}}$, imposing a mass cut leads to no statistically significant change in the constraint on $\sigma_8$, with a marginal degradation of $1\%^{+18\%}_{-17\%}$, while the constraint on $\Omega_{\mathrm{m}}$ weakens by $36\%^{+27\%}_{-23\%}$. This behavior reflects the loss of abundance-related information combined with an increase in  shot-noise, whose impact is not fully compensated by the higher bias of massive tracers, particularly for parameters sensitive to the overall matter content.
For the $\mathbf{MFs}^{(15)}+\mathbf{CMD}^{(15)}$ statistic, the constraint on $\sigma_8$ remains stable, degrading by only $4\%^{+25\%}_{-13\%}$. Remarkably, however, the constraint on $\Omega_{\mathrm{m}}$ improves significantly, by $42\%^{+6\%}_{-6\%}$ relative to the all-halos baseline. This enhancement indicates that, for (weighted) morphology-based estimators, the increased coherence and nonlinear bias of massive halos outweigh the loss of tracer abundance. By suppressing small-scale noise and enhancing the contrast of large-scale structures, the mass-selected sample amplifies geometric and topological features of the density field that are directly sensitive to $\Omega_{\mathrm{m}}$ at the smoothing scale considered.

These trends are further highlighted by directly comparing the two statistics in the fixed mass threshold case. The $\mathbf{MFs}^{(15)}+\mathbf{CMD}^{(15)}$ estimator outperforms the $\mathbf{PS^{(\ell=0,2,4)}}$ by $45\%^{+20\%}_{-9\%}$ for $\sigma_8$, consistent with the all-halos result, and by $43\%^{+10\%}_{-7\%}$ for $\Omega_{\mathrm{m}}$. In contrast to the baseline analysis, where the power spectrum provides slightly tighter constraints on $\Omega_{\mathrm{m}}$, the mass-selected sample reveals a clear advantage for weighted morphological statistics. This inversion demonstrates that the constraining power of $\mathbf{MFs}^{(15)}+\mathbf{CMD}^{(15)}$ does not rely on low-mass or marginally resolved halos, nor on abundance-driven effects, but instead arises from physically meaningful non-Gaussian large-scale structural information traced more cleanly by well-resolved, coherently biased halos.

\section{Summary and Conclusion}
\label{sec:summary}

Motivated by the imperative to establish stringent constraints on cosmological parameters in the big data era, we proposed a simulation‑based forecasting pipeline (Figure~\ref{fig:fig_1}) that employs two classes of higher‑order summary statistics—the scalar morphology (MFs) and its directional weighted extension (CMD)— to extract non-linear and anisotropy‑induced information embedded within the LSS redshift survey. Relying on halo catalogs from the BSQ simulations at $z=0.5$, the inference of $(\Omega_{\mathrm{m}}, \sigma_8)$ was performed within a likelihood‑free framework using neural posterior estimation (Figure~\ref{fig:fig_3}). The halo samples exhibit number densities ranging from $\sim[0.34,5.70]\times10^{-4}\,h^{3}\mathrm{Mpc}^{-3}$ (Figure~\ref{fig:fig_2}), including their representations in redshift-space where the halo density field has been smoothed over multiple smoothing scales.
We employed the simulation‑based calibration technique based on rank statistics, applied to the validation dataset, to verifying the calibration of the inferred posterior distributions. Our results demonstrated that the best calibration achieved for $\Omega_{\mathrm{m}}$ and $\sigma_8$ confirmed by the uniform shape of rank distribution (Figure~\ref{fig:fig_4}).

We performed the forecast uncertainty for single fiducial value over the $\Omega_{\mathrm{m}}$ and $\sigma_8$ by our summary statistics ($\{\mathbfcal{D}\}: \{  \mathbf{PS^{(\ell=0,2,4)}}, \bold {V}_{\diamond}, \bold{CMD}, \bold {MFs}, \bold {MFs} + \bold {CMD} \}$). Carrying the directional information through  $\bold {CMD}$ originated by RSDs, enhanced the inference significance of cosmological parameters. The joint analysis of scalar morphological measures and $\bold{CMD}$ improved forecast precision by $(26\%^{+7\%}_{-5\%}, 27\%^{+9\%}_{-5\%})$ for ($\Omega_{\mathrm{m}}$, $\sigma_{8}$) compared to the $\bold{MFs}$ for smoothing scale of $R=15\,h^{-1}\mathrm{Mpc}$ (Figures~ \ref{fig:fig_6} and \ref{fig:fig_7}).

Extending the analysis beyond the single fiducial cosmology, we performed forecasts over a continuous range of $(\Omega_{\mathrm{m}},\,\sigma_8)$ values using 2000 independent BSQ realizations.  We advocated  that however the  constringing capability of our summary statistics depends on the fiducial values, nevertheless, the ratio of forecast uncertainty are almost constant in the adopted range of $\Omega_{\mathrm{m}}$ and $\sigma_8$. The combined $\mathbf{MFs}+\mathbf{CMD}$ configuration consistently provided the narrowest scatter of forecast precision, underscoring its stability and reduced realization‑to‑realization variance (Figure~\ref{fig:fig_8}).

We also evaluated the influence of the smoothing scale, spanning the range $R \in [15\text{--}25]\,h^{-1}\mathrm{Mpc}$, on the constraining power of the adopted summary statistics. Our results elucidated that the uncertainties in constraining $\Omega_{\mathrm{m}}$ and $\sigma_8$ from the MFs varied by about 10\% and 40\%, respectively. The joint analysis of MFs and weighted morphology (CMD) exhibits even stronger sensitivity to the smoothing scale, showing that for smaller $R$ values the CMD yields tighter constraints on cosmological parameters, as it captures non‑linear information in redshift-space more effectively (Figure~\ref{fig:fig_9}).

We further benchmarked our weighted morphological framework against the halo power spectrum in redshift space, restricting both analyses to matched effective scales corresponding to $R=15\,h^{-1}\mathrm{Mpc}$ (or $k_{\max}\simeq0.16\,h\,\mathrm{Mpc^{-1}}$). 
As shown in Figure~\ref{fig:fig_10}, the combined $\mathbf{MFs}^{(15)}+\mathbf{CMD}^{(15)}$ statistic improves the constraints on $\sigma_8$ by $\sim47\%^{+6\%}_{-4\%}$ relative to the full power spectrum multipole analysis $\mathbf{PS^{(\ell=0,2,4)}}$, while no statistically significant improvement is observed for $\Omega_{\mathrm{m}}$ on these mildly non-linear scales.

We finally assessed the impact of tracer selection by fixing the halo number density and by imposing a minimum halo mass threshold (Figure~\ref{fig:fig_11}). Fixing the number density weakens constraints for both statistics, but substantially more for the power spectrum, increasing the relative gain of $\mathbf{MFs}^{(15)}+\mathbf{CMD}^{(15)}$ on $\sigma_8$ from $\sim47\%^{+6\%}_{-4\%}$ to $\sim57\%^{+3\%}_{-7\%}$. In contrast, applying a mass threshold leads to a significant improvement of morphological constraints on $\Omega_{\mathrm{m}}$ by $42\%^{+6\%}_{-6\%}$, while the power spectrum degrades, resulting in morphology outperforming the power spectrum for both $\sigma_8$ and $\Omega_{\mathrm{m}}$. Overally, these tests demonstrate that weighted morphological statistics extract robust cosmological information that is less dependent on tracer abundance and benefits from coherently biased halo populations.

Our focus in this study has been predominantly on the anisotropy‑sensitive features that emerge in redshift-space, whereas substantial non‑linear information—available only at smaller scales—was intentionally excluded by our smoothing choice. Consequently, the results presented here prioritize the directional effects of redshift‑space distortions over the quasi‑linear clustering regime.

In a forthcoming extension of this work, we plan to perform a direct comparison between the constraining power of the morphology‑based descriptors used here and that of the non‑linear matter power spectrum up to $k_{\max}=0.5\,h\,\mathrm{Mpc^{-1}}$. This can be achieved either by employing the matter density field in BSQ simulations—thus permitting a reduction of the smoothing radius to $R=5\,h^{-1}\mathrm{Mpc}$, equivalently $k_{\max}=0.5\,h\,\mathrm{Mpc^{-1}}$—or by constructing mock galaxy catalogs derived from halo distributions. Alternatively, high‑resolution simulations could be exploited to probe smaller‑scale structures. These avenues will be explored in future studies aimed at integrating non‑linear and anisotropic information in a unified likelihood‑free framework.

While the analysis presented here is restricted to the idealized simulation conditions, assessing the impact of observational systematics is an important step toward applications to real galaxy surveys. Previous observational studies have shown that MFs and related morphological statistics can be robust to survey masks and moderate redshift uncertainties when these effects are consistently incorporated through masking schemes or mock-based forward modeling (e.g. \cite{hikage2003minkowski, appleby2018minkowski, liu2025cosmological}). Redshift errors mainly act as an additional smoothing along the LOS, while survey masks primarily affect boundary terms, both of them can be mitigated at sufficiently large smoothing scales. Sampling incompleteness effects, such as fiber collisions, effectively introduce scale-dependent shot-noise and have been shown to limit the extraction of small scale non-Gaussian information, motivating careful scale choices and number-density tests. A comprehensive treatment of these observational effects, including realistic survey geometry, selection functions, and fiber collision modeling within a SBI framework, is left for future work. Nevertheless, our results indicated that (weighted) morphological statistics are promising candidates for application to observational data, with dominant systematics expected to enter through controlled smoothing and sampling effects rather than by generating spurious cosmological signals.

\section*{Data Availability}
The results presented in this work are based on the publicly available Quijote simulation suite~\cite{Quijote_sims}. 
\section*{ACKNOWLEDGMENTS}
This work is based upon research funded by Iran National Science Foundation (INSF) under project No.4040049


\begin{thebibliography}{117}%
\makeatletter
\providecommand \@ifxundefined [1]{%
 \@ifx{#1\undefined}
}%
\providecommand \@ifnum [1]{%
 \ifnum #1\expandafter \@firstoftwo
 \else \expandafter \@secondoftwo
 \fi
}%
\providecommand \@ifx [1]{%
 \ifx #1\expandafter \@firstoftwo
 \else \expandafter \@secondoftwo
 \fi
}%
\providecommand \natexlab [1]{#1}%
\providecommand \enquote  [1]{``#1''}%
\providecommand \bibnamefont  [1]{#1}%
\providecommand \bibfnamefont [1]{#1}%
\providecommand \citenamefont [1]{#1}%
\providecommand \href@noop [0]{\@secondoftwo}%
\providecommand \href [0]{\begingroup \@sanitize@url \@href}%
\providecommand \@href[1]{\@@startlink{#1}\@@href}%
\providecommand \@@href[1]{\endgroup#1\@@endlink}%
\providecommand \@sanitize@url [0]{\catcode `\\12\catcode `\$12\catcode
  `\&12\catcode `\#12\catcode `\^12\catcode `\_12\catcode `\%12\relax}%
\providecommand \@@startlink[1]{}%
\providecommand \@@endlink[0]{}%
\providecommand \url  [0]{\begingroup\@sanitize@url \@url }%
\providecommand \@url [1]{\endgroup\@href {#1}{\urlprefix }}%
\providecommand \urlprefix  [0]{URL }%
\providecommand \Eprint [0]{\href }%
\providecommand \doibase [0]{https://doi.org/}%
\providecommand \selectlanguage [0]{\@gobble}%
\providecommand \bibinfo  [0]{\@secondoftwo}%
\providecommand \bibfield  [0]{\@secondoftwo}%
\providecommand \translation [1]{[#1]}%
\providecommand \BibitemOpen [0]{}%
\providecommand \bibitemStop [0]{}%
\providecommand \bibitemNoStop [0]{.\EOS\space}%
\providecommand \EOS [0]{\spacefactor3000\relax}%
\providecommand \BibitemShut  [1]{\csname bibitem#1\endcsname}%
\let\auto@bib@innerbib\@empty
\bibitem [{\citenamefont {Tegmark}(1997)}]{Tegmark1997}%
  \BibitemOpen
  \bibfield  {author} {\bibinfo {author} {\bibfnamefont {M.}~\bibnamefont
  {Tegmark}},\ }\href {https://doi.org/10.1103/PhysRevLett.79.3806} {\bibfield
  {journal} {\bibinfo  {journal} {Physical Review Letters}\ }\textbf {\bibinfo
  {volume} {79}},\ \bibinfo {pages} {3806} (\bibinfo {year}
  {1997})}\BibitemShut {NoStop}%
\bibitem [{\citenamefont {Peebles}(1980)}]{Peebles1980}%
  \BibitemOpen
  \bibfield  {author} {\bibinfo {author} {\bibfnamefont {P.~J.~E.}\
  \bibnamefont {Peebles}},\ }\href
  {https://press.princeton.edu/books/paperback/9780691082400/the-large-scale-structure-of-the-universe}
  {\emph {\bibinfo {title} {The Large-Scale Structure of the Universe}}}\
  (\bibinfo  {publisher} {Princeton University Press},\ \bibinfo {address}
  {Princeton, NJ},\ \bibinfo {year} {1980})\BibitemShut {NoStop}%
\bibitem [{\citenamefont {Cole}\ \emph {et~al.}(2005)\citenamefont {Cole},
  \citenamefont {Percival}, \citenamefont {Peacock} \emph {et~al.}}]{Cole2005}%
  \BibitemOpen
  \bibfield  {author} {\bibinfo {author} {\bibfnamefont {S.}~\bibnamefont
  {Cole}}, \bibinfo {author} {\bibfnamefont {W.~J.}\ \bibnamefont {Percival}},
  \bibinfo {author} {\bibfnamefont {J.~A.}\ \bibnamefont {Peacock}}, \emph
  {et~al.},\ }\href {https://doi.org/10.1111/j.1365-2966.2005.09318.x}
  {\bibfield  {journal} {\bibinfo  {journal} {Monthly Notices of the Royal
  Astronomical Society}\ }\textbf {\bibinfo {volume} {362}},\ \bibinfo {pages}
  {505} (\bibinfo {year} {2005})}\BibitemShut {NoStop}%
\bibitem [{\citenamefont {Alam}\ \emph {et~al.}(2021)\citenamefont {Alam},
  \citenamefont {Aubert}, \citenamefont {Avila} \emph
  {et~al.}}]{PhysRevD.103.083533}%
  \BibitemOpen
  \bibfield  {author} {\bibinfo {author} {\bibfnamefont {S.}~\bibnamefont
  {Alam}}, \bibinfo {author} {\bibfnamefont {M.}~\bibnamefont {Aubert}},
  \bibinfo {author} {\bibfnamefont {S.}~\bibnamefont {Avila}}, \emph {et~al.},\
  }\href {https://doi.org/10.1103/PhysRevD.103.083533} {\bibfield  {journal}
  {\bibinfo  {journal} {Phys. Rev. D}\ }\textbf {\bibinfo {volume} {103}},\
  \bibinfo {pages} {083533} (\bibinfo {year} {2021})}\BibitemShut {NoStop}%
\bibitem [{\citenamefont {Abbott}\ \emph {et~al.}(2022)\citenamefont {Abbott},
  \citenamefont {Aguena}, \citenamefont {Alarcon} \emph
  {et~al.}}]{PhysRevD.105.023520}%
  \BibitemOpen
  \bibfield  {author} {\bibinfo {author} {\bibfnamefont {T.~M.~C.}\
  \bibnamefont {Abbott}}, \bibinfo {author} {\bibfnamefont {M.}~\bibnamefont
  {Aguena}}, \bibinfo {author} {\bibfnamefont {A.}~\bibnamefont {Alarcon}},
  \emph {et~al.} (\bibinfo {collaboration} {DES Collaboration}),\ }\href
  {https://doi.org/10.1103/PhysRevD.105.023520} {\bibfield  {journal} {\bibinfo
   {journal} {Phys. Rev. D}\ }\textbf {\bibinfo {volume} {105}},\ \bibinfo
  {pages} {023520} (\bibinfo {year} {2022})}\BibitemShut {NoStop}%
\bibitem [{\citenamefont {Trotta}(2017)}]{Trotta2017}%
  \BibitemOpen
  \bibfield  {author} {\bibinfo {author} {\bibfnamefont {R.}~\bibnamefont
  {Trotta}},\ }\bibfield  {journal} {\bibinfo  {journal} {arXiv e-prints}\
  }\textbf {\bibinfo {volume} {arXiv:1701.01467}},\ \href
  {https://doi.org/10.48550/arXiv.1701.01467} {10.48550/arXiv.1701.01467}
  (\bibinfo {year} {2017}),\ \Eprint {https://arxiv.org/abs/1701.01467}
  {arXiv:1701.01467 [astro-ph.IM]} \BibitemShut {NoStop}%
\bibitem [{\citenamefont {Bernardeau}\ \emph {et~al.}(2002)\citenamefont
  {Bernardeau}, \citenamefont {Colombi}, \citenamefont {Gazta{\~n}aga},\ and\
  \citenamefont {Scoccimarro}}]{Bernardeau2002}%
  \BibitemOpen
  \bibfield  {author} {\bibinfo {author} {\bibfnamefont {F.}~\bibnamefont
  {Bernardeau}}, \bibinfo {author} {\bibfnamefont {S.}~\bibnamefont {Colombi}},
  \bibinfo {author} {\bibfnamefont {E.}~\bibnamefont {Gazta{\~n}aga}},\ and\
  \bibinfo {author} {\bibfnamefont {R.}~\bibnamefont {Scoccimarro}},\ }\href
  {https://doi.org/10.1016/S0370-1573(02)00135-7} {\bibfield  {journal}
  {\bibinfo  {journal} {Physics Reports}\ }\textbf {\bibinfo {volume} {367}},\
  \bibinfo {pages} {1} (\bibinfo {year} {2002})}\BibitemShut {NoStop}%
\bibitem [{\citenamefont {Scoccimarro}(2000)}]{Scoccimarro2000}%
  \BibitemOpen
  \bibfield  {author} {\bibinfo {author} {\bibfnamefont {R.}~\bibnamefont
  {Scoccimarro}},\ }\href {https://doi.org/10.1086/317248} {\bibfield
  {journal} {\bibinfo  {journal} {The Astrophysical Journal}\ }\textbf
  {\bibinfo {volume} {544}},\ \bibinfo {pages} {597} (\bibinfo {year}
  {2000})}\BibitemShut {NoStop}%
\bibitem [{\citenamefont {Cranmer}\ \emph {et~al.}(2020)\citenamefont
  {Cranmer}, \citenamefont {Brehmer},\ and\ \citenamefont
  {Louppe}}]{Cranmer2020}%
  \BibitemOpen
  \bibfield  {author} {\bibinfo {author} {\bibfnamefont {K.}~\bibnamefont
  {Cranmer}}, \bibinfo {author} {\bibfnamefont {J.}~\bibnamefont {Brehmer}},\
  and\ \bibinfo {author} {\bibfnamefont {G.}~\bibnamefont {Louppe}},\ }\href
  {https://doi.org/10.1073/pnas.1912789117} {\bibfield  {journal} {\bibinfo
  {journal} {Proceedings of the National Academy of Sciences}\ }\textbf
  {\bibinfo {volume} {117}},\ \bibinfo {pages} {30055} (\bibinfo {year}
  {2020})}\BibitemShut {NoStop}%
\bibitem [{\citenamefont {Alsing}\ \emph {et~al.}(2019)\citenamefont {Alsing},
  \citenamefont {Charnock}, \citenamefont {Feeney},\ and\ \citenamefont
  {Wandelt}}]{Alsing2019}%
  \BibitemOpen
  \bibfield  {author} {\bibinfo {author} {\bibfnamefont {J.}~\bibnamefont
  {Alsing}}, \bibinfo {author} {\bibfnamefont {T.}~\bibnamefont {Charnock}},
  \bibinfo {author} {\bibfnamefont {S.}~\bibnamefont {Feeney}},\ and\ \bibinfo
  {author} {\bibfnamefont {B.}~\bibnamefont {Wandelt}},\ }\href
  {https://doi.org/10.1093/mnras/stz1960} {\bibfield  {journal} {\bibinfo
  {journal} {Monthly Notices of the Royal Astronomical Society}\ }\textbf
  {\bibinfo {volume} {488}},\ \bibinfo {pages} {4440} (\bibinfo {year}
  {2019})}\BibitemShut {NoStop}%
\bibitem [{\citenamefont {{DESI Collaboration}}\ \emph
  {et~al.}(2016)\citenamefont {{DESI Collaboration}}, \citenamefont
  {{Aghamousa}} \emph {et~al.}}]{DESI2016}%
  \BibitemOpen
  \bibfield  {author} {\bibinfo {author} {\bibnamefont {{DESI Collaboration}}},
  \bibinfo {author} {\bibfnamefont {A.}~\bibnamefont {{Aghamousa}}}, \emph
  {et~al.},\ }\href {https://doi.org/10.48550/arXiv.1611.00036} {\bibfield
  {journal} {\bibinfo  {journal} {arXiv e-prints}\ ,\ \bibinfo {eid}
  {arXiv:1611.00036}} (\bibinfo {year} {2016})},\ \Eprint
  {https://arxiv.org/abs/1611.00036} {arXiv:1611.00036 [astro-ph.IM]}
  \BibitemShut {NoStop}%
\bibitem [{\citenamefont {Tamura}\ \emph {et~al.}(2016)\citenamefont {Tamura},
  \citenamefont {Takato}, \citenamefont {Shimono}, \citenamefont {Moritani},
  \citenamefont {Yabe}, \citenamefont {Ishizuka}, \citenamefont {Ueda},
  \citenamefont {Kamata}, \citenamefont {Aghazarian}, \citenamefont {Arnouts}
  \emph {et~al.}}]{tamura2016prime}%
  \BibitemOpen
  \bibfield  {author} {\bibinfo {author} {\bibfnamefont {N.}~\bibnamefont
  {Tamura}}, \bibinfo {author} {\bibfnamefont {N.}~\bibnamefont {Takato}},
  \bibinfo {author} {\bibfnamefont {A.}~\bibnamefont {Shimono}}, \bibinfo
  {author} {\bibfnamefont {Y.}~\bibnamefont {Moritani}}, \bibinfo {author}
  {\bibfnamefont {K.}~\bibnamefont {Yabe}}, \bibinfo {author} {\bibfnamefont
  {Y.}~\bibnamefont {Ishizuka}}, \bibinfo {author} {\bibfnamefont
  {A.}~\bibnamefont {Ueda}}, \bibinfo {author} {\bibfnamefont {Y.}~\bibnamefont
  {Kamata}}, \bibinfo {author} {\bibfnamefont {H.}~\bibnamefont {Aghazarian}},
  \bibinfo {author} {\bibfnamefont {S.}~\bibnamefont {Arnouts}}, \emph
  {et~al.},\ }\href@noop {} {\bibfield  {journal} {\bibinfo  {journal}
  {Ground-based and airborne instrumentation for astronomy VI}\ }\textbf
  {\bibinfo {volume} {9908}},\ \bibinfo {pages} {456} (\bibinfo {year}
  {2016})}\BibitemShut {NoStop}%
\bibitem [{\citenamefont {Spergel}\ \emph {et~al.}(2015)\citenamefont
  {Spergel}, \citenamefont {Gehrels}, \citenamefont {Baltay}, \citenamefont
  {Bennett}, \citenamefont {Breckinridge}, \citenamefont {Donahue},
  \citenamefont {Dressler}, \citenamefont {Gaudi}, \citenamefont {Greene},
  \citenamefont {Guyon} \emph {et~al.}}]{spergel2015wide}%
  \BibitemOpen
  \bibfield  {author} {\bibinfo {author} {\bibfnamefont {D.}~\bibnamefont
  {Spergel}}, \bibinfo {author} {\bibfnamefont {N.}~\bibnamefont {Gehrels}},
  \bibinfo {author} {\bibfnamefont {C.}~\bibnamefont {Baltay}}, \bibinfo
  {author} {\bibfnamefont {D.}~\bibnamefont {Bennett}}, \bibinfo {author}
  {\bibfnamefont {J.}~\bibnamefont {Breckinridge}}, \bibinfo {author}
  {\bibfnamefont {M.}~\bibnamefont {Donahue}}, \bibinfo {author} {\bibfnamefont
  {A.}~\bibnamefont {Dressler}}, \bibinfo {author} {\bibfnamefont {B.~S.}\
  \bibnamefont {Gaudi}}, \bibinfo {author} {\bibfnamefont {T.}~\bibnamefont
  {Greene}}, \bibinfo {author} {\bibfnamefont {O.}~\bibnamefont {Guyon}}, \emph
  {et~al.},\ }\href@noop {} {\bibfield  {journal} {\bibinfo  {journal} {arXiv
  preprint arXiv:1503.03757}\ } (\bibinfo {year} {2015})}\BibitemShut {NoStop}%
\bibitem [{\citenamefont {Wang}\ \emph {et~al.}(2022)\citenamefont {Wang},
  \citenamefont {Zhai}, \citenamefont {Alavi}, \citenamefont {Massara},
  \citenamefont {Pisani}, \citenamefont {Benson}, \citenamefont {Hirata},
  \citenamefont {Samushia}, \citenamefont {Weinberg}, \citenamefont {Colbert}
  \emph {et~al.}}]{wang2022high}%
  \BibitemOpen
  \bibfield  {author} {\bibinfo {author} {\bibfnamefont {Y.}~\bibnamefont
  {Wang}}, \bibinfo {author} {\bibfnamefont {Z.}~\bibnamefont {Zhai}}, \bibinfo
  {author} {\bibfnamefont {A.}~\bibnamefont {Alavi}}, \bibinfo {author}
  {\bibfnamefont {E.}~\bibnamefont {Massara}}, \bibinfo {author} {\bibfnamefont
  {A.}~\bibnamefont {Pisani}}, \bibinfo {author} {\bibfnamefont
  {A.}~\bibnamefont {Benson}}, \bibinfo {author} {\bibfnamefont {C.~M.}\
  \bibnamefont {Hirata}}, \bibinfo {author} {\bibfnamefont {L.}~\bibnamefont
  {Samushia}}, \bibinfo {author} {\bibfnamefont {D.~H.}\ \bibnamefont
  {Weinberg}}, \bibinfo {author} {\bibfnamefont {J.}~\bibnamefont {Colbert}},
  \emph {et~al.},\ }\href@noop {} {\bibfield  {journal} {\bibinfo  {journal}
  {The Astrophysical Journal}\ }\textbf {\bibinfo {volume} {928}},\ \bibinfo
  {pages} {1} (\bibinfo {year} {2022})}\BibitemShut {NoStop}%
\bibitem [{\citenamefont {Laureijs}\ \emph {et~al.}(2011)\citenamefont
  {Laureijs} \emph {et~al.}}]{Euclid2016}%
  \BibitemOpen
  \bibfield  {author} {\bibinfo {author} {\bibfnamefont {R.}~\bibnamefont
  {Laureijs}} \emph {et~al.},\ }\bibfield  {journal} {\bibinfo  {journal}
  {arXiv preprint}\ }\href {https://doi.org/10.48550/arXiv.1110.3193}
  {10.48550/arXiv.1110.3193} (\bibinfo {year} {2011}),\ \Eprint
  {https://arxiv.org/abs/1110.3193} {arXiv:1110.3193 [astro-ph.CO]}
  \BibitemShut {NoStop}%
\bibitem [{\citenamefont {Sefusatti}\ \emph {et~al.}(2006)\citenamefont
  {Sefusatti}, \citenamefont {Crocce}, \citenamefont {Pueblas},\ and\
  \citenamefont {Scoccimarro}}]{Sefusatti2007}%
  \BibitemOpen
  \bibfield  {author} {\bibinfo {author} {\bibfnamefont {E.}~\bibnamefont
  {Sefusatti}}, \bibinfo {author} {\bibfnamefont {M.}~\bibnamefont {Crocce}},
  \bibinfo {author} {\bibfnamefont {S.}~\bibnamefont {Pueblas}},\ and\ \bibinfo
  {author} {\bibfnamefont {R.}~\bibnamefont {Scoccimarro}},\ }\href
  {https://doi.org/10.1103/PhysRevD.74.023522} {\bibfield  {journal} {\bibinfo
  {journal} {Phys. Rev. D}\ }\textbf {\bibinfo {volume} {74}},\ \bibinfo
  {pages} {023522} (\bibinfo {year} {2006})},\ \Eprint
  {https://arxiv.org/abs/astro-ph/0604505} {arXiv:astro-ph/0604505}
  \BibitemShut {NoStop}%
\bibitem [{\citenamefont {Gil-Marín}\ \emph {et~al.}(2016)\citenamefont
  {Gil-Marín}, \citenamefont {Percival}, \citenamefont {Verde}, \citenamefont
  {Brownstein}, \citenamefont {Chuang}, \citenamefont {Kitaura}, \citenamefont
  {Rodríguez-Torres},\ and\ \citenamefont {Olmstead}}]{GilMarin2016}%
  \BibitemOpen
  \bibfield  {author} {\bibinfo {author} {\bibfnamefont {H.}~\bibnamefont
  {Gil-Marín}}, \bibinfo {author} {\bibfnamefont {W.~J.}\ \bibnamefont
  {Percival}}, \bibinfo {author} {\bibfnamefont {L.}~\bibnamefont {Verde}},
  \bibinfo {author} {\bibfnamefont {J.~R.}\ \bibnamefont {Brownstein}},
  \bibinfo {author} {\bibfnamefont {C.-H.}\ \bibnamefont {Chuang}}, \bibinfo
  {author} {\bibfnamefont {F.-S.}\ \bibnamefont {Kitaura}}, \bibinfo {author}
  {\bibfnamefont {S.~A.}\ \bibnamefont {Rodríguez-Torres}},\ and\ \bibinfo
  {author} {\bibfnamefont {M.~D.}\ \bibnamefont {Olmstead}},\ }\href
  {https://doi.org/10.1093/mnras/stw2679} {\bibfield  {journal} {\bibinfo
  {journal} {Monthly Notices of the Royal Astronomical Society}\ }\textbf
  {\bibinfo {volume} {465}},\ \bibinfo {pages} {1757} (\bibinfo {year}
  {2016})}\BibitemShut {NoStop}%
\bibitem [{\citenamefont {Fry}(1985)}]{fry1985cosmological}%
  \BibitemOpen
  \bibfield  {author} {\bibinfo {author} {\bibfnamefont {J.}~\bibnamefont
  {Fry}},\ }\href@noop {} {\bibfield  {journal} {\bibinfo  {journal}
  {Astrophysical Journal, Part 1 (ISSN 0004-637X), vol. 289, Feb. 1, 1985, p.
  10-17. NASA-supported research.}\ }\textbf {\bibinfo {volume} {289}},\
  \bibinfo {pages} {10} (\bibinfo {year} {1985})}\BibitemShut {NoStop}%
\bibitem [{\citenamefont {Bouchet}(1993)}]{bouchet1993moments}%
  \BibitemOpen
  \bibfield  {author} {\bibinfo {author} {\bibfnamefont {F.~R.}\ \bibnamefont
  {Bouchet}},\ }\href@noop {} {\bibfield  {journal} {\bibinfo  {journal} {arXiv
  preprint astro-ph/9305018}\ } (\bibinfo {year} {1993})}\BibitemShut {NoStop}%
\bibitem [{\citenamefont {Bernardeau}(1993)}]{bernardeau1993skewness}%
  \BibitemOpen
  \bibfield  {author} {\bibinfo {author} {\bibfnamefont {F.}~\bibnamefont
  {Bernardeau}},\ }\href@noop {} {\bibfield  {journal} {\bibinfo  {journal}
  {arXiv preprint astro-ph/9312026}\ } (\bibinfo {year} {1993})}\BibitemShut
  {NoStop}%
\bibitem [{\citenamefont {Croton}\ \emph {et~al.}(2004)\citenamefont {Croton},
  \citenamefont {Gaztanaga}, \citenamefont {Baugh}, \citenamefont {Norberg},
  \citenamefont {Colless}, \citenamefont {Baldry}, \citenamefont
  {Bland-Hawthorn}, \citenamefont {Bridges}, \citenamefont {Cannon},
  \citenamefont {Cole} \emph {et~al.}}]{croton20042df}%
  \BibitemOpen
  \bibfield  {author} {\bibinfo {author} {\bibfnamefont {D.~J.}\ \bibnamefont
  {Croton}}, \bibinfo {author} {\bibfnamefont {E.}~\bibnamefont {Gaztanaga}},
  \bibinfo {author} {\bibfnamefont {C.~M.}\ \bibnamefont {Baugh}}, \bibinfo
  {author} {\bibfnamefont {P.}~\bibnamefont {Norberg}}, \bibinfo {author}
  {\bibfnamefont {M.}~\bibnamefont {Colless}}, \bibinfo {author} {\bibfnamefont
  {I.~K.}\ \bibnamefont {Baldry}}, \bibinfo {author} {\bibfnamefont
  {J.}~\bibnamefont {Bland-Hawthorn}}, \bibinfo {author} {\bibfnamefont
  {T.}~\bibnamefont {Bridges}}, \bibinfo {author} {\bibfnamefont
  {R.}~\bibnamefont {Cannon}}, \bibinfo {author} {\bibfnamefont
  {S.}~\bibnamefont {Cole}}, \emph {et~al.},\ }\href@noop {} {\bibfield
  {journal} {\bibinfo  {journal} {Monthly Notices of the Royal Astronomical
  Society}\ }\textbf {\bibinfo {volume} {352}},\ \bibinfo {pages} {1232}
  (\bibinfo {year} {2004})}\BibitemShut {NoStop}%
\bibitem [{\citenamefont {Cappi}\ \emph {et~al.}(2015)\citenamefont {Cappi},
  \citenamefont {Marulli}, \citenamefont {Bel}, \citenamefont {Cucciati},
  \citenamefont {Branchini}, \citenamefont {de~La~Torre}, \citenamefont
  {Moscardini}, \citenamefont {Bolzonella}, \citenamefont {Guzzo},
  \citenamefont {Abbas} \emph {et~al.}}]{cappi2015vimos}%
  \BibitemOpen
  \bibfield  {author} {\bibinfo {author} {\bibfnamefont {A.}~\bibnamefont
  {Cappi}}, \bibinfo {author} {\bibfnamefont {F.}~\bibnamefont {Marulli}},
  \bibinfo {author} {\bibfnamefont {J.}~\bibnamefont {Bel}}, \bibinfo {author}
  {\bibfnamefont {O.}~\bibnamefont {Cucciati}}, \bibinfo {author}
  {\bibfnamefont {E.}~\bibnamefont {Branchini}}, \bibinfo {author}
  {\bibfnamefont {S.}~\bibnamefont {de~La~Torre}}, \bibinfo {author}
  {\bibfnamefont {L.}~\bibnamefont {Moscardini}}, \bibinfo {author}
  {\bibfnamefont {M.}~\bibnamefont {Bolzonella}}, \bibinfo {author}
  {\bibfnamefont {L.}~\bibnamefont {Guzzo}}, \bibinfo {author} {\bibfnamefont
  {U.}~\bibnamefont {Abbas}}, \emph {et~al.},\ }\href@noop {} {\bibfield
  {journal} {\bibinfo  {journal} {Astronomy \& Astrophysics}\ }\textbf
  {\bibinfo {volume} {579}},\ \bibinfo {pages} {A70} (\bibinfo {year}
  {2015})}\BibitemShut {NoStop}%
\bibitem [{\citenamefont {Sabiu}\ \emph {et~al.}(2019)\citenamefont {Sabiu},
  \citenamefont {Hoyle}, \citenamefont {Kim},\ and\ \citenamefont
  {Li}}]{sabiu2019graph}%
  \BibitemOpen
  \bibfield  {author} {\bibinfo {author} {\bibfnamefont {C.~G.}\ \bibnamefont
  {Sabiu}}, \bibinfo {author} {\bibfnamefont {B.}~\bibnamefont {Hoyle}},
  \bibinfo {author} {\bibfnamefont {J.}~\bibnamefont {Kim}},\ and\ \bibinfo
  {author} {\bibfnamefont {X.-D.}\ \bibnamefont {Li}},\ }\href@noop {}
  {\bibfield  {journal} {\bibinfo  {journal} {The Astrophysical Journal
  Supplement Series}\ }\textbf {\bibinfo {volume} {242}},\ \bibinfo {pages}
  {29} (\bibinfo {year} {2019})}\BibitemShut {NoStop}%
\bibitem [{\citenamefont {Philcox}(2022)}]{philcox2022probing}%
  \BibitemOpen
  \bibfield  {author} {\bibinfo {author} {\bibfnamefont {O.~H.}\ \bibnamefont
  {Philcox}},\ }\href@noop {} {\bibfield  {journal} {\bibinfo  {journal}
  {Physical Review D}\ }\textbf {\bibinfo {volume} {106}},\ \bibinfo {pages}
  {063501} (\bibinfo {year} {2022})}\BibitemShut {NoStop}%
\bibitem [{\citenamefont {White}(2016)}]{White2016}%
  \BibitemOpen
  \bibfield  {author} {\bibinfo {author} {\bibfnamefont {M.}~\bibnamefont
  {White}},\ }\href {https://doi.org/10.1088/1475-7516/2016/11/057} {\bibfield
  {journal} {\bibinfo  {journal} {JCAP}\ }\textbf {\bibinfo {volume} {11}},\
  \bibinfo {pages} {057}},\ \Eprint {https://arxiv.org/abs/1609.08632}
  {arXiv:1609.08632 [astro-ph.CO]} \BibitemShut {NoStop}%
\bibitem [{\citenamefont {Massara}\ \emph {et~al.}(2023)\citenamefont
  {Massara}, \citenamefont {Villaescusa-Navarro}, \citenamefont {Hahn},
  \citenamefont {Abidi}, \citenamefont {Eickenberg}, \citenamefont {Ho},
  \citenamefont {Lemos}, \citenamefont {Dizgah},\ and\ \citenamefont
  {Blancard}}]{Massara_2023}%
  \BibitemOpen
  \bibfield  {author} {\bibinfo {author} {\bibfnamefont {E.}~\bibnamefont
  {Massara}}, \bibinfo {author} {\bibfnamefont {F.}~\bibnamefont
  {Villaescusa-Navarro}}, \bibinfo {author} {\bibfnamefont {C.}~\bibnamefont
  {Hahn}}, \bibinfo {author} {\bibfnamefont {M.~M.}\ \bibnamefont {Abidi}},
  \bibinfo {author} {\bibfnamefont {M.}~\bibnamefont {Eickenberg}}, \bibinfo
  {author} {\bibfnamefont {S.}~\bibnamefont {Ho}}, \bibinfo {author}
  {\bibfnamefont {P.}~\bibnamefont {Lemos}}, \bibinfo {author} {\bibfnamefont
  {A.~M.}\ \bibnamefont {Dizgah}},\ and\ \bibinfo {author} {\bibfnamefont
  {B.~R.-S.}\ \bibnamefont {Blancard}},\ }\href
  {https://doi.org/10.3847/1538-4357/acd44d} {\bibfield  {journal} {\bibinfo
  {journal} {The Astrophysical Journal}\ }\textbf {\bibinfo {volume} {951}},\
  \bibinfo {pages} {70} (\bibinfo {year} {2023})}\BibitemShut {NoStop}%
\bibitem [{\citenamefont {Cowell}\ \emph {et~al.}(2024)\citenamefont {Cowell},
  \citenamefont {Alonso},\ and\ \citenamefont {Liu}}]{Cowell2024}%
  \BibitemOpen
  \bibfield  {author} {\bibinfo {author} {\bibfnamefont {J.~A.}\ \bibnamefont
  {Cowell}}, \bibinfo {author} {\bibfnamefont {D.}~\bibnamefont {Alonso}},\
  and\ \bibinfo {author} {\bibfnamefont {J.}~\bibnamefont {Liu}},\ }\href
  {https://doi.org/10.1093/mnras/stae2492} {\bibfield  {journal} {\bibinfo
  {journal} {Monthly Notices of the Royal Astronomical Society}\ }\textbf
  {\bibinfo {volume} {535}},\ \bibinfo {pages} {3129} (\bibinfo {year}
  {2024})},\ \Eprint
  {https://arxiv.org/abs/https://academic.oup.com/mnras/article-pdf/535/4/3129/60881005/stae2492.pdf}
  {https://academic.oup.com/mnras/article-pdf/535/4/3129/60881005/stae2492.pdf}
  \BibitemShut {NoStop}%
\bibitem [{\citenamefont {Valogiannis}\ and\ \citenamefont
  {Dvorkin}(2022)}]{valogiannis2022towards}%
  \BibitemOpen
  \bibfield  {author} {\bibinfo {author} {\bibfnamefont {G.}~\bibnamefont
  {Valogiannis}}\ and\ \bibinfo {author} {\bibfnamefont {C.}~\bibnamefont
  {Dvorkin}},\ }\href@noop {} {\bibfield  {journal} {\bibinfo  {journal}
  {Physical Review D}\ }\textbf {\bibinfo {volume} {105}},\ \bibinfo {pages}
  {103534} (\bibinfo {year} {2022})}\BibitemShut {NoStop}%
\bibitem [{\citenamefont {Eickenberg}\ \emph {et~al.}(2022)\citenamefont
  {Eickenberg}, \citenamefont {Allys}, \citenamefont {Dizgah}, \citenamefont
  {Lemos}, \citenamefont {Massara}, \citenamefont {Abidi}, \citenamefont
  {Hahn}, \citenamefont {Hassan}, \citenamefont {Blancard}, \citenamefont {Ho}
  \emph {et~al.}}]{eickenberg2022wavelet}%
  \BibitemOpen
  \bibfield  {author} {\bibinfo {author} {\bibfnamefont {M.}~\bibnamefont
  {Eickenberg}}, \bibinfo {author} {\bibfnamefont {E.}~\bibnamefont {Allys}},
  \bibinfo {author} {\bibfnamefont {A.~M.}\ \bibnamefont {Dizgah}}, \bibinfo
  {author} {\bibfnamefont {P.}~\bibnamefont {Lemos}}, \bibinfo {author}
  {\bibfnamefont {E.}~\bibnamefont {Massara}}, \bibinfo {author} {\bibfnamefont
  {M.}~\bibnamefont {Abidi}}, \bibinfo {author} {\bibfnamefont
  {C.}~\bibnamefont {Hahn}}, \bibinfo {author} {\bibfnamefont {S.}~\bibnamefont
  {Hassan}}, \bibinfo {author} {\bibfnamefont {B.~R.-S.}\ \bibnamefont
  {Blancard}}, \bibinfo {author} {\bibfnamefont {S.}~\bibnamefont {Ho}}, \emph
  {et~al.},\ }\href@noop {} {\bibfield  {journal} {\bibinfo  {journal} {arXiv
  preprint arXiv:2204.07646}\ } (\bibinfo {year} {2022})}\BibitemShut {NoStop}%
\bibitem [{\citenamefont {Paillas}\ \emph {et~al.}(2024)\citenamefont
  {Paillas}, \citenamefont {Cuesta-Lazaro}, \citenamefont {Percival},
  \citenamefont {Nadathur}, \citenamefont {Cai}, \citenamefont {Yuan},
  \citenamefont {Beutler}, \citenamefont {De~Mattia}, \citenamefont
  {Eisenstein}, \citenamefont {Forero-Sanchez} \emph
  {et~al.}}]{paillas2024cosmological}%
  \BibitemOpen
  \bibfield  {author} {\bibinfo {author} {\bibfnamefont {E.}~\bibnamefont
  {Paillas}}, \bibinfo {author} {\bibfnamefont {C.}~\bibnamefont
  {Cuesta-Lazaro}}, \bibinfo {author} {\bibfnamefont {W.~J.}\ \bibnamefont
  {Percival}}, \bibinfo {author} {\bibfnamefont {S.}~\bibnamefont {Nadathur}},
  \bibinfo {author} {\bibfnamefont {Y.-C.}\ \bibnamefont {Cai}}, \bibinfo
  {author} {\bibfnamefont {S.}~\bibnamefont {Yuan}}, \bibinfo {author}
  {\bibfnamefont {F.}~\bibnamefont {Beutler}}, \bibinfo {author} {\bibfnamefont
  {A.}~\bibnamefont {De~Mattia}}, \bibinfo {author} {\bibfnamefont {D.~J.}\
  \bibnamefont {Eisenstein}}, \bibinfo {author} {\bibfnamefont
  {D.}~\bibnamefont {Forero-Sanchez}}, \emph {et~al.},\ }\href@noop {}
  {\bibfield  {journal} {\bibinfo  {journal} {Monthly Notices of the Royal
  Astronomical Society}\ }\textbf {\bibinfo {volume} {531}},\ \bibinfo {pages}
  {898} (\bibinfo {year} {2024})}\BibitemShut {NoStop}%
\bibitem [{\citenamefont {Paillas}\ \emph {et~al.}(2023)\citenamefont
  {Paillas}, \citenamefont {Cuesta-Lazaro}, \citenamefont {Zarrouk},
  \citenamefont {Cai}, \citenamefont {Percival}, \citenamefont {Nadathur},
  \citenamefont {Pinon}, \citenamefont {de~Mattia},\ and\ \citenamefont
  {Beutler}}]{paillas2023constraining}%
  \BibitemOpen
  \bibfield  {author} {\bibinfo {author} {\bibfnamefont {E.}~\bibnamefont
  {Paillas}}, \bibinfo {author} {\bibfnamefont {C.}~\bibnamefont
  {Cuesta-Lazaro}}, \bibinfo {author} {\bibfnamefont {P.}~\bibnamefont
  {Zarrouk}}, \bibinfo {author} {\bibfnamefont {Y.-C.}\ \bibnamefont {Cai}},
  \bibinfo {author} {\bibfnamefont {W.~J.}\ \bibnamefont {Percival}}, \bibinfo
  {author} {\bibfnamefont {S.}~\bibnamefont {Nadathur}}, \bibinfo {author}
  {\bibfnamefont {M.}~\bibnamefont {Pinon}}, \bibinfo {author} {\bibfnamefont
  {A.}~\bibnamefont {de~Mattia}},\ and\ \bibinfo {author} {\bibfnamefont
  {F.}~\bibnamefont {Beutler}},\ }\href@noop {} {\bibfield  {journal} {\bibinfo
   {journal} {Monthly Notices of the Royal Astronomical Society}\ }\textbf
  {\bibinfo {volume} {522}},\ \bibinfo {pages} {606} (\bibinfo {year}
  {2023})}\BibitemShut {NoStop}%
\bibitem [{\citenamefont {Cuesta-Lazaro}\ \emph {et~al.}(2024)\citenamefont
  {Cuesta-Lazaro}, \citenamefont {Paillas}, \citenamefont {Yuan}, \citenamefont
  {Cai}, \citenamefont {Nadathur}, \citenamefont {Percival}, \citenamefont
  {Beutler}, \citenamefont {de~Mattia}, \citenamefont {Eisenstein},
  \citenamefont {Forero-Sanchez} \emph {et~al.}}]{cuesta2024sunbird}%
  \BibitemOpen
  \bibfield  {author} {\bibinfo {author} {\bibfnamefont {C.}~\bibnamefont
  {Cuesta-Lazaro}}, \bibinfo {author} {\bibfnamefont {E.}~\bibnamefont
  {Paillas}}, \bibinfo {author} {\bibfnamefont {S.}~\bibnamefont {Yuan}},
  \bibinfo {author} {\bibfnamefont {Y.-C.}\ \bibnamefont {Cai}}, \bibinfo
  {author} {\bibfnamefont {S.}~\bibnamefont {Nadathur}}, \bibinfo {author}
  {\bibfnamefont {W.~J.}\ \bibnamefont {Percival}}, \bibinfo {author}
  {\bibfnamefont {F.}~\bibnamefont {Beutler}}, \bibinfo {author} {\bibfnamefont
  {A.}~\bibnamefont {de~Mattia}}, \bibinfo {author} {\bibfnamefont {D.~J.}\
  \bibnamefont {Eisenstein}}, \bibinfo {author} {\bibfnamefont
  {D.}~\bibnamefont {Forero-Sanchez}}, \emph {et~al.},\ }\href@noop {}
  {\bibfield  {journal} {\bibinfo  {journal} {Monthly Notices of the Royal
  Astronomical Society}\ }\textbf {\bibinfo {volume} {531}},\ \bibinfo {pages}
  {3336} (\bibinfo {year} {2024})}\BibitemShut {NoStop}%
\bibitem [{\citenamefont {Hadwiger}(2013)}]{hadwiger2013vorlesungen}%
  \BibitemOpen
  \bibfield  {author} {\bibinfo {author} {\bibfnamefont {H.}~\bibnamefont
  {Hadwiger}},\ }\href@noop {} {\emph {\bibinfo {title} {Vorlesungen {\"u}ber
  inhalt, oberfl{\"a}che und isoperimetrie}}},\ Vol.~\bibinfo {volume} {93}\
  (\bibinfo  {publisher} {Springer-Verlag},\ \bibinfo {year}
  {2013})\BibitemShut {NoStop}%
\bibitem [{\citenamefont {{Matsubara}}(2003)}]{matsubara2003}%
  \BibitemOpen
  \bibfield  {author} {\bibinfo {author} {\bibfnamefont {T.}~\bibnamefont
  {{Matsubara}}},\ }\href {https://doi.org/10.1086/345521} {\bibfield
  {journal} {\bibinfo  {journal} {\apj}\ }\textbf {\bibinfo {volume} {584}},\
  \bibinfo {pages} {1} (\bibinfo {year} {2003})}\BibitemShut {NoStop}%
\bibitem [{\citenamefont {{Gay}}\ \emph {et~al.}(2012)\citenamefont {{Gay}},
  \citenamefont {{Pichon}},\ and\ \citenamefont {{Pogosyan}}}]{gay2012non}%
  \BibitemOpen
  \bibfield  {author} {\bibinfo {author} {\bibfnamefont {C.}~\bibnamefont
  {{Gay}}}, \bibinfo {author} {\bibfnamefont {C.}~\bibnamefont {{Pichon}}},\
  and\ \bibinfo {author} {\bibfnamefont {D.}~\bibnamefont {{Pogosyan}}},\
  }\href {https://doi.org/10.1103/PhysRevD.85.023011} {\bibfield  {journal}
  {\bibinfo  {journal} {\prd}\ }\textbf {\bibinfo {volume} {85}},\ \bibinfo
  {eid} {023011} (\bibinfo {year} {2012})},\ \Eprint
  {https://arxiv.org/abs/1110.0261} {arXiv:1110.0261 [astro-ph.CO]}
  \BibitemShut {NoStop}%
\bibitem [{\citenamefont {{Codis}}\ \emph {et~al.}(2013)\citenamefont
  {{Codis}}, \citenamefont {{Pichon}}, \citenamefont {{Pogosyan}},
  \citenamefont {{Bernardeau}},\ and\ \citenamefont
  {{Matsubara}}}]{codis2013non}%
  \BibitemOpen
  \bibfield  {author} {\bibinfo {author} {\bibfnamefont {S.}~\bibnamefont
  {{Codis}}}, \bibinfo {author} {\bibfnamefont {C.}~\bibnamefont {{Pichon}}},
  \bibinfo {author} {\bibfnamefont {D.}~\bibnamefont {{Pogosyan}}}, \bibinfo
  {author} {\bibfnamefont {F.}~\bibnamefont {{Bernardeau}}},\ and\ \bibinfo
  {author} {\bibfnamefont {T.}~\bibnamefont {{Matsubara}}},\ }\href
  {https://doi.org/10.1093/mnras/stt1316} {\bibfield  {journal} {\bibinfo
  {journal} {Monthly Notices of the RAS}\ }\textbf {\bibinfo {volume} {435}},\
  \bibinfo {pages} {531} (\bibinfo {year} {2013})},\ \Eprint
  {https://arxiv.org/abs/1305.7402} {arXiv:1305.7402} \BibitemShut {NoStop}%
\bibitem [{\citenamefont {Sousbie}\ \emph {et~al.}(2008)\citenamefont
  {Sousbie}, \citenamefont {Pichon}, \citenamefont {Colombi}, \citenamefont
  {Novikov},\ and\ \citenamefont {Pogosyan}}]{sousbie20083d}%
  \BibitemOpen
  \bibfield  {author} {\bibinfo {author} {\bibfnamefont {T.}~\bibnamefont
  {Sousbie}}, \bibinfo {author} {\bibfnamefont {C.}~\bibnamefont {Pichon}},
  \bibinfo {author} {\bibfnamefont {S.}~\bibnamefont {Colombi}}, \bibinfo
  {author} {\bibfnamefont {D.}~\bibnamefont {Novikov}},\ and\ \bibinfo {author}
  {\bibfnamefont {D.}~\bibnamefont {Pogosyan}},\ }\href@noop {} {\bibfield
  {journal} {\bibinfo  {journal} {Monthly Notices of the Royal Astronomical
  Society}\ }\textbf {\bibinfo {volume} {383}},\ \bibinfo {pages} {1655}
  (\bibinfo {year} {2008})}\BibitemShut {NoStop}%
\bibitem [{\citenamefont {Sousbie}\ \emph {et~al.}(2009)\citenamefont
  {Sousbie}, \citenamefont {Colombi},\ and\ \citenamefont
  {Pichon}}]{sousbie2009fully}%
  \BibitemOpen
  \bibfield  {author} {\bibinfo {author} {\bibfnamefont {T.}~\bibnamefont
  {Sousbie}}, \bibinfo {author} {\bibfnamefont {S.}~\bibnamefont {Colombi}},\
  and\ \bibinfo {author} {\bibfnamefont {C.}~\bibnamefont {Pichon}},\
  }\href@noop {} {\bibfield  {journal} {\bibinfo  {journal} {Monthly Notices of
  the Royal Astronomical Society}\ }\textbf {\bibinfo {volume} {393}},\
  \bibinfo {pages} {457} (\bibinfo {year} {2009})}\BibitemShut {NoStop}%
\bibitem [{\citenamefont {Sousbie}\ \emph {et~al.}(2011)\citenamefont
  {Sousbie}, \citenamefont {Pichon},\ and\ \citenamefont
  {Kawahara}}]{sousbie2011persistent}%
  \BibitemOpen
  \bibfield  {author} {\bibinfo {author} {\bibfnamefont {T.}~\bibnamefont
  {Sousbie}}, \bibinfo {author} {\bibfnamefont {C.}~\bibnamefont {Pichon}},\
  and\ \bibinfo {author} {\bibfnamefont {H.}~\bibnamefont {Kawahara}},\
  }\href@noop {} {\bibfield  {journal} {\bibinfo  {journal} {Monthly Notices of
  the Royal Astronomical Society}\ }\textbf {\bibinfo {volume} {414}},\
  \bibinfo {pages} {384} (\bibinfo {year} {2011})}\BibitemShut {NoStop}%
\bibitem [{\citenamefont {{Mecke}}\ \emph {et~al.}(1994)\citenamefont
  {{Mecke}}, \citenamefont {{Buchert}},\ and\ \citenamefont
  {{Wagner}}}]{mecke1993robust}%
  \BibitemOpen
  \bibfield  {author} {\bibinfo {author} {\bibfnamefont {K.~R.}\ \bibnamefont
  {{Mecke}}}, \bibinfo {author} {\bibfnamefont {T.}~\bibnamefont {{Buchert}}},\
  and\ \bibinfo {author} {\bibfnamefont {H.}~\bibnamefont {{Wagner}}},\ }\href
  {https://doi.org/10.48550/arXiv.astro-ph/9312028} {\bibfield  {journal}
  {\bibinfo  {journal} {Astronomy and Astrophysics}\ }\textbf {\bibinfo
  {volume} {288}},\ \bibinfo {pages} {697} (\bibinfo {year} {1994})},\ \Eprint
  {https://arxiv.org/abs/astro-ph/9312028} {arXiv:astro-ph/9312028}
  \BibitemShut {NoStop}%
\bibitem [{\citenamefont {Schmalzing}\ \emph {et~al.}(1995)\citenamefont
  {Schmalzing}, \citenamefont {Buchert},\ and\ \citenamefont
  {Kerscher}}]{schmalzing1995minkowski}%
  \BibitemOpen
  \bibfield  {author} {\bibinfo {author} {\bibfnamefont {J.}~\bibnamefont
  {Schmalzing}}, \bibinfo {author} {\bibfnamefont {T.}~\bibnamefont
  {Buchert}},\ and\ \bibinfo {author} {\bibfnamefont {M.}~\bibnamefont
  {Kerscher}},\ }\href@noop {} {\bibfield  {journal} {\bibinfo  {journal}
  {Proc. Int. Sch. Phys. Fermi}\ }\textbf {\bibinfo {volume} {132}},\ \bibinfo
  {pages} {281} (\bibinfo {year} {1995})}\BibitemShut {NoStop}%
\bibitem [{\citenamefont {{Matsubara}}(1996)}]{matsubara1996statistics}%
  \BibitemOpen
  \bibfield  {author} {\bibinfo {author} {\bibfnamefont {T.}~\bibnamefont
  {{Matsubara}}},\ }\href {https://doi.org/10.1086/176708} {\bibfield
  {journal} {\bibinfo  {journal} {\apj}\ }\textbf {\bibinfo {volume} {457}},\
  \bibinfo {pages} {13} (\bibinfo {year} {1996})},\ \Eprint
  {https://arxiv.org/abs/astro-ph/9501055} {arXiv:astro-ph/9501055 [astro-ph]}
  \BibitemShut {NoStop}%
\bibitem [{\citenamefont {Matsubara}\ \emph {et~al.}(2022)\citenamefont
  {Matsubara}, \citenamefont {Hikage},\ and\ \citenamefont
  {Kuriki}}]{matsubara2022minkowski}%
  \BibitemOpen
  \bibfield  {author} {\bibinfo {author} {\bibfnamefont {T.}~\bibnamefont
  {Matsubara}}, \bibinfo {author} {\bibfnamefont {C.}~\bibnamefont {Hikage}},\
  and\ \bibinfo {author} {\bibfnamefont {S.}~\bibnamefont {Kuriki}},\
  }\href@noop {} {\bibfield  {journal} {\bibinfo  {journal} {Physical Review
  D}\ }\textbf {\bibinfo {volume} {105}},\ \bibinfo {pages} {023527} (\bibinfo
  {year} {2022})}\BibitemShut {NoStop}%
\bibitem [{\citenamefont {Jiang}\ \emph {et~al.}(2022)\citenamefont {Jiang},
  \citenamefont {Liu}, \citenamefont {Fang},\ and\ \citenamefont
  {Zhao}}]{jiang2022effects}%
  \BibitemOpen
  \bibfield  {author} {\bibinfo {author} {\bibfnamefont {A.}~\bibnamefont
  {Jiang}}, \bibinfo {author} {\bibfnamefont {W.}~\bibnamefont {Liu}}, \bibinfo
  {author} {\bibfnamefont {W.}~\bibnamefont {Fang}},\ and\ \bibinfo {author}
  {\bibfnamefont {W.}~\bibnamefont {Zhao}},\ }\href@noop {} {\bibfield
  {journal} {\bibinfo  {journal} {Physical Review D}\ }\textbf {\bibinfo
  {volume} {105}},\ \bibinfo {pages} {103028} (\bibinfo {year}
  {2022})}\BibitemShut {NoStop}%
\bibitem [{\citenamefont {{Liu}}\ \emph {et~al.}(2022)\citenamefont {{Liu}},
  \citenamefont {{Jiang}},\ and\ \citenamefont {{Fang}}}]{Liu2022ProbingMN}%
  \BibitemOpen
  \bibfield  {author} {\bibinfo {author} {\bibfnamefont {W.}~\bibnamefont
  {{Liu}}}, \bibinfo {author} {\bibfnamefont {A.}~\bibnamefont {{Jiang}}},\
  and\ \bibinfo {author} {\bibfnamefont {W.}~\bibnamefont {{Fang}}},\ }\href
  {https://doi.org/10.1088/1475-7516/2022/07/045} {\bibfield  {journal}
  {\bibinfo  {journal} {Journal of Cosmology and Astroparticle Physics}\
  }\textbf {\bibinfo {volume} {2022}}\bibfield  {number} {\bibinfo  {number} {
  (7)},\ \bibinfo {eid} {045}},\ }\Eprint {https://arxiv.org/abs/2204.02945}
  {arXiv:2204.02945} \BibitemShut {NoStop}%
\bibitem [{\citenamefont {Liu}\ \emph {et~al.}(2023)\citenamefont {Liu},
  \citenamefont {Jiang},\ and\ \citenamefont {Fang}}]{Liu2023ProbingMN}%
  \BibitemOpen
  \bibfield  {author} {\bibinfo {author} {\bibfnamefont {W.}~\bibnamefont
  {Liu}}, \bibinfo {author} {\bibfnamefont {A.}~\bibnamefont {Jiang}},\ and\
  \bibinfo {author} {\bibfnamefont {W.}~\bibnamefont {Fang}},\ }\href
  {https://doi.org/10.1088/1475-7516/2023/09/037} {\bibfield  {journal}
  {\bibinfo  {journal} {Journal of Cosmology and Astroparticle Physics}\
  }\textbf {\bibinfo {volume} {2023}}\bibfield  {number} {\bibinfo  {number} {
  (9)},\ \bibinfo {eid} {037}},\ }\Eprint {https://arxiv.org/abs/2302.08162}
  {arXiv:2302.08162} \BibitemShut {NoStop}%
\bibitem [{\citenamefont {{Appleby}}\ \emph {et~al.}(2018)\citenamefont
  {{Appleby}}, \citenamefont {{Chingangbam}}, \citenamefont {{Park}},
  \citenamefont {{Hong}}, \citenamefont {{Kim}},\ and\ \citenamefont
  {{Ganesan}}}]{appleby2018minkowski}%
  \BibitemOpen
  \bibfield  {author} {\bibinfo {author} {\bibfnamefont {S.}~\bibnamefont
  {{Appleby}}}, \bibinfo {author} {\bibfnamefont {P.}~\bibnamefont
  {{Chingangbam}}}, \bibinfo {author} {\bibfnamefont {C.}~\bibnamefont
  {{Park}}}, \bibinfo {author} {\bibfnamefont {S.~E.}\ \bibnamefont {{Hong}}},
  \bibinfo {author} {\bibfnamefont {J.}~\bibnamefont {{Kim}}},\ and\ \bibinfo
  {author} {\bibfnamefont {V.}~\bibnamefont {{Ganesan}}},\ }\href
  {https://doi.org/10.3847/1538-4357/aabb53} {\bibfield  {journal} {\bibinfo
  {journal} {Astrophysical Journal}\ }\textbf {\bibinfo {volume} {858}},\
  \bibinfo {eid} {87} (\bibinfo {year} {2018})},\ \Eprint
  {https://arxiv.org/abs/1712.07466} {arXiv:1712.07466} \BibitemShut {NoStop}%
\bibitem [{\citenamefont {{Appleby}}\ \emph {et~al.}(2019)\citenamefont
  {{Appleby}}, \citenamefont {{Kochappan}}, \citenamefont {{Chingangbam}},\
  and\ \citenamefont {{Park}}}]{appleby2019ensemble}%
  \BibitemOpen
  \bibfield  {author} {\bibinfo {author} {\bibfnamefont {S.}~\bibnamefont
  {{Appleby}}}, \bibinfo {author} {\bibfnamefont {J.~P.}\ \bibnamefont
  {{Kochappan}}}, \bibinfo {author} {\bibfnamefont {P.}~\bibnamefont
  {{Chingangbam}}},\ and\ \bibinfo {author} {\bibfnamefont {C.}~\bibnamefont
  {{Park}}},\ }\href {https://doi.org/10.3847/1538-4357/ab5057} {\bibfield
  {journal} {\bibinfo  {journal} {Astrophysical Journal}\ }\textbf {\bibinfo
  {volume} {887}},\ \bibinfo {eid} {128} (\bibinfo {year} {2019})},\ \Eprint
  {https://arxiv.org/abs/1908.02440} {arXiv:1908.02440} \BibitemShut {NoStop}%
\bibitem [{\citenamefont {{Appleby}}\ \emph {et~al.}(2023)\citenamefont
  {{Appleby}}, \citenamefont {{Kochappan}}, \citenamefont {{Chingangbam}},\
  and\ \citenamefont {{Park}}}]{Appleby2023}%
  \BibitemOpen
  \bibfield  {author} {\bibinfo {author} {\bibfnamefont {S.}~\bibnamefont
  {{Appleby}}}, \bibinfo {author} {\bibfnamefont {J.~P.}\ \bibnamefont
  {{Kochappan}}}, \bibinfo {author} {\bibfnamefont {P.}~\bibnamefont
  {{Chingangbam}}},\ and\ \bibinfo {author} {\bibfnamefont {C.}~\bibnamefont
  {{Park}}},\ }\href {https://doi.org/10.3847/1538-4357/aca530} {\bibfield
  {journal} {\bibinfo  {journal} {Astrophysical Journal}\ }\textbf {\bibinfo
  {volume} {942}},\ \bibinfo {eid} {110} (\bibinfo {year} {2023})},\ \Eprint
  {https://arxiv.org/abs/2208.10164} {arXiv:2208.10164} \BibitemShut {NoStop}%
\bibitem [{\citenamefont {Liu}\ \emph {et~al.}(2025{\natexlab{a}})\citenamefont
  {Liu}, \citenamefont {Wu}, \citenamefont {Villaescusa-Navarro}, \citenamefont
  {Baldi}, \citenamefont {Valogiannis},\ and\ \citenamefont
  {Fang}}]{liu2025probing}%
  \BibitemOpen
  \bibfield  {author} {\bibinfo {author} {\bibfnamefont {W.}~\bibnamefont
  {Liu}}, \bibinfo {author} {\bibfnamefont {L.}~\bibnamefont {Wu}}, \bibinfo
  {author} {\bibfnamefont {F.}~\bibnamefont {Villaescusa-Navarro}}, \bibinfo
  {author} {\bibfnamefont {M.}~\bibnamefont {Baldi}}, \bibinfo {author}
  {\bibfnamefont {G.}~\bibnamefont {Valogiannis}},\ and\ \bibinfo {author}
  {\bibfnamefont {W.}~\bibnamefont {Fang}},\ }\href@noop {} {\bibfield
  {journal} {\bibinfo  {journal} {Journal of Cosmology and Astroparticle
  Physics}\ }\textbf {\bibinfo {volume} {2025}}\bibinfo  {number} { (04)},\
  \bibinfo {pages} {088}}\BibitemShut {NoStop}%
\bibitem [{\citenamefont {Kanafi}\ and\ \citenamefont
  {Movahed}(2024)}]{kanafi2024probing}%
  \BibitemOpen
\bibfield  {number} {  }\bibfield  {author} {\bibinfo {author} {\bibfnamefont
  {M.~J.}\ \bibnamefont {Kanafi}}\ and\ \bibinfo {author} {\bibfnamefont
  {S.}~\bibnamefont {Movahed}},\ }\href@noop {} {\bibfield  {journal} {\bibinfo
   {journal} {The Astrophysical Journal}\ }\textbf {\bibinfo {volume} {963}},\
  \bibinfo {pages} {31} (\bibinfo {year} {2024})}\BibitemShut {NoStop}%
\bibitem [{\citenamefont {Afzal}\ \emph {et~al.}(2025)\citenamefont {Afzal},
  \citenamefont {Alakhras}, \citenamefont {Kanafi},\ and\ \citenamefont
  {Movahed}}]{afzal2025cosmic}%
  \BibitemOpen
  \bibfield  {author} {\bibinfo {author} {\bibfnamefont {A.}~\bibnamefont
  {Afzal}}, \bibinfo {author} {\bibfnamefont {M.}~\bibnamefont {Alakhras}},
  \bibinfo {author} {\bibfnamefont {M.~J.}\ \bibnamefont {Kanafi}},\ and\
  \bibinfo {author} {\bibfnamefont {S.}~\bibnamefont {Movahed}},\ }\href@noop
  {} {\bibfield  {journal} {\bibinfo  {journal} {Monthly Notices of the Royal
  Astronomical Society}\ }\textbf {\bibinfo {volume} {541}},\ \bibinfo {pages}
  {3851} (\bibinfo {year} {2025})}\BibitemShut {NoStop}%
\bibitem [{\citenamefont {Pranav}\ \emph {et~al.}(2017)\citenamefont {Pranav},
  \citenamefont {Edelsbrunner}, \citenamefont {Van~de Weygaert}, \citenamefont
  {Vegter}, \citenamefont {Kerber}, \citenamefont {Jones},\ and\ \citenamefont
  {Wintraecken}}]{pranav2017topology}%
  \BibitemOpen
  \bibfield  {author} {\bibinfo {author} {\bibfnamefont {P.}~\bibnamefont
  {Pranav}}, \bibinfo {author} {\bibfnamefont {H.}~\bibnamefont
  {Edelsbrunner}}, \bibinfo {author} {\bibfnamefont {R.}~\bibnamefont {Van~de
  Weygaert}}, \bibinfo {author} {\bibfnamefont {G.}~\bibnamefont {Vegter}},
  \bibinfo {author} {\bibfnamefont {M.}~\bibnamefont {Kerber}}, \bibinfo
  {author} {\bibfnamefont {B.~J.}\ \bibnamefont {Jones}},\ and\ \bibinfo
  {author} {\bibfnamefont {M.}~\bibnamefont {Wintraecken}},\ }\href@noop {}
  {\bibfield  {journal} {\bibinfo  {journal} {Monthly Notices of the Royal
  Astronomical Society}\ }\textbf {\bibinfo {volume} {465}},\ \bibinfo {pages}
  {4281} (\bibinfo {year} {2017})}\BibitemShut {NoStop}%
\bibitem [{\citenamefont {{Xu}}\ \emph {et~al.}(2019)\citenamefont {{Xu}},
  \citenamefont {{Cisewski-Kehe}}, \citenamefont {{Green}},\ and\ \citenamefont
  {{Nagai}}}]{Xu2019}%
  \BibitemOpen
  \bibfield  {author} {\bibinfo {author} {\bibfnamefont {X.}~\bibnamefont
  {{Xu}}}, \bibinfo {author} {\bibfnamefont {J.}~\bibnamefont
  {{Cisewski-Kehe}}}, \bibinfo {author} {\bibfnamefont {S.~B.}\ \bibnamefont
  {{Green}}},\ and\ \bibinfo {author} {\bibfnamefont {D.}~\bibnamefont
  {{Nagai}}},\ }\href {https://doi.org/10.1016/j.ascom.2019.02.003} {\bibfield
  {journal} {\bibinfo  {journal} {Astronomy and Computing}\ }\textbf {\bibinfo
  {volume} {27}},\ \bibinfo {eid} {34} (\bibinfo {year} {2019})},\ \Eprint
  {https://arxiv.org/abs/1811.08450} {arXiv:1811.08450} \BibitemShut {NoStop}%
\bibitem [{\citenamefont {{Wilding}}\ \emph {et~al.}(2021)\citenamefont
  {{Wilding}}, \citenamefont {{Nevenzeel}}, \citenamefont {{van de Weygaert}},
  \citenamefont {{Vegter}}, \citenamefont {{Pranav}}, \citenamefont {{Jones}},
  \citenamefont {{Efstathiou}},\ and\ \citenamefont
  {{Feldbrugge}}}]{2021MNRAS.507.2968W}%
  \BibitemOpen
  \bibfield  {author} {\bibinfo {author} {\bibfnamefont {G.}~\bibnamefont
  {{Wilding}}}, \bibinfo {author} {\bibfnamefont {K.}~\bibnamefont
  {{Nevenzeel}}}, \bibinfo {author} {\bibfnamefont {R.}~\bibnamefont {{van de
  Weygaert}}}, \bibinfo {author} {\bibfnamefont {G.}~\bibnamefont {{Vegter}}},
  \bibinfo {author} {\bibfnamefont {P.}~\bibnamefont {{Pranav}}}, \bibinfo
  {author} {\bibfnamefont {B.~J.~T.}\ \bibnamefont {{Jones}}}, \bibinfo
  {author} {\bibfnamefont {K.}~\bibnamefont {{Efstathiou}}},\ and\ \bibinfo
  {author} {\bibfnamefont {J.}~\bibnamefont {{Feldbrugge}}},\ }\href
  {https://doi.org/10.1093/mnras/stab2326} {\bibfield  {journal} {\bibinfo
  {journal} {Monthly Notices of the RAS}\ }\textbf {\bibinfo {volume} {507}},\
  \bibinfo {pages} {2968} (\bibinfo {year} {2021})},\ \Eprint
  {https://arxiv.org/abs/2011.12851} {arXiv:2011.12851} \BibitemShut {NoStop}%
\bibitem [{\citenamefont {Biagetti}\ \emph {et~al.}(2022)\citenamefont
  {Biagetti}, \citenamefont {Calles}, \citenamefont {Castiblanco},
  \citenamefont {Cole},\ and\ \citenamefont {Nore{\~n}a}}]{biagetti2022fisher}%
  \BibitemOpen
  \bibfield  {author} {\bibinfo {author} {\bibfnamefont {M.}~\bibnamefont
  {Biagetti}}, \bibinfo {author} {\bibfnamefont {J.}~\bibnamefont {Calles}},
  \bibinfo {author} {\bibfnamefont {L.}~\bibnamefont {Castiblanco}}, \bibinfo
  {author} {\bibfnamefont {A.}~\bibnamefont {Cole}},\ and\ \bibinfo {author}
  {\bibfnamefont {J.}~\bibnamefont {Nore{\~n}a}},\ }\href@noop {} {\bibfield
  {journal} {\bibinfo  {journal} {Journal of Cosmology and Astroparticle
  Physics}\ }\textbf {\bibinfo {volume} {2022}}\bibinfo  {number} { (10)},\
  \bibinfo {pages} {002}}\BibitemShut {NoStop}%
\bibitem [{\citenamefont {Yip}\ \emph {et~al.}(2024)\citenamefont {Yip},
  \citenamefont {Biagetti}, \citenamefont {Cole}, \citenamefont {Viswanathan},\
  and\ \citenamefont {Shiu}}]{yip2024cosmology}%
  \BibitemOpen
\bibfield  {number} {  }\bibfield  {author} {\bibinfo {author} {\bibfnamefont
  {J.~H.}\ \bibnamefont {Yip}}, \bibinfo {author} {\bibfnamefont
  {M.}~\bibnamefont {Biagetti}}, \bibinfo {author} {\bibfnamefont
  {A.}~\bibnamefont {Cole}}, \bibinfo {author} {\bibfnamefont {K.}~\bibnamefont
  {Viswanathan}},\ and\ \bibinfo {author} {\bibfnamefont {G.}~\bibnamefont
  {Shiu}},\ }\href@noop {} {\bibfield  {journal} {\bibinfo  {journal} {Journal
  of Cosmology and Astroparticle Physics}\ }\textbf {\bibinfo {volume}
  {2024}}\bibinfo  {number} { (09)},\ \bibinfo {pages} {034}}\BibitemShut
  {NoStop}%
\bibitem [{\citenamefont {Jalali~Kanafi}\ \emph {et~al.}(2024)\citenamefont
  {Jalali~Kanafi}, \citenamefont {Ansarifard},\ and\ \citenamefont
  {Movahed}}]{jalali2024imprint}%
  \BibitemOpen
\bibfield  {number} {  }\bibfield  {author} {\bibinfo {author} {\bibfnamefont
  {M.}~\bibnamefont {Jalali~Kanafi}}, \bibinfo {author} {\bibfnamefont
  {S.}~\bibnamefont {Ansarifard}},\ and\ \bibinfo {author} {\bibfnamefont
  {S.}~\bibnamefont {Movahed}},\ }\href@noop {} {\bibfield  {journal} {\bibinfo
   {journal} {Monthly Notices of the Royal Astronomical Society}\ }\textbf
  {\bibinfo {volume} {535}},\ \bibinfo {pages} {657} (\bibinfo {year}
  {2024})}\BibitemShut {NoStop}%
\bibitem [{\citenamefont {Abedi}\ \emph {et~al.}(2024)\citenamefont {Abedi},
  \citenamefont {Kanafi},\ and\ \citenamefont {Movahed}}]{abedi2024impact}%
  \BibitemOpen
  \bibfield  {author} {\bibinfo {author} {\bibfnamefont {F.}~\bibnamefont
  {Abedi}}, \bibinfo {author} {\bibfnamefont {M.~H.~J.}\ \bibnamefont
  {Kanafi}},\ and\ \bibinfo {author} {\bibfnamefont {S.~M.~S.}\ \bibnamefont
  {Movahed}},\ }\href@noop {} {\bibfield  {journal} {\bibinfo  {journal} {arXiv
  preprint arXiv:2410.01751}\ } (\bibinfo {year} {2024})}\BibitemShut {NoStop}%
\bibitem [{\citenamefont {Prat}\ \emph {et~al.}(2025)\citenamefont {Prat},
  \citenamefont {Gatti}, \citenamefont {Doux}, \citenamefont {Pranav},
  \citenamefont {Chang}, \citenamefont {Jeffrey}, \citenamefont {Whiteway},
  \citenamefont {Anbajagane}, \citenamefont {Sugiyama}, \citenamefont {Thomsen}
  \emph {et~al.}}]{prat2025dark}%
  \BibitemOpen
  \bibfield  {author} {\bibinfo {author} {\bibfnamefont {J.}~\bibnamefont
  {Prat}}, \bibinfo {author} {\bibfnamefont {M.}~\bibnamefont {Gatti}},
  \bibinfo {author} {\bibfnamefont {C.}~\bibnamefont {Doux}}, \bibinfo {author}
  {\bibfnamefont {P.}~\bibnamefont {Pranav}}, \bibinfo {author} {\bibfnamefont
  {C.}~\bibnamefont {Chang}}, \bibinfo {author} {\bibfnamefont
  {N.}~\bibnamefont {Jeffrey}}, \bibinfo {author} {\bibfnamefont
  {L.}~\bibnamefont {Whiteway}}, \bibinfo {author} {\bibfnamefont
  {D.}~\bibnamefont {Anbajagane}}, \bibinfo {author} {\bibfnamefont
  {S.}~\bibnamefont {Sugiyama}}, \bibinfo {author} {\bibfnamefont
  {A.}~\bibnamefont {Thomsen}}, \emph {et~al.},\ }\href@noop {} {\bibfield
  {journal} {\bibinfo  {journal} {arXiv preprint arXiv:2506.13439}\ } (\bibinfo
  {year} {2025})}\BibitemShut {NoStop}%
\bibitem [{\citenamefont {Lin}\ \emph {et~al.}(2023)\citenamefont {Lin},
  \citenamefont {von wietersheim Kramsta}, \citenamefont {Joachimi},\ and\
  \citenamefont {Feeney}}]{10.1093/mnras/stad2262}%
  \BibitemOpen
  \bibfield  {author} {\bibinfo {author} {\bibfnamefont {K.}~\bibnamefont
  {Lin}}, \bibinfo {author} {\bibfnamefont {M.}~\bibnamefont {von wietersheim
  Kramsta}}, \bibinfo {author} {\bibfnamefont {B.}~\bibnamefont {Joachimi}},\
  and\ \bibinfo {author} {\bibfnamefont {S.}~\bibnamefont {Feeney}},\ }\href
  {https://doi.org/10.1093/mnras/stad2262} {\bibfield  {journal} {\bibinfo
  {journal} {Monthly Notices of the Royal Astronomical Society}\ }\textbf
  {\bibinfo {volume} {524}},\ \bibinfo {pages} {6167} (\bibinfo {year}
  {2023})}\BibitemShut {NoStop}%
\bibitem [{\citenamefont {von Wietersheim-Kramsta}\ \emph
  {et~al.}(2025)\citenamefont {von Wietersheim-Kramsta}, \citenamefont {Lin},
  \citenamefont {Tessore}, \citenamefont {Joachimi}, \citenamefont {Loureiro},
  \citenamefont {Reischke},\ and\ \citenamefont {Wright}}]{von2025kids}%
  \BibitemOpen
  \bibfield  {author} {\bibinfo {author} {\bibfnamefont {M.}~\bibnamefont {von
  Wietersheim-Kramsta}}, \bibinfo {author} {\bibfnamefont {K.}~\bibnamefont
  {Lin}}, \bibinfo {author} {\bibfnamefont {N.}~\bibnamefont {Tessore}},
  \bibinfo {author} {\bibfnamefont {B.}~\bibnamefont {Joachimi}}, \bibinfo
  {author} {\bibfnamefont {A.}~\bibnamefont {Loureiro}}, \bibinfo {author}
  {\bibfnamefont {R.}~\bibnamefont {Reischke}},\ and\ \bibinfo {author}
  {\bibfnamefont {A.~H.}\ \bibnamefont {Wright}},\ }\href@noop {} {\bibfield
  {journal} {\bibinfo  {journal} {Astronomy \& Astrophysics}\ }\textbf
  {\bibinfo {volume} {694}},\ \bibinfo {pages} {A223} (\bibinfo {year}
  {2025})}\BibitemShut {NoStop}%
\bibitem [{\citenamefont {Novaes}\ \emph {et~al.}(2025)\citenamefont {Novaes},
  \citenamefont {Thiele}, \citenamefont {Armijo}, \citenamefont {Cheng},
  \citenamefont {Cowell}, \citenamefont {Marques}, \citenamefont {Ferreira},
  \citenamefont {Shirasaki}, \citenamefont {Osato},\ and\ \citenamefont
  {Liu}}]{novaes2025cosmology}%
  \BibitemOpen
  \bibfield  {author} {\bibinfo {author} {\bibfnamefont {C.~P.}\ \bibnamefont
  {Novaes}}, \bibinfo {author} {\bibfnamefont {L.}~\bibnamefont {Thiele}},
  \bibinfo {author} {\bibfnamefont {J.}~\bibnamefont {Armijo}}, \bibinfo
  {author} {\bibfnamefont {S.}~\bibnamefont {Cheng}}, \bibinfo {author}
  {\bibfnamefont {J.~A.}\ \bibnamefont {Cowell}}, \bibinfo {author}
  {\bibfnamefont {G.~A.}\ \bibnamefont {Marques}}, \bibinfo {author}
  {\bibfnamefont {E.~G.}\ \bibnamefont {Ferreira}}, \bibinfo {author}
  {\bibfnamefont {M.}~\bibnamefont {Shirasaki}}, \bibinfo {author}
  {\bibfnamefont {K.}~\bibnamefont {Osato}},\ and\ \bibinfo {author}
  {\bibfnamefont {J.}~\bibnamefont {Liu}},\ }\href@noop {} {\bibfield
  {journal} {\bibinfo  {journal} {Physical Review D}\ }\textbf {\bibinfo
  {volume} {111}},\ \bibinfo {pages} {083510} (\bibinfo {year}
  {2025})}\BibitemShut {NoStop}%
\bibitem [{\citenamefont {Jeffrey}\ \emph {et~al.}(2021)\citenamefont
  {Jeffrey}, \citenamefont {Alsing},\ and\ \citenamefont
  {Lanusse}}]{jeffrey2021likelihood}%
  \BibitemOpen
  \bibfield  {author} {\bibinfo {author} {\bibfnamefont {N.}~\bibnamefont
  {Jeffrey}}, \bibinfo {author} {\bibfnamefont {J.}~\bibnamefont {Alsing}},\
  and\ \bibinfo {author} {\bibfnamefont {F.}~\bibnamefont {Lanusse}},\
  }\href@noop {} {\bibfield  {journal} {\bibinfo  {journal} {Monthly Notices of
  the Royal Astronomical Society}\ }\textbf {\bibinfo {volume} {501}},\
  \bibinfo {pages} {954} (\bibinfo {year} {2021})}\BibitemShut {NoStop}%
\bibitem [{\citenamefont {Gatti}\ \emph {et~al.}(2025)\citenamefont {Gatti},
  \citenamefont {Campailla}, \citenamefont {Jeffrey} \emph
  {et~al.}}]{Gatti2024}%
  \BibitemOpen
  \bibfield  {author} {\bibinfo {author} {\bibfnamefont {M.}~\bibnamefont
  {Gatti}}, \bibinfo {author} {\bibfnamefont {G.}~\bibnamefont {Campailla}},
  \bibinfo {author} {\bibfnamefont {N.}~\bibnamefont {Jeffrey}}, \emph {et~al.}
  (\bibinfo {collaboration} {Dark Energy Survey}),\ }\href
  {https://doi.org/10.1103/PhysRevD.111.063504} {\bibfield  {journal} {\bibinfo
   {journal} {Phys. Rev. D}\ }\textbf {\bibinfo {volume} {111}},\ \bibinfo
  {pages} {063504} (\bibinfo {year} {2025})}\BibitemShut {NoStop}%
\bibitem [{\citenamefont {Jeffrey}\ \emph {et~al.}(2025)\citenamefont
  {Jeffrey}, \citenamefont {Whiteway}, \citenamefont {Gatti}, \citenamefont
  {Williamson}, \citenamefont {Alsing}, \citenamefont {Porredon}, \citenamefont
  {Prat}, \citenamefont {Doux}, \citenamefont {Jain}, \citenamefont {Chang}
  \emph {et~al.}}]{jeffrey2025dark}%
  \BibitemOpen
  \bibfield  {author} {\bibinfo {author} {\bibfnamefont {N.}~\bibnamefont
  {Jeffrey}}, \bibinfo {author} {\bibfnamefont {L.}~\bibnamefont {Whiteway}},
  \bibinfo {author} {\bibfnamefont {M.}~\bibnamefont {Gatti}}, \bibinfo
  {author} {\bibfnamefont {J.}~\bibnamefont {Williamson}}, \bibinfo {author}
  {\bibfnamefont {J.}~\bibnamefont {Alsing}}, \bibinfo {author} {\bibfnamefont
  {A.}~\bibnamefont {Porredon}}, \bibinfo {author} {\bibfnamefont
  {J.}~\bibnamefont {Prat}}, \bibinfo {author} {\bibfnamefont {C.}~\bibnamefont
  {Doux}}, \bibinfo {author} {\bibfnamefont {B.}~\bibnamefont {Jain}}, \bibinfo
  {author} {\bibfnamefont {C.}~\bibnamefont {Chang}}, \emph {et~al.},\
  }\href@noop {} {\bibfield  {journal} {\bibinfo  {journal} {Monthly Notices of
  the Royal Astronomical Society}\ }\textbf {\bibinfo {volume} {536}},\
  \bibinfo {pages} {1303} (\bibinfo {year} {2025})}\BibitemShut {NoStop}%
\bibitem [{\citenamefont {Lanzieri}\ \emph {et~al.}(2025)\citenamefont
  {Lanzieri}, \citenamefont {Zeghal}, \citenamefont {Makinen}, \citenamefont
  {Boucaud}, \citenamefont {Starck},\ and\ \citenamefont
  {Lanusse}}]{lanzieri2025optimal}%
  \BibitemOpen
  \bibfield  {author} {\bibinfo {author} {\bibfnamefont {D.}~\bibnamefont
  {Lanzieri}}, \bibinfo {author} {\bibfnamefont {J.}~\bibnamefont {Zeghal}},
  \bibinfo {author} {\bibfnamefont {T.~L.}\ \bibnamefont {Makinen}}, \bibinfo
  {author} {\bibfnamefont {A.}~\bibnamefont {Boucaud}}, \bibinfo {author}
  {\bibfnamefont {J.-L.}\ \bibnamefont {Starck}},\ and\ \bibinfo {author}
  {\bibfnamefont {F.}~\bibnamefont {Lanusse}},\ }\href@noop {} {\bibfield
  {journal} {\bibinfo  {journal} {Astronomy \& Astrophysics}\ }\textbf
  {\bibinfo {volume} {697}},\ \bibinfo {pages} {A162} (\bibinfo {year}
  {2025})}\BibitemShut {NoStop}%
\bibitem [{\citenamefont {Zeghal}\ \emph {et~al.}(2025)\citenamefont {Zeghal},
  \citenamefont {Lanzieri}, \citenamefont {Lanusse}, \citenamefont {Boucaud},
  \citenamefont {Louppe}, \citenamefont {Aubourg},\ and\ \citenamefont
  {Bayer}}]{zeghal2025simulation}%
  \BibitemOpen
  \bibfield  {author} {\bibinfo {author} {\bibfnamefont {J.}~\bibnamefont
  {Zeghal}}, \bibinfo {author} {\bibfnamefont {D.}~\bibnamefont {Lanzieri}},
  \bibinfo {author} {\bibfnamefont {F.}~\bibnamefont {Lanusse}}, \bibinfo
  {author} {\bibfnamefont {A.}~\bibnamefont {Boucaud}}, \bibinfo {author}
  {\bibfnamefont {G.}~\bibnamefont {Louppe}}, \bibinfo {author} {\bibfnamefont
  {E.}~\bibnamefont {Aubourg}},\ and\ \bibinfo {author} {\bibfnamefont {A.~E.}\
  \bibnamefont {Bayer}},\ }\href@noop {} {\bibfield  {journal} {\bibinfo
  {journal} {Astronomy \& Astrophysics}\ }\textbf {\bibinfo {volume} {699}},\
  \bibinfo {pages} {A327} (\bibinfo {year} {2025})}\BibitemShut {NoStop}%
\bibitem [{\citenamefont {Lemos}\ \emph {et~al.}(2024)\citenamefont {Lemos},
  \citenamefont {Parker}, \citenamefont {Hahn}, \citenamefont {Ho},
  \citenamefont {Eickenberg}, \citenamefont {Hou}, \citenamefont {Massara},
  \citenamefont {Modi}, \citenamefont {Dizgah}, \citenamefont {Blancard},\ and\
  \citenamefont {Spergel}}]{PhysRevD.109.083536}%
  \BibitemOpen
  \bibfield  {author} {\bibinfo {author} {\bibfnamefont {P.}~\bibnamefont
  {Lemos}}, \bibinfo {author} {\bibfnamefont {L.}~\bibnamefont {Parker}},
  \bibinfo {author} {\bibfnamefont {C.}~\bibnamefont {Hahn}}, \bibinfo {author}
  {\bibfnamefont {S.}~\bibnamefont {Ho}}, \bibinfo {author} {\bibfnamefont
  {M.}~\bibnamefont {Eickenberg}}, \bibinfo {author} {\bibfnamefont
  {J.}~\bibnamefont {Hou}}, \bibinfo {author} {\bibfnamefont {E.}~\bibnamefont
  {Massara}}, \bibinfo {author} {\bibfnamefont {C.}~\bibnamefont {Modi}},
  \bibinfo {author} {\bibfnamefont {A.~M.}\ \bibnamefont {Dizgah}}, \bibinfo
  {author} {\bibfnamefont {B.~R.-S.}\ \bibnamefont {Blancard}},\ and\ \bibinfo
  {author} {\bibfnamefont {D.}~\bibnamefont {Spergel}} (\bibinfo
  {collaboration} {SimBIG Collaboration}),\ }\href
  {https://doi.org/10.1103/PhysRevD.109.083536} {\bibfield  {journal} {\bibinfo
   {journal} {Phys. Rev. D}\ }\textbf {\bibinfo {volume} {109}},\ \bibinfo
  {pages} {083536} (\bibinfo {year} {2024})}\BibitemShut {NoStop}%
\bibitem [{\citenamefont {Massara}\ \emph {et~al.}(2024)\citenamefont
  {Massara}, \citenamefont {Hahn}, \citenamefont {Eickenberg}, \citenamefont
  {Ho}, \citenamefont {Hou}, \citenamefont {Lemos}, \citenamefont {Modi},
  \citenamefont {Dizgah}, \citenamefont {Parker},\ and\ \citenamefont
  {Blancard}}]{massara2024sc}%
  \BibitemOpen
  \bibfield  {author} {\bibinfo {author} {\bibfnamefont {E.}~\bibnamefont
  {Massara}}, \bibinfo {author} {\bibfnamefont {C.}~\bibnamefont {Hahn}},
  \bibinfo {author} {\bibfnamefont {M.}~\bibnamefont {Eickenberg}}, \bibinfo
  {author} {\bibfnamefont {S.}~\bibnamefont {Ho}}, \bibinfo {author}
  {\bibfnamefont {J.}~\bibnamefont {Hou}}, \bibinfo {author} {\bibfnamefont
  {P.}~\bibnamefont {Lemos}}, \bibinfo {author} {\bibfnamefont
  {C.}~\bibnamefont {Modi}}, \bibinfo {author} {\bibfnamefont {A.~M.}\
  \bibnamefont {Dizgah}}, \bibinfo {author} {\bibfnamefont {L.}~\bibnamefont
  {Parker}},\ and\ \bibinfo {author} {\bibfnamefont {B.~R.-S.}\ \bibnamefont
  {Blancard}},\ }\href@noop {} {\bibfield  {journal} {\bibinfo  {journal}
  {arXiv preprint arXiv:2404.04228}\ } (\bibinfo {year} {2024})}\BibitemShut
  {NoStop}%
\bibitem [{\citenamefont {Hahn}\ \emph {et~al.}(2024)\citenamefont {Hahn},
  \citenamefont {Eickenberg}, \citenamefont {Ho}, \citenamefont {Hou},
  \citenamefont {Lemos}, \citenamefont {Massara}, \citenamefont {Modi},
  \citenamefont {Dizgah}, \citenamefont {Parker}, \citenamefont {Blancard}
  \emph {et~al.}}]{hahn2024cosmological}%
  \BibitemOpen
  \bibfield  {author} {\bibinfo {author} {\bibfnamefont {C.}~\bibnamefont
  {Hahn}}, \bibinfo {author} {\bibfnamefont {M.}~\bibnamefont {Eickenberg}},
  \bibinfo {author} {\bibfnamefont {S.}~\bibnamefont {Ho}}, \bibinfo {author}
  {\bibfnamefont {J.}~\bibnamefont {Hou}}, \bibinfo {author} {\bibfnamefont
  {P.}~\bibnamefont {Lemos}}, \bibinfo {author} {\bibfnamefont
  {E.}~\bibnamefont {Massara}}, \bibinfo {author} {\bibfnamefont
  {C.}~\bibnamefont {Modi}}, \bibinfo {author} {\bibfnamefont {A.~M.}\
  \bibnamefont {Dizgah}}, \bibinfo {author} {\bibfnamefont {L.}~\bibnamefont
  {Parker}}, \bibinfo {author} {\bibfnamefont {B.~R.-S.}\ \bibnamefont
  {Blancard}}, \emph {et~al.},\ }\href@noop {} {\bibfield  {journal} {\bibinfo
  {journal} {Physical Review D}\ }\textbf {\bibinfo {volume} {109}},\ \bibinfo
  {pages} {083534} (\bibinfo {year} {2024})}\BibitemShut {NoStop}%
\bibitem [{\citenamefont {Hou}\ \emph {et~al.}(2024)\citenamefont {Hou},
  \citenamefont {Dizgah}, \citenamefont {Hahn}, \citenamefont {Eickenberg},
  \citenamefont {Ho}, \citenamefont {Lemos}, \citenamefont {Massara},
  \citenamefont {Modi}, \citenamefont {Parker},\ and\ \citenamefont
  {Blancard}}]{Hou2024}%
  \BibitemOpen
  \bibfield  {author} {\bibinfo {author} {\bibfnamefont {J.}~\bibnamefont
  {Hou}}, \bibinfo {author} {\bibfnamefont {A.~M.}\ \bibnamefont {Dizgah}},
  \bibinfo {author} {\bibfnamefont {C.}~\bibnamefont {Hahn}}, \bibinfo {author}
  {\bibfnamefont {M.}~\bibnamefont {Eickenberg}}, \bibinfo {author}
  {\bibfnamefont {S.}~\bibnamefont {Ho}}, \bibinfo {author} {\bibfnamefont
  {P.}~\bibnamefont {Lemos}}, \bibinfo {author} {\bibfnamefont
  {E.}~\bibnamefont {Massara}}, \bibinfo {author} {\bibfnamefont
  {C.}~\bibnamefont {Modi}}, \bibinfo {author} {\bibfnamefont {L.}~\bibnamefont
  {Parker}},\ and\ \bibinfo {author} {\bibfnamefont {B.~R.-S.}\ \bibnamefont
  {Blancard}},\ }\href {https://doi.org/10.1103/PhysRevD.109.103528} {\bibfield
   {journal} {\bibinfo  {journal} {Phys. Rev. D}\ }\textbf {\bibinfo {volume}
  {109}},\ \bibinfo {pages} {103528} (\bibinfo {year} {2024})}\BibitemShut
  {NoStop}%
\bibitem [{\citenamefont {Mancini}\ \emph {et~al.}(2024)\citenamefont
  {Mancini}, \citenamefont {Lin},\ and\ \citenamefont
  {McEwen}}]{mancini2024field}%
  \BibitemOpen
  \bibfield  {author} {\bibinfo {author} {\bibfnamefont {A.~S.}\ \bibnamefont
  {Mancini}}, \bibinfo {author} {\bibfnamefont {K.}~\bibnamefont {Lin}},\ and\
  \bibinfo {author} {\bibfnamefont {J.~D.}\ \bibnamefont {McEwen}},\
  }\href@noop {} {\bibfield  {journal} {\bibinfo  {journal} {arXiv preprint
  arXiv:2410.10616}\ } (\bibinfo {year} {2024})}\BibitemShut {NoStop}%
\bibitem [{\citenamefont {Tucci}\ and\ \citenamefont
  {Schmidt}(2024)}]{tucci2024eftoflss}%
  \BibitemOpen
  \bibfield  {author} {\bibinfo {author} {\bibfnamefont {B.}~\bibnamefont
  {Tucci}}\ and\ \bibinfo {author} {\bibfnamefont {F.}~\bibnamefont
  {Schmidt}},\ }\href@noop {} {\bibfield  {journal} {\bibinfo  {journal}
  {Journal of Cosmology and Astroparticle Physics}\ }\textbf {\bibinfo {volume}
  {2024}}\bibinfo  {number} { (05)},\ \bibinfo {pages} {063}}\BibitemShut
  {NoStop}%
\bibitem [{\citenamefont {Modi}\ \emph {et~al.}(2025)\citenamefont {Modi},
  \citenamefont {Pandey}, \citenamefont {Ho}, \citenamefont {Hahn},
  \citenamefont {R{\'e}galdo-Saint~Blancard},\ and\ \citenamefont
  {Wandelt}}]{modi2025sensitivity}%
  \BibitemOpen
\bibfield  {number} {  }\bibfield  {author} {\bibinfo {author} {\bibfnamefont
  {C.}~\bibnamefont {Modi}}, \bibinfo {author} {\bibfnamefont {S.}~\bibnamefont
  {Pandey}}, \bibinfo {author} {\bibfnamefont {M.}~\bibnamefont {Ho}}, \bibinfo
  {author} {\bibfnamefont {C.}~\bibnamefont {Hahn}}, \bibinfo {author}
  {\bibfnamefont {B.}~\bibnamefont {R{\'e}galdo-Saint~Blancard}},\ and\
  \bibinfo {author} {\bibfnamefont {B.}~\bibnamefont {Wandelt}},\ }\href@noop
  {} {\bibfield  {journal} {\bibinfo  {journal} {Monthly Notices of the Royal
  Astronomical Society}\ }\textbf {\bibinfo {volume} {536}},\ \bibinfo {pages}
  {254} (\bibinfo {year} {2025})}\BibitemShut {NoStop}%
\bibitem [{\citenamefont {Reza}\ \emph {et~al.}(2024)\citenamefont {Reza},
  \citenamefont {Zhang}, \citenamefont {Avestruz}, \citenamefont {Strigari},
  \citenamefont {Shevchuk},\ and\ \citenamefont
  {Villaescusa-Navarro}}]{reza2024constraining}%
  \BibitemOpen
  \bibfield  {author} {\bibinfo {author} {\bibfnamefont {M.}~\bibnamefont
  {Reza}}, \bibinfo {author} {\bibfnamefont {Y.}~\bibnamefont {Zhang}},
  \bibinfo {author} {\bibfnamefont {C.}~\bibnamefont {Avestruz}}, \bibinfo
  {author} {\bibfnamefont {L.~E.}\ \bibnamefont {Strigari}}, \bibinfo {author}
  {\bibfnamefont {S.}~\bibnamefont {Shevchuk}},\ and\ \bibinfo {author}
  {\bibfnamefont {F.}~\bibnamefont {Villaescusa-Navarro}},\ }\href@noop {}
  {\bibfield  {journal} {\bibinfo  {journal} {arXiv preprint arXiv:2409.20507}\
  } (\bibinfo {year} {2024})}\BibitemShut {NoStop}%
\bibitem [{\citenamefont {Zubeldia}\ \emph {et~al.}(2025)\citenamefont
  {Zubeldia}, \citenamefont {Bolliet}, \citenamefont {Challinor},\ and\
  \citenamefont {Handley}}]{zubeldia2025extracting}%
  \BibitemOpen
  \bibfield  {author} {\bibinfo {author} {\bibfnamefont {{\'I}.}~\bibnamefont
  {Zubeldia}}, \bibinfo {author} {\bibfnamefont {B.}~\bibnamefont {Bolliet}},
  \bibinfo {author} {\bibfnamefont {A.}~\bibnamefont {Challinor}},\ and\
  \bibinfo {author} {\bibfnamefont {W.}~\bibnamefont {Handley}},\ }\href@noop
  {} {\bibfield  {journal} {\bibinfo  {journal} {arXiv preprint
  arXiv:2504.10230}\ } (\bibinfo {year} {2025})}\BibitemShut {NoStop}%
\bibitem [{\citenamefont {Su}\ \emph {et~al.}(2025)\citenamefont {Su},
  \citenamefont {Shan}, \citenamefont {Zhao}, \citenamefont {Xu},\ and\
  \citenamefont {Zhang}}]{su2025cosmological}%
  \BibitemOpen
  \bibfield  {author} {\bibinfo {author} {\bibfnamefont {C.}~\bibnamefont
  {Su}}, \bibinfo {author} {\bibfnamefont {H.}~\bibnamefont {Shan}}, \bibinfo
  {author} {\bibfnamefont {C.}~\bibnamefont {Zhao}}, \bibinfo {author}
  {\bibfnamefont {W.}~\bibnamefont {Xu}},\ and\ \bibinfo {author}
  {\bibfnamefont {J.}~\bibnamefont {Zhang}},\ }\href@noop {} {\bibfield
  {journal} {\bibinfo  {journal} {arXiv preprint arXiv:2504.15149}\ } (\bibinfo
  {year} {2025})}\BibitemShut {NoStop}%
\bibitem [{\citenamefont {Prelogovi{\'c}}\ and\ \citenamefont
  {Mesinger}(2023)}]{prelogovic2023exploring}%
  \BibitemOpen
  \bibfield  {author} {\bibinfo {author} {\bibfnamefont {D.}~\bibnamefont
  {Prelogovi{\'c}}}\ and\ \bibinfo {author} {\bibfnamefont {A.}~\bibnamefont
  {Mesinger}},\ }\href@noop {} {\bibfield  {journal} {\bibinfo  {journal}
  {Monthly Notices of the Royal Astronomical Society}\ }\textbf {\bibinfo
  {volume} {524}},\ \bibinfo {pages} {4239} (\bibinfo {year}
  {2023})}\BibitemShut {NoStop}%
\bibitem [{\citenamefont {Bairagi}\ \emph {et~al.}(2025)\citenamefont
  {Bairagi}, \citenamefont {Wandelt},\ and\ \citenamefont
  {Villaescusa-Navarro}}]{bairagi2025many}%
  \BibitemOpen
  \bibfield  {author} {\bibinfo {author} {\bibfnamefont {A.}~\bibnamefont
  {Bairagi}}, \bibinfo {author} {\bibfnamefont {B.}~\bibnamefont {Wandelt}},\
  and\ \bibinfo {author} {\bibfnamefont {F.}~\bibnamefont
  {Villaescusa-Navarro}},\ }\href@noop {} {\bibfield  {journal} {\bibinfo
  {journal} {arXiv preprint arXiv:2503.13755}\ } (\bibinfo {year}
  {2025})}\BibitemShut {NoStop}%
\bibitem [{\citenamefont {Lemos}\ \emph {et~al.}(2023)\citenamefont {Lemos},
  \citenamefont {Cranmer}, \citenamefont {Abidi}, \citenamefont {Hahn},
  \citenamefont {Eickenberg}, \citenamefont {Massara}, \citenamefont {Yallup},\
  and\ \citenamefont {Ho}}]{lemos2023robust}%
  \BibitemOpen
  \bibfield  {author} {\bibinfo {author} {\bibfnamefont {P.}~\bibnamefont
  {Lemos}}, \bibinfo {author} {\bibfnamefont {M.}~\bibnamefont {Cranmer}},
  \bibinfo {author} {\bibfnamefont {M.}~\bibnamefont {Abidi}}, \bibinfo
  {author} {\bibfnamefont {C.}~\bibnamefont {Hahn}}, \bibinfo {author}
  {\bibfnamefont {M.}~\bibnamefont {Eickenberg}}, \bibinfo {author}
  {\bibfnamefont {E.}~\bibnamefont {Massara}}, \bibinfo {author} {\bibfnamefont
  {D.}~\bibnamefont {Yallup}},\ and\ \bibinfo {author} {\bibfnamefont
  {S.}~\bibnamefont {Ho}},\ }\href@noop {} {\bibfield  {journal} {\bibinfo
  {journal} {Machine Learning: Science and Technology}\ }\textbf {\bibinfo
  {volume} {4}},\ \bibinfo {pages} {01LT01} (\bibinfo {year}
  {2023})}\BibitemShut {NoStop}%
\bibitem [{\citenamefont {de~Santi}\ \emph {et~al.}(2025)\citenamefont
  {de~Santi}, \citenamefont {Villaescusa-Navarro}, \citenamefont {Abramo},
  \citenamefont {Shao}, \citenamefont {Perez}, \citenamefont {Castro},
  \citenamefont {Ni}, \citenamefont {Lovell}, \citenamefont
  {Hern{\'a}ndez-Mart{\'\i}nez}, \citenamefont {Marinacci} \emph
  {et~al.}}]{de2025field}%
  \BibitemOpen
  \bibfield  {author} {\bibinfo {author} {\bibfnamefont {N.~S.}\ \bibnamefont
  {de~Santi}}, \bibinfo {author} {\bibfnamefont {F.}~\bibnamefont
  {Villaescusa-Navarro}}, \bibinfo {author} {\bibfnamefont {L.~R.}\
  \bibnamefont {Abramo}}, \bibinfo {author} {\bibfnamefont {H.}~\bibnamefont
  {Shao}}, \bibinfo {author} {\bibfnamefont {L.~A.}\ \bibnamefont {Perez}},
  \bibinfo {author} {\bibfnamefont {T.}~\bibnamefont {Castro}}, \bibinfo
  {author} {\bibfnamefont {Y.}~\bibnamefont {Ni}}, \bibinfo {author}
  {\bibfnamefont {C.~C.}\ \bibnamefont {Lovell}}, \bibinfo {author}
  {\bibfnamefont {E.}~\bibnamefont {Hern{\'a}ndez-Mart{\'\i}nez}}, \bibinfo
  {author} {\bibfnamefont {F.}~\bibnamefont {Marinacci}}, \emph {et~al.},\
  }\href@noop {} {\bibfield  {journal} {\bibinfo  {journal} {Journal of
  Cosmology and Astroparticle Physics}\ }\textbf {\bibinfo {volume}
  {2025}}\bibinfo  {number} { (01)},\ \bibinfo {pages} {082}}\BibitemShut
  {NoStop}%
\bibitem [{\citenamefont {Lemos}\ \emph {et~al.}(2021)\citenamefont {Lemos},
  \citenamefont {Jeffrey}, \citenamefont {Whiteway}, \citenamefont {Lahav},
  \citenamefont {Libeskind},\ and\ \citenamefont {Hoffman}}]{lemos2021sum}%
  \BibitemOpen
\bibfield  {number} {  }\bibfield  {author} {\bibinfo {author} {\bibfnamefont
  {P.}~\bibnamefont {Lemos}}, \bibinfo {author} {\bibfnamefont
  {N.}~\bibnamefont {Jeffrey}}, \bibinfo {author} {\bibfnamefont
  {L.}~\bibnamefont {Whiteway}}, \bibinfo {author} {\bibfnamefont
  {O.}~\bibnamefont {Lahav}}, \bibinfo {author} {\bibfnamefont
  {N.}~\bibnamefont {Libeskind}},\ and\ \bibinfo {author} {\bibfnamefont
  {Y.}~\bibnamefont {Hoffman}},\ }\href@noop {} {\bibfield  {journal} {\bibinfo
   {journal} {Physical Review D}\ }\textbf {\bibinfo {volume} {103}},\ \bibinfo
  {pages} {023009} (\bibinfo {year} {2021})}\BibitemShut {NoStop}%
\bibitem [{\citenamefont {Hermans}\ \emph {et~al.}(2021)\citenamefont
  {Hermans}, \citenamefont {Banik}, \citenamefont {Weniger}, \citenamefont
  {Bertone},\ and\ \citenamefont {Louppe}}]{hermans2021towards}%
  \BibitemOpen
  \bibfield  {author} {\bibinfo {author} {\bibfnamefont {J.}~\bibnamefont
  {Hermans}}, \bibinfo {author} {\bibfnamefont {N.}~\bibnamefont {Banik}},
  \bibinfo {author} {\bibfnamefont {C.}~\bibnamefont {Weniger}}, \bibinfo
  {author} {\bibfnamefont {G.}~\bibnamefont {Bertone}},\ and\ \bibinfo {author}
  {\bibfnamefont {G.}~\bibnamefont {Louppe}},\ }\href@noop {} {\bibfield
  {journal} {\bibinfo  {journal} {Monthly Notices of the Royal Astronomical
  Society}\ }\textbf {\bibinfo {volume} {507}},\ \bibinfo {pages} {1999}
  (\bibinfo {year} {2021})}\BibitemShut {NoStop}%
\bibitem [{\citenamefont {Brehmer}\ \emph {et~al.}(2019)\citenamefont
  {Brehmer}, \citenamefont {Mishra-Sharma}, \citenamefont {Hermans},
  \citenamefont {Louppe},\ and\ \citenamefont {Cranmer}}]{brehmer2019mining}%
  \BibitemOpen
  \bibfield  {author} {\bibinfo {author} {\bibfnamefont {J.}~\bibnamefont
  {Brehmer}}, \bibinfo {author} {\bibfnamefont {S.}~\bibnamefont
  {Mishra-Sharma}}, \bibinfo {author} {\bibfnamefont {J.}~\bibnamefont
  {Hermans}}, \bibinfo {author} {\bibfnamefont {G.}~\bibnamefont {Louppe}},\
  and\ \bibinfo {author} {\bibfnamefont {K.}~\bibnamefont {Cranmer}},\
  }\href@noop {} {\bibfield  {journal} {\bibinfo  {journal} {The Astrophysical
  Journal}\ }\textbf {\bibinfo {volume} {886}},\ \bibinfo {pages} {49}
  (\bibinfo {year} {2019})}\BibitemShut {NoStop}%
\bibitem [{\citenamefont {Coogan}\ \emph {et~al.}(2024)\citenamefont {Coogan},
  \citenamefont {Anau~Montel}, \citenamefont {Karchev}, \citenamefont
  {Grootes}, \citenamefont {Nattino},\ and\ \citenamefont
  {Weniger}}]{coogan2024effect}%
  \BibitemOpen
  \bibfield  {author} {\bibinfo {author} {\bibfnamefont {A.}~\bibnamefont
  {Coogan}}, \bibinfo {author} {\bibfnamefont {N.}~\bibnamefont {Anau~Montel}},
  \bibinfo {author} {\bibfnamefont {K.}~\bibnamefont {Karchev}}, \bibinfo
  {author} {\bibfnamefont {M.~W.}\ \bibnamefont {Grootes}}, \bibinfo {author}
  {\bibfnamefont {F.}~\bibnamefont {Nattino}},\ and\ \bibinfo {author}
  {\bibfnamefont {C.}~\bibnamefont {Weniger}},\ }\href@noop {} {\bibfield
  {journal} {\bibinfo  {journal} {Monthly Notices of the Royal Astronomical
  Society}\ }\textbf {\bibinfo {volume} {527}},\ \bibinfo {pages} {66}
  (\bibinfo {year} {2024})}\BibitemShut {NoStop}%
\bibitem [{\citenamefont {Hahn}\ \emph {et~al.}(2017)\citenamefont {Hahn},
  \citenamefont {Vakili}, \citenamefont {Walsh}, \citenamefont {Hearin},
  \citenamefont {Hogg},\ and\ \citenamefont {Campbell}}]{hahn2017approximate}%
  \BibitemOpen
  \bibfield  {author} {\bibinfo {author} {\bibfnamefont {C.}~\bibnamefont
  {Hahn}}, \bibinfo {author} {\bibfnamefont {M.}~\bibnamefont {Vakili}},
  \bibinfo {author} {\bibfnamefont {K.}~\bibnamefont {Walsh}}, \bibinfo
  {author} {\bibfnamefont {A.~P.}\ \bibnamefont {Hearin}}, \bibinfo {author}
  {\bibfnamefont {D.~W.}\ \bibnamefont {Hogg}},\ and\ \bibinfo {author}
  {\bibfnamefont {D.}~\bibnamefont {Campbell}},\ }\href@noop {} {\bibfield
  {journal} {\bibinfo  {journal} {Monthly Notices of the Royal Astronomical
  Society}\ }\textbf {\bibinfo {volume} {469}},\ \bibinfo {pages} {2791}
  (\bibinfo {year} {2017})}\BibitemShut {NoStop}%
\bibitem [{\citenamefont {Legin}\ \emph {et~al.}(2021)\citenamefont {Legin},
  \citenamefont {Hezaveh}, \citenamefont {Levasseur},\ and\ \citenamefont
  {Wandelt}}]{legin2021simulation}%
  \BibitemOpen
  \bibfield  {author} {\bibinfo {author} {\bibfnamefont {R.}~\bibnamefont
  {Legin}}, \bibinfo {author} {\bibfnamefont {Y.}~\bibnamefont {Hezaveh}},
  \bibinfo {author} {\bibfnamefont {L.~P.}\ \bibnamefont {Levasseur}},\ and\
  \bibinfo {author} {\bibfnamefont {B.}~\bibnamefont {Wandelt}},\ }\href@noop
  {} {\bibfield  {journal} {\bibinfo  {journal} {arXiv preprint
  arXiv:2112.05278}\ } (\bibinfo {year} {2021})}\BibitemShut {NoStop}%
\bibitem [{\citenamefont {Wagner-Carena}\ \emph {et~al.}(2021)\citenamefont
  {Wagner-Carena}, \citenamefont {Park}, \citenamefont {Birrer}, \citenamefont
  {Marshall}, \citenamefont {Roodman}, \citenamefont {Wechsler}, \citenamefont
  {Collaboration} \emph {et~al.}}]{wagner2021hierarchical}%
  \BibitemOpen
  \bibfield  {author} {\bibinfo {author} {\bibfnamefont {S.}~\bibnamefont
  {Wagner-Carena}}, \bibinfo {author} {\bibfnamefont {J.~W.}\ \bibnamefont
  {Park}}, \bibinfo {author} {\bibfnamefont {S.}~\bibnamefont {Birrer}},
  \bibinfo {author} {\bibfnamefont {P.~J.}\ \bibnamefont {Marshall}}, \bibinfo
  {author} {\bibfnamefont {A.}~\bibnamefont {Roodman}}, \bibinfo {author}
  {\bibfnamefont {R.~H.}\ \bibnamefont {Wechsler}}, \bibinfo {author}
  {\bibfnamefont {L.~D. E.~S.}\ \bibnamefont {Collaboration}}, \emph {et~al.},\
  }\href@noop {} {\bibfield  {journal} {\bibinfo  {journal} {The Astrophysical
  Journal}\ }\textbf {\bibinfo {volume} {909}},\ \bibinfo {pages} {187}
  (\bibinfo {year} {2021})}\BibitemShut {NoStop}%
\bibitem [{\citenamefont {Dax}\ \emph {et~al.}(2021)\citenamefont {Dax},
  \citenamefont {Green}, \citenamefont {Gair}, \citenamefont {Macke},
  \citenamefont {Buonanno},\ and\ \citenamefont {Sch\"olkopf}}]{Dax2021}%
  \BibitemOpen
  \bibfield  {author} {\bibinfo {author} {\bibfnamefont {M.}~\bibnamefont
  {Dax}}, \bibinfo {author} {\bibfnamefont {S.~R.}\ \bibnamefont {Green}},
  \bibinfo {author} {\bibfnamefont {J.}~\bibnamefont {Gair}}, \bibinfo {author}
  {\bibfnamefont {J.~H.}\ \bibnamefont {Macke}}, \bibinfo {author}
  {\bibfnamefont {A.}~\bibnamefont {Buonanno}},\ and\ \bibinfo {author}
  {\bibfnamefont {B.}~\bibnamefont {Sch\"olkopf}},\ }\href
  {https://doi.org/10.1103/PhysRevLett.127.241103} {\bibfield  {journal}
  {\bibinfo  {journal} {Phys. Rev. Lett.}\ }\textbf {\bibinfo {volume} {127}},\
  \bibinfo {pages} {241103} (\bibinfo {year} {2021})}\BibitemShut {NoStop}%
\bibitem [{\citenamefont {Bhardwaj}\ \emph {et~al.}(2023)\citenamefont
  {Bhardwaj}, \citenamefont {Alvey}, \citenamefont {Miller}, \citenamefont
  {Nissanke},\ and\ \citenamefont {Weniger}}]{Bhardwaj2024}%
  \BibitemOpen
  \bibfield  {author} {\bibinfo {author} {\bibfnamefont {U.}~\bibnamefont
  {Bhardwaj}}, \bibinfo {author} {\bibfnamefont {J.}~\bibnamefont {Alvey}},
  \bibinfo {author} {\bibfnamefont {B.~K.}\ \bibnamefont {Miller}}, \bibinfo
  {author} {\bibfnamefont {S.}~\bibnamefont {Nissanke}},\ and\ \bibinfo
  {author} {\bibfnamefont {C.}~\bibnamefont {Weniger}},\ }\href
  {https://doi.org/10.1103/PhysRevD.108.042004} {\bibfield  {journal} {\bibinfo
   {journal} {Phys. Rev. D}\ }\textbf {\bibinfo {volume} {108}},\ \bibinfo
  {pages} {042004} (\bibinfo {year} {2023})}\BibitemShut {NoStop}%
\bibitem [{\citenamefont {Cheung}\ \emph {et~al.}(2022)\citenamefont {Cheung},
  \citenamefont {Wong}, \citenamefont {Hannuksela}, \citenamefont {Li},\ and\
  \citenamefont {Ho}}]{PhysRevD.106.083014}%
  \BibitemOpen
  \bibfield  {author} {\bibinfo {author} {\bibfnamefont {D.~H.~T.}\
  \bibnamefont {Cheung}}, \bibinfo {author} {\bibfnamefont {K.~W.~K.}\
  \bibnamefont {Wong}}, \bibinfo {author} {\bibfnamefont {O.~A.}\ \bibnamefont
  {Hannuksela}}, \bibinfo {author} {\bibfnamefont {T.~G.~F.}\ \bibnamefont
  {Li}},\ and\ \bibinfo {author} {\bibfnamefont {S.}~\bibnamefont {Ho}},\
  }\href {https://doi.org/10.1103/PhysRevD.106.083014} {\bibfield  {journal}
  {\bibinfo  {journal} {Phys. Rev. D}\ }\textbf {\bibinfo {volume} {106}},\
  \bibinfo {pages} {083014} (\bibinfo {year} {2022})}\BibitemShut {NoStop}%
\bibitem [{\citenamefont {Alvey}\ \emph {et~al.}(2024)\citenamefont {Alvey},
  \citenamefont {Bhardwaj}, \citenamefont {Domcke}, \citenamefont {Pieroni},\
  and\ \citenamefont {Weniger}}]{Alvey2024}%
  \BibitemOpen
  \bibfield  {author} {\bibinfo {author} {\bibfnamefont {J.}~\bibnamefont
  {Alvey}}, \bibinfo {author} {\bibfnamefont {U.}~\bibnamefont {Bhardwaj}},
  \bibinfo {author} {\bibfnamefont {V.}~\bibnamefont {Domcke}}, \bibinfo
  {author} {\bibfnamefont {M.}~\bibnamefont {Pieroni}},\ and\ \bibinfo {author}
  {\bibfnamefont {C.}~\bibnamefont {Weniger}},\ }\href
  {https://doi.org/10.1103/PhysRevD.109.083008} {\bibfield  {journal} {\bibinfo
   {journal} {Phys. Rev. D}\ }\textbf {\bibinfo {volume} {109}},\ \bibinfo
  {pages} {083008} (\bibinfo {year} {2024})}\BibitemShut {NoStop}%
\bibitem [{\citenamefont {{Villaescusa-Navarro}}\ \emph
  {et~al.}(2020)\citenamefont {{Villaescusa-Navarro}}, \citenamefont {{Hahn}},
  \citenamefont {{Massara}}, \citenamefont {{Banerjee}}, \citenamefont
  {{Delgado}}, \citenamefont {{Ramanah}}, \citenamefont {{Charnock}},
  \citenamefont {{Giusarma}}, \citenamefont {{Li}}, \citenamefont {{Allys}},
  \citenamefont {{Brochard}}, \citenamefont {{Uhlemann}}, \citenamefont
  {{Chiang}}, \citenamefont {{He}}, \citenamefont {{Pisani}}, \citenamefont
  {{Obuljen}}, \citenamefont {{Feng}}, \citenamefont {{Castorina}},
  \citenamefont {{Contardo}}, \citenamefont {{Kreisch}}, \citenamefont
  {{Nicola}}, \citenamefont {{Alsing}}, \citenamefont {{Scoccimarro}},
  \citenamefont {{Verde}}, \citenamefont {{Viel}}, \citenamefont {{Ho}},
  \citenamefont {{Mallat}}, \citenamefont {{Wandelt}},\ and\ \citenamefont
  {{Spergel}}}]{Quijote_sims}%
  \BibitemOpen
  \bibfield  {author} {\bibinfo {author} {\bibfnamefont {F.}~\bibnamefont
  {{Villaescusa-Navarro}}}, \bibinfo {author} {\bibfnamefont {C.}~\bibnamefont
  {{Hahn}}}, \bibinfo {author} {\bibfnamefont {E.}~\bibnamefont {{Massara}}},
  \bibinfo {author} {\bibfnamefont {A.}~\bibnamefont {{Banerjee}}}, \bibinfo
  {author} {\bibfnamefont {A.~M.}\ \bibnamefont {{Delgado}}}, \bibinfo {author}
  {\bibfnamefont {D.~K.}\ \bibnamefont {{Ramanah}}}, \bibinfo {author}
  {\bibfnamefont {T.}~\bibnamefont {{Charnock}}}, \bibinfo {author}
  {\bibfnamefont {E.}~\bibnamefont {{Giusarma}}}, \bibinfo {author}
  {\bibfnamefont {Y.}~\bibnamefont {{Li}}}, \bibinfo {author} {\bibfnamefont
  {E.}~\bibnamefont {{Allys}}}, \bibinfo {author} {\bibfnamefont
  {A.}~\bibnamefont {{Brochard}}}, \bibinfo {author} {\bibfnamefont
  {C.}~\bibnamefont {{Uhlemann}}}, \bibinfo {author} {\bibfnamefont {C.-T.}\
  \bibnamefont {{Chiang}}}, \bibinfo {author} {\bibfnamefont {S.}~\bibnamefont
  {{He}}}, \bibinfo {author} {\bibfnamefont {A.}~\bibnamefont {{Pisani}}},
  \bibinfo {author} {\bibfnamefont {A.}~\bibnamefont {{Obuljen}}}, \bibinfo
  {author} {\bibfnamefont {Y.}~\bibnamefont {{Feng}}}, \bibinfo {author}
  {\bibfnamefont {E.}~\bibnamefont {{Castorina}}}, \bibinfo {author}
  {\bibfnamefont {G.}~\bibnamefont {{Contardo}}}, \bibinfo {author}
  {\bibfnamefont {C.~D.}\ \bibnamefont {{Kreisch}}}, \bibinfo {author}
  {\bibfnamefont {A.}~\bibnamefont {{Nicola}}}, \bibinfo {author}
  {\bibfnamefont {J.}~\bibnamefont {{Alsing}}}, \bibinfo {author}
  {\bibfnamefont {R.}~\bibnamefont {{Scoccimarro}}}, \bibinfo {author}
  {\bibfnamefont {L.}~\bibnamefont {{Verde}}}, \bibinfo {author} {\bibfnamefont
  {M.}~\bibnamefont {{Viel}}}, \bibinfo {author} {\bibfnamefont
  {S.}~\bibnamefont {{Ho}}}, \bibinfo {author} {\bibfnamefont {S.}~\bibnamefont
  {{Mallat}}}, \bibinfo {author} {\bibfnamefont {B.}~\bibnamefont
  {{Wandelt}}},\ and\ \bibinfo {author} {\bibfnamefont {D.~N.}\ \bibnamefont
  {{Spergel}}},\ }\href {https://doi.org/10.3847/1538-4365/ab9d82} {\bibfield
  {journal} {\bibinfo  {journal} {Astrophysical Journal s}\ }\textbf {\bibinfo
  {volume} {250}},\ \bibinfo {eid} {2} (\bibinfo {year} {2020})},\ \Eprint
  {https://arxiv.org/abs/1909.05273} {arXiv:1909.05273} \BibitemShut {NoStop}%
\bibitem [{\citenamefont {Springel}(2005)}]{Springel2005TheCS}%
  \BibitemOpen
  \bibfield  {author} {\bibinfo {author} {\bibfnamefont {V.}~\bibnamefont
  {Springel}},\ }\href {https://api.semanticscholar.org/CorpusID:16378825}
  {\bibfield  {journal} {\bibinfo  {journal} {Monthly Notices of the Royal
  Astronomical Society}\ }\textbf {\bibinfo {volume} {364}},\ \bibinfo {pages}
  {1105} (\bibinfo {year} {2005})}\BibitemShut {NoStop}%
\bibitem [{\citenamefont {Desjacques}\ \emph {et~al.}(2018)\citenamefont
  {Desjacques}, \citenamefont {Jeong},\ and\ \citenamefont
  {Schmidt}}]{desjacques2018large}%
  \BibitemOpen
  \bibfield  {author} {\bibinfo {author} {\bibfnamefont {V.}~\bibnamefont
  {Desjacques}}, \bibinfo {author} {\bibfnamefont {D.}~\bibnamefont {Jeong}},\
  and\ \bibinfo {author} {\bibfnamefont {F.}~\bibnamefont {Schmidt}},\
  }\href@noop {} {\bibfield  {journal} {\bibinfo  {journal} {Physics reports}\
  }\textbf {\bibinfo {volume} {733}},\ \bibinfo {pages} {1} (\bibinfo {year}
  {2018})}\BibitemShut {NoStop}%
\bibitem [{\citenamefont {Kaiser}(1987)}]{kaiser1987clustering}%
  \BibitemOpen
  \bibfield  {author} {\bibinfo {author} {\bibfnamefont {N.}~\bibnamefont
  {Kaiser}},\ }\href@noop {} {\bibfield  {journal} {\bibinfo  {journal}
  {Monthly Notices of the Royal Astronomical Society}\ }\textbf {\bibinfo
  {volume} {227}},\ \bibinfo {pages} {1} (\bibinfo {year} {1987})}\BibitemShut
  {NoStop}%
\bibitem [{\citenamefont {Hamilton}(1998)}]{hamilton1998linear}%
  \BibitemOpen
  \bibfield  {author} {\bibinfo {author} {\bibfnamefont {A.}~\bibnamefont
  {Hamilton}},\ }\href@noop {} {\bibfield  {journal} {\bibinfo  {journal} {The
  Evolving Universe: Selected Topics on Large-Scale Structure and on the
  Properties of Galaxies}\ ,\ \bibinfo {pages} {185}} (\bibinfo {year}
  {1998})}\BibitemShut {NoStop}%
\bibitem [{\citenamefont {Scoccimarro}(2004)}]{scoccimarro2004redshift}%
  \BibitemOpen
  \bibfield  {author} {\bibinfo {author} {\bibfnamefont {R.}~\bibnamefont
  {Scoccimarro}},\ }\href@noop {} {\bibfield  {journal} {\bibinfo  {journal}
  {Physical Review D—Particles, Fields, Gravitation, and Cosmology}\ }\textbf
  {\bibinfo {volume} {70}},\ \bibinfo {pages} {083007} (\bibinfo {year}
  {2004})}\BibitemShut {NoStop}%
\bibitem [{\citenamefont {Hockney}\ and\ \citenamefont
  {Eastwood}(2021)}]{hockney2021computer}%
  \BibitemOpen
  \bibfield  {author} {\bibinfo {author} {\bibfnamefont {R.~W.}\ \bibnamefont
  {Hockney}}\ and\ \bibinfo {author} {\bibfnamefont {J.~W.}\ \bibnamefont
  {Eastwood}},\ }\href@noop {} {\emph {\bibinfo {title} {Computer simulation
  using particles}}}\ (\bibinfo  {publisher} {crc Press},\ \bibinfo {year}
  {2021})\BibitemShut {NoStop}%
\bibitem [{\citenamefont {{Villaescusa-Navarro}}(2018)}]{Pylians}%
  \BibitemOpen
  \bibfield  {author} {\bibinfo {author} {\bibfnamefont {F.}~\bibnamefont
  {{Villaescusa-Navarro}}},\ }\href@noop {} {\bibinfo {title} {{Pylians: Python
  libraries for the analysis of numerical simulations}}},\ \bibinfo
  {howpublished} {Astrophysics Source Code Library, record ascl:1811.008}
  (\bibinfo {year} {2018}),\ \Eprint {https://arxiv.org/abs/1811.008}
  {ascl:1811.008} \BibitemShut {NoStop}%
\bibitem [{\citenamefont {Feldman}\ \emph {et~al.}(1993)\citenamefont
  {Feldman}, \citenamefont {Kaiser},\ and\ \citenamefont
  {Peacock}}]{feldman1993power}%
  \BibitemOpen
  \bibfield  {author} {\bibinfo {author} {\bibfnamefont {H.~A.}\ \bibnamefont
  {Feldman}}, \bibinfo {author} {\bibfnamefont {N.}~\bibnamefont {Kaiser}},\
  and\ \bibinfo {author} {\bibfnamefont {J.~A.}\ \bibnamefont {Peacock}},\
  }\href@noop {} {\bibfield  {journal} {\bibinfo  {journal} {arXiv preprint
  astro-ph/9304022}\ } (\bibinfo {year} {1993})}\BibitemShut {NoStop}%
\bibitem [{\citenamefont {Schmalzing}\ and\ \citenamefont
  {Buchert}(1997)}]{schmalzing1997beyond}%
  \BibitemOpen
  \bibfield  {author} {\bibinfo {author} {\bibfnamefont {J.}~\bibnamefont
  {Schmalzing}}\ and\ \bibinfo {author} {\bibfnamefont {T.}~\bibnamefont
  {Buchert}},\ }\href@noop {} {\bibfield  {journal} {\bibinfo  {journal} {The
  Astrophysical Journal}\ }\textbf {\bibinfo {volume} {482}},\ \bibinfo {pages}
  {L1} (\bibinfo {year} {1997})}\BibitemShut {NoStop}%
\bibitem [{\citenamefont {Beisbart}\ \emph {et~al.}(2002)\citenamefont
  {Beisbart}, \citenamefont {Dahlke}, \citenamefont {Mecke},\ and\
  \citenamefont {Wagner}}]{beisbart2002vector}%
  \BibitemOpen
  \bibfield  {author} {\bibinfo {author} {\bibfnamefont {C.}~\bibnamefont
  {Beisbart}}, \bibinfo {author} {\bibfnamefont {R.}~\bibnamefont {Dahlke}},
  \bibinfo {author} {\bibfnamefont {K.}~\bibnamefont {Mecke}},\ and\ \bibinfo
  {author} {\bibfnamefont {H.}~\bibnamefont {Wagner}},\ }in\ \href@noop {}
  {\emph {\bibinfo {booktitle} {Morphology of Condensed Matter: Physics and
  Geometry of Spatially Complex Systems}}}\ (\bibinfo  {publisher} {Springer},\
  \bibinfo {year} {2002})\ pp.\ \bibinfo {pages} {238--260}\BibitemShut
  {NoStop}%
\bibitem [{\citenamefont {Trotta}(2008)}]{trotta2008bayes}%
  \BibitemOpen
  \bibfield  {author} {\bibinfo {author} {\bibfnamefont {R.}~\bibnamefont
  {Trotta}},\ }\href@noop {} {\bibfield  {journal} {\bibinfo  {journal}
  {Contemporary Physics}\ }\textbf {\bibinfo {volume} {49}},\ \bibinfo {pages}
  {71} (\bibinfo {year} {2008})}\BibitemShut {NoStop}%
\bibitem [{\citenamefont {Papamakarios}\ and\ \citenamefont
  {Murray}(2016)}]{papamakarios2016fast}%
  \BibitemOpen
  \bibfield  {author} {\bibinfo {author} {\bibfnamefont {G.}~\bibnamefont
  {Papamakarios}}\ and\ \bibinfo {author} {\bibfnamefont {I.}~\bibnamefont
  {Murray}},\ }\href@noop {} {\bibfield  {journal} {\bibinfo  {journal}
  {Advances in neural information processing systems}\ }\textbf {\bibinfo
  {volume} {29}} (\bibinfo {year} {2016})}\BibitemShut {NoStop}%
\bibitem [{\citenamefont {Papamakarios}\ \emph {et~al.}(2021)\citenamefont
  {Papamakarios}, \citenamefont {Nalisnick}, \citenamefont {Rezende},
  \citenamefont {Mohamed},\ and\ \citenamefont
  {Lakshminarayanan}}]{papamakarios2021normalizing}%
  \BibitemOpen
  \bibfield  {author} {\bibinfo {author} {\bibfnamefont {G.}~\bibnamefont
  {Papamakarios}}, \bibinfo {author} {\bibfnamefont {E.}~\bibnamefont
  {Nalisnick}}, \bibinfo {author} {\bibfnamefont {D.~J.}\ \bibnamefont
  {Rezende}}, \bibinfo {author} {\bibfnamefont {S.}~\bibnamefont {Mohamed}},\
  and\ \bibinfo {author} {\bibfnamefont {B.}~\bibnamefont {Lakshminarayanan}},\
  }\href@noop {} {\bibfield  {journal} {\bibinfo  {journal} {Journal of Machine
  Learning Research}\ }\textbf {\bibinfo {volume} {22}},\ \bibinfo {pages} {1}
  (\bibinfo {year} {2021})}\BibitemShut {NoStop}%
\bibitem [{\citenamefont {Durkan}\ \emph {et~al.}(2019)\citenamefont {Durkan},
  \citenamefont {Bekasov}, \citenamefont {Murray},\ and\ \citenamefont
  {Papamakarios}}]{durkan2019neural}%
  \BibitemOpen
  \bibfield  {author} {\bibinfo {author} {\bibfnamefont {C.}~\bibnamefont
  {Durkan}}, \bibinfo {author} {\bibfnamefont {A.}~\bibnamefont {Bekasov}},
  \bibinfo {author} {\bibfnamefont {I.}~\bibnamefont {Murray}},\ and\ \bibinfo
  {author} {\bibfnamefont {G.}~\bibnamefont {Papamakarios}},\ }\href@noop {}
  {\bibfield  {journal} {\bibinfo  {journal} {Advances in neural information
  processing systems}\ }\textbf {\bibinfo {volume} {32}} (\bibinfo {year}
  {2019})}\BibitemShut {NoStop}%
\bibitem [{\citenamefont {Tejero-Cantero}\ \emph {et~al.}(2020)\citenamefont
  {Tejero-Cantero}, \citenamefont {Boelts}, \citenamefont {Deistler},
  \citenamefont {Lueckmann}, \citenamefont {Durkan}, \citenamefont
  {Gon{\c{c}}alves}, \citenamefont {Greenberg},\ and\ \citenamefont
  {Macke}}]{tejero2020sbi}%
  \BibitemOpen
  \bibfield  {author} {\bibinfo {author} {\bibfnamefont {A.}~\bibnamefont
  {Tejero-Cantero}}, \bibinfo {author} {\bibfnamefont {J.}~\bibnamefont
  {Boelts}}, \bibinfo {author} {\bibfnamefont {M.}~\bibnamefont {Deistler}},
  \bibinfo {author} {\bibfnamefont {J.-M.}\ \bibnamefont {Lueckmann}}, \bibinfo
  {author} {\bibfnamefont {C.}~\bibnamefont {Durkan}}, \bibinfo {author}
  {\bibfnamefont {P.~J.}\ \bibnamefont {Gon{\c{c}}alves}}, \bibinfo {author}
  {\bibfnamefont {D.~S.}\ \bibnamefont {Greenberg}},\ and\ \bibinfo {author}
  {\bibfnamefont {J.~H.}\ \bibnamefont {Macke}},\ }\href@noop {} {\bibfield
  {journal} {\bibinfo  {journal} {arXiv preprint arXiv:2007.09114}\ } (\bibinfo
  {year} {2020})}\BibitemShut {NoStop}%
\bibitem [{\citenamefont {Akiba}\ \emph {et~al.}(2019)\citenamefont {Akiba},
  \citenamefont {Sano}, \citenamefont {Yanase}, \citenamefont {Ohta},\ and\
  \citenamefont {Koyama}}]{akiba2019optuna}%
  \BibitemOpen
  \bibfield  {author} {\bibinfo {author} {\bibfnamefont {T.}~\bibnamefont
  {Akiba}}, \bibinfo {author} {\bibfnamefont {S.}~\bibnamefont {Sano}},
  \bibinfo {author} {\bibfnamefont {T.}~\bibnamefont {Yanase}}, \bibinfo
  {author} {\bibfnamefont {T.}~\bibnamefont {Ohta}},\ and\ \bibinfo {author}
  {\bibfnamefont {M.}~\bibnamefont {Koyama}},\ }in\ \href@noop {} {\emph
  {\bibinfo {booktitle} {Proceedings of the 25th ACM SIGKDD international
  conference on knowledge discovery \& data mining}}}\ (\bibinfo {year}
  {2019})\ pp.\ \bibinfo {pages} {2623--2631}\BibitemShut {NoStop}%
\bibitem [{\citenamefont {Talts}\ \emph {et~al.}(2018)\citenamefont {Talts},
  \citenamefont {Betancourt}, \citenamefont {Simpson}, \citenamefont
  {Vehtari},\ and\ \citenamefont {Gelman}}]{talts2018validating}%
  \BibitemOpen
  \bibfield  {author} {\bibinfo {author} {\bibfnamefont {S.}~\bibnamefont
  {Talts}}, \bibinfo {author} {\bibfnamefont {M.}~\bibnamefont {Betancourt}},
  \bibinfo {author} {\bibfnamefont {D.}~\bibnamefont {Simpson}}, \bibinfo
  {author} {\bibfnamefont {A.}~\bibnamefont {Vehtari}},\ and\ \bibinfo {author}
  {\bibfnamefont {A.}~\bibnamefont {Gelman}},\ }\href@noop {} {\bibfield
  {journal} {\bibinfo  {journal} {arXiv preprint arXiv:1804.06788}\ } (\bibinfo
  {year} {2018})}\BibitemShut {NoStop}%
\bibitem [{\citenamefont {Tirapongprasert}\ and\ \citenamefont
  {Ho}(2026)}]{tirapongprasert2026learning}%
  \BibitemOpen
  \bibfield  {author} {\bibinfo {author} {\bibfnamefont {C.}~\bibnamefont
  {Tirapongprasert}}\ and\ \bibinfo {author} {\bibfnamefont {M.}~\bibnamefont
  {Ho}},\ }\href@noop {} {\bibfield  {journal} {\bibinfo  {journal} {arXiv
  preprint arXiv:2601.17120}\ } (\bibinfo {year} {2026})}\BibitemShut {NoStop}%
\bibitem [{\citenamefont {{Lewis}}\ \emph {et~al.}(2000)\citenamefont
  {{Lewis}}, \citenamefont {{Challinor}},\ and\ \citenamefont
  {{Lasenby}}}]{Lewis:1999bs}%
  \BibitemOpen
  \bibfield  {author} {\bibinfo {author} {\bibfnamefont {A.}~\bibnamefont
  {{Lewis}}}, \bibinfo {author} {\bibfnamefont {A.}~\bibnamefont
  {{Challinor}}},\ and\ \bibinfo {author} {\bibfnamefont {A.}~\bibnamefont
  {{Lasenby}}},\ }\href {https://doi.org/10.1086/309179} {\bibfield  {journal}
  {\bibinfo  {journal} {\apj}\ }\textbf {\bibinfo {volume} {538}},\ \bibinfo
  {pages} {473} (\bibinfo {year} {2000})},\ \Eprint
  {https://arxiv.org/abs/astro-ph/9911177} {arXiv:astro-ph/9911177 [astro-ph]}
  \BibitemShut {NoStop}%
\bibitem [{\citenamefont {Vogeley}\ \emph {et~al.}(1994)\citenamefont
  {Vogeley}, \citenamefont {Park}, \citenamefont {Geller}, \citenamefont
  {Huchra},\ and\ \citenamefont {Gott~III}}]{vogeley1994topological}%
  \BibitemOpen
  \bibfield  {author} {\bibinfo {author} {\bibfnamefont {M.~S.}\ \bibnamefont
  {Vogeley}}, \bibinfo {author} {\bibfnamefont {C.}~\bibnamefont {Park}},
  \bibinfo {author} {\bibfnamefont {M.~J.}\ \bibnamefont {Geller}}, \bibinfo
  {author} {\bibfnamefont {J.~P.}\ \bibnamefont {Huchra}},\ and\ \bibinfo
  {author} {\bibfnamefont {J.~R.}\ \bibnamefont {Gott~III}},\ }\href@noop {}
  {\bibfield  {journal} {\bibinfo  {journal} {The Astrophysical Journal, Part 1
  (ISSN 0004-637X), vol. 420, no. 2, p. 525-544}\ }\textbf {\bibinfo {volume}
  {420}},\ \bibinfo {pages} {525} (\bibinfo {year} {1994})}\BibitemShut
  {NoStop}%
\bibitem [{\citenamefont {Gott~III}\ \emph {et~al.}(1989)\citenamefont
  {Gott~III}, \citenamefont {Miller}, \citenamefont {Thuan}, \citenamefont
  {Schneider}, \citenamefont {Weinberg}, \citenamefont {Gammie}, \citenamefont
  {Polk}, \citenamefont {Vogeley}, \citenamefont {Jeffrey}, \citenamefont
  {Bhavsar} \emph {et~al.}}]{gott1989topology}%
  \BibitemOpen
  \bibfield  {author} {\bibinfo {author} {\bibfnamefont {J.~R.}\ \bibnamefont
  {Gott~III}}, \bibinfo {author} {\bibfnamefont {J.}~\bibnamefont {Miller}},
  \bibinfo {author} {\bibfnamefont {T.~X.}\ \bibnamefont {Thuan}}, \bibinfo
  {author} {\bibfnamefont {S.~E.}\ \bibnamefont {Schneider}}, \bibinfo {author}
  {\bibfnamefont {D.~H.}\ \bibnamefont {Weinberg}}, \bibinfo {author}
  {\bibfnamefont {C.}~\bibnamefont {Gammie}}, \bibinfo {author} {\bibfnamefont
  {K.}~\bibnamefont {Polk}}, \bibinfo {author} {\bibfnamefont {M.}~\bibnamefont
  {Vogeley}}, \bibinfo {author} {\bibfnamefont {S.}~\bibnamefont {Jeffrey}},
  \bibinfo {author} {\bibfnamefont {S.~P.}\ \bibnamefont {Bhavsar}}, \emph
  {et~al.},\ }\href@noop {} {\bibfield  {journal} {\bibinfo  {journal}
  {Astrophysical Journal, Part 1 (ISSN 0004-637X), vol. 340, May 15, 1989, p.
  625-646.}\ }\textbf {\bibinfo {volume} {340}},\ \bibinfo {pages} {625}
  (\bibinfo {year} {1989})}\BibitemShut {NoStop}%
\bibitem [{\citenamefont {Hikage}\ \emph {et~al.}(2003)\citenamefont {Hikage},
  \citenamefont {Schmalzing}, \citenamefont {Buchert}, \citenamefont {Suto},
  \citenamefont {Kayo}, \citenamefont {Taruya}, \citenamefont {Vogeley},
  \citenamefont {Hoyle}, \citenamefont {Gott~III},\ and\ \citenamefont
  {Brinkmann}}]{hikage2003minkowski}%
  \BibitemOpen
  \bibfield  {author} {\bibinfo {author} {\bibfnamefont {C.}~\bibnamefont
  {Hikage}}, \bibinfo {author} {\bibfnamefont {J.}~\bibnamefont {Schmalzing}},
  \bibinfo {author} {\bibfnamefont {T.}~\bibnamefont {Buchert}}, \bibinfo
  {author} {\bibfnamefont {Y.}~\bibnamefont {Suto}}, \bibinfo {author}
  {\bibfnamefont {I.}~\bibnamefont {Kayo}}, \bibinfo {author} {\bibfnamefont
  {A.}~\bibnamefont {Taruya}}, \bibinfo {author} {\bibfnamefont {M.~S.}\
  \bibnamefont {Vogeley}}, \bibinfo {author} {\bibfnamefont {F.}~\bibnamefont
  {Hoyle}}, \bibinfo {author} {\bibfnamefont {J.~R.}\ \bibnamefont
  {Gott~III}},\ and\ \bibinfo {author} {\bibfnamefont {J.}~\bibnamefont
  {Brinkmann}},\ }\href@noop {} {\bibfield  {journal} {\bibinfo  {journal}
  {Publications of the Astronomical Society of Japan}\ }\textbf {\bibinfo
  {volume} {55}},\ \bibinfo {pages} {911} (\bibinfo {year} {2003})}\BibitemShut
  {NoStop}%
\bibitem [{\citenamefont {Liu}\ \emph {et~al.}(2025{\natexlab{b}})\citenamefont
  {Liu}, \citenamefont {Paillas}, \citenamefont {Cuesta-Lazaro}, \citenamefont
  {Valogiannis},\ and\ \citenamefont {Fang}}]{liu2025cosmological}%
  \BibitemOpen
  \bibfield  {author} {\bibinfo {author} {\bibfnamefont {W.}~\bibnamefont
  {Liu}}, \bibinfo {author} {\bibfnamefont {E.}~\bibnamefont {Paillas}},
  \bibinfo {author} {\bibfnamefont {C.}~\bibnamefont {Cuesta-Lazaro}}, \bibinfo
  {author} {\bibfnamefont {G.}~\bibnamefont {Valogiannis}},\ and\ \bibinfo
  {author} {\bibfnamefont {W.}~\bibnamefont {Fang}},\ }\href@noop {} {\bibfield
   {journal} {\bibinfo  {journal} {Journal of Cosmology and Astroparticle
  Physics}\ }\textbf {\bibinfo {volume} {2025}}\bibinfo  {number} { (05)},\
  \bibinfo {pages} {064}}\BibitemShut {NoStop}%
\end{thebibliography}
%

\end{document}